\documentclass[letter,11pt]{article}
\usepackage{jheppub_2} 

\usepackage{xcolor}
\usepackage[draft]{changes} 

\definecolor{Flamingo}{rgb}{0.94,0.32,0.23}
\definecolor{Neptuno}{rgb}{0.5,0.75,0.74}
\definecolor{Color}{rgb}{0.28, 0.24, 0.55}
\definecolor{Orange}{RGB}{255, 108, 12}

\hypersetup{
    colorlinks = true,
    citecolor = Orange,
    linkcolor = Color,
    urlcolor  = Orange,
}

\def\SU{\rm{SU}}

\usepackage{placeins}
\usepackage[normalem]{ulem}

\usepackage{tikz}
\usetikzlibrary{"arrows", "automata", "backgrounds", "calendar", "chains", "matrix", "mindmap", "patterns", "petri", "shadows", "shapes.geometric", "shapes.misc", "spy", "trees"}
\usetikzlibrary{arrows,shapes}
\usetikzlibrary{trees}
\usetikzlibrary{matrix,arrows} 				
\usetikzlibrary{positioning}	
\usetikzlibrary{calc,through}
\usetikzlibrary{decorations.pathreplacing}  
\usepackage{pgffor}

\usetikzlibrary{decorations.pathmorphing}
\usetikzlibrary{decorations.markings}
\tikzset{
    vector/.style={decorate, decoration={snake}, draw},
	provector/.style={decorate, decoration={snake,amplitude=2.5pt}, draw},
	antivector/.style={decorate, decoration={snake,amplitude=-2.5pt}, draw},
    fermion/.style={draw=black, postaction={decorate},
        decoration={markings,mark=at position .55 with {\arrow[draw=black]{>}}}},
    fermioncyan/.style={draw=black, postaction={decorate},
        decoration={markings,mark=at position .55 with {\arrow[draw=cyan]{<}}}},
        fermionpp/.style={draw=black, postaction={decorate},
        decoration={markings,mark=at position 0.27 with {\arrow[draw=black]{>}}}},
        fermionp/.style={draw=black, postaction={decorate},
        decoration={markings,mark=at position .78 with {\arrow[draw=black]{>}}}},
    fermionppp/.style={draw=black, postaction={decorate},
        decoration={markings,mark=at position 1.0 with {\arrow[draw=black]{>}}}},
    fermiondif/.style={draw=black, postaction={decorate},
        decoration={markings,mark=at position .7 with {\arrow[draw=black]{>}}}},
            fermiondif2/.style={draw=black, postaction={decorate},
        decoration={markings,mark=at position .7 with {\arrow[draw=black]{<}}}},
         scalarloop/.style={dashed,draw=black, postaction={decorate},
        decoration={markings,mark=at position .75 with {\arrow[draw=black]{<}}}},
            scalarloop1/.style={dashed,draw=black, postaction={decorate},
        decoration={markings,mark=at position .25 with {\arrow[draw=black]{<}}}},
    fermionend/.style={draw=black, postaction={decorate},
        decoration={markings,mark=at position 1 with {\arrow[draw=black]{>}}}},
    fermionuchannel2/.style={draw=black, postaction={decorate},
        decoration={markings,mark=at position .4 with {\arrow[draw=black]{>}}}},
    scalardif/.style={dashed,draw=black, postaction={decorate},
        decoration={markings,mark=at position .7 with {\arrow[draw=black]{>}}}},
    scalarend/.style={dashed,draw=black, postaction={decorate},
        decoration={markings,mark=at position 1 with {\arrow[draw=black]{>}}}},
    fermionbar/.style={draw=black, postaction={decorate},
        decoration={markings,mark=at position .55 with {\arrow[draw=black]{<}}}},
    fermiondif/.style={draw=black, postaction={decorate},
        decoration={markings,mark=at position .7 with {\arrow[draw=black]{>}}}},
            fermiondif2/.style={draw=black, postaction={decorate},
        decoration={markings,mark=at position .7 with {\arrow[draw=black]{<}}}},
    fermionend/.style={draw=black, postaction={decorate},
        decoration={markings,mark=at position 1 with {\arrow[draw=black]{>}}}},
    fermionuchannel2/.style={draw=black, postaction={decorate},
        decoration={markings,mark=at position .4 with {\arrow[draw=black]{>}}}},
    scalardif/.style={dashed,draw=black, postaction={decorate},
        decoration={markings,mark=at position .7 with {\arrow[draw=black]{>}}}},
    scalarend/.style={dashed,draw=black, postaction={decorate},
        decoration={markings,mark=at position 1 with {\arrow[draw=black]{>}}}},
    fermionbar/.style={draw=black, postaction={decorate},
        decoration={markings,mark=at position .55 with {\arrow[draw=black]{<}}}},
    fermionnoarrow/.style={draw=black},
    gluon/.style={decorate, draw=black,
        decoration={coil,amplitude=4pt, segment length=5pt}},
    scalar/.style={dashed,draw=black, postaction={decorate},
        decoration={markings,mark=at position .55 with {\arrow[draw=black]{>}}}},
    scalarcyan/.style={dashed,draw=black, postaction={decorate},
        decoration={markings,mark=at position .55 with {\arrow[draw=cyan]{>}}}},
    scalaruchannel1/.style={dashed,draw=black, postaction={decorate},
        decoration={markings,mark=at position .7 with {\arrow[draw=black]{>}}}},
                  scalaruchannel2/.style={dashed,draw=black, postaction={decorate},
        decoration={markings,mark=at position .4 with {\arrow[draw=black]{>}}}},
    scalarbar/.style={dashed,draw=black, postaction={decorate},
        decoration={markings,mark=at position .55 with {\arrow[draw=black]{<}}}},
    scalarnoarrow/.style={dashed,draw=black},
    electron/.style={draw=black, postaction={decorate},
        decoration={markings,mark=at position .55 with {\arrow[draw=black]{>}}}},
	bigvector/.style={decorate, decoration={snake,amplitude=4pt}, draw},
}

\tikzstyle{block} = [draw, rectangle, 
    minimum height=3em, minimum width=6em]

\usetikzlibrary{fit}
\tikzset{%
  highlight/.style={rectangle,rounded corners,color=granate,draw,text opacity =1,
    fill opacity=0.5,thick,inner sep=0pt}
}

\usepackage{placeins}

\tikzset{
    cross/.pic = {
    \draw[rotate = 45] (-#1,0) -- (#1,0);
    \draw[rotate = 45] (0,-#1) -- (0, #1);
    }
}

\tikzset{
    square/.style={%
        draw=none,
        circle,
        append after command={%
            \pgfextra \draw[#1] (\tikzlastnode.north-|\tikzlastnode.west) rectangle 
                (\tikzlastnode.south-|\tikzlastnode.east);\endpgfextra}
    },
    square/.default=black
}

\tikzstyle{block} = [draw, rectangle, 
    minimum height=3em, minimum width=6em]

\usepackage{xparse}
\NewDocumentCommand\semiloop{O{black}mmmO{}O{above}}
{%
\draw[#1] let \p1 = ($(#3)-(#2)$) in (#3) arc (#4:({#4+180}):({0.5*veclen(\x1,\y1)})node[midway, #6] {#5};)
}

\NewDocumentCommand\hello{O{black}mmmO{}O{above}}
{%
\draw[#1] let \p1 = ($(#3)-(#2)$) in (#3) arc (#4:({#4-180}):({0.5*veclen(\x1,\y1)})node[midway, #6] {#5};)
}

\title{\boldmath Leptogenesis in automatic Nelson-Barr models}

\author[a]{C. Murgui,}
\author[b]{S. Patrone}
\affiliation[a]{Theoretical Physics Department, CERN, 1 Esplanade des Particules, CH-1211 Geneva 23, Switzerland}
\affiliation[b]{Walter Burke Institute for Theoretical Physics,
California Institute of Technology, Pasadena, CA 91125}
\emailAdd{spatrone@caltech.edu}
\emailAdd{clara.murgui@cern.ch}

\abstract{
In this study, we numerically show that automatic Nelson-Barr models with new chiral fermions can simultaneously solve the strong CP problem and generate the observed baryon asymmetry via high-scale leptogenesis. In these models, all CP violation arises from a single spontaneous symmetry-breaking scale, linking the origin of quark and lepton CP phases. Using conservative assumptions and minimal dynamics, we identify a viable parameter window where successful leptogenesis occurs without spoiling the quality of the strong CP solution. Models with vector-like fermions face tension within this leptogenesis scenario. A key prediction is a correlation between the baryon asymmetry and the induced QCD vacuum angle shift. Remarkably, we find that the majority of the available parameter space is within reach of current and future nucleon EDM experiments.
}

\preprint{CALT-TH/2025-020, CERN-TH-2025-118} 

 \notoc
\begin{document}
\maketitle
\flushbottom

\newpage 
\section{Introduction}
\label{sec:intro}
In spite of the enormous success of the Standard Model (SM) of particle physics in explaining nature, there is experimental evidence that cannot be accommodated within the model and calls for an extension either in field content, gauge symmetries, or both. Among those, measurements of the Cosmic Microwave Background  (CMB)~\cite{Planck:2018vyg} and Big Bang Nucleosynthesis (BBN)~\cite{Cyburt:2015mya,Workman:2022ynf} manifest a primordial baryon asymmetry
\begin{equation}\label{eq:etaB}
   \eta_B  \equiv\frac{n^{0}_B-n^{0}_{\bar B}}{n^{0}_\gamma}=  (6.12 \pm 0.04) \times 10^{-10},
\end{equation} 
where $n^{0}_B$ ($n^{0}_{\bar B}$) and $n^{0}_\gamma$ are the (anti-)baryon and photon number density today, respectively. 
A mechanism is required to dynamically generate this amount of baryon asymmetry, which demands ingredients beyond the SM~\cite{Sakharov:1967dj}. It is noteworthy that the same new physics may also play a role in generating neutrino masses. At least two generations of active neutrinos with nondegenerate mass, in particular between 0.04 and 0.12 eV~\cite{Gonzalez-Garcia:2021dve,Planck:2018vyg}, are a requisite to address the neutrino oscillation data~\cite{SNO:2002tuh, Super-Kamiokande:1998kpq, KamLAND:2002uet}. These oscillations reflect lepton flavor violation, parametrized by the Pontecorvo-Maki-Nakagawa-Sakata (PMNS) matrix. Analogously, in the quark sector, the weak mixing is described by the Cabibbo-Kobayashi-Maskawa (CKM) matrix, which contains an imaginary phase that violates a combination of charge-parity (CP) and parity (P) symmetries and, in general, contributes to the base-independent $\bar \theta_{\rm QCD}$ entering in the nucleon electric dipole moment (EDM),
\begin{equation}\label{eq:thetabar}
    \bar \theta_{\rm QCD} = \theta_{\rm QCD} + \text{arg}\{ \text{Det}({\cal M}_u {\cal M}_d ) \} < 10^{-10}.
\end{equation}
Here, ${\cal M}_u$ and ${\cal M}_d$ are the mass matrices of the up-type and down-type quarks, respectively, in the weak basis, and  $\theta_{\rm QCD}$ is the QCD vacuum angle, sourced by the topological term  
${\cal L}_{\rm QCD} \supset \alpha_s \theta_{\rm QCD} G_{\mu \nu}^a \tilde G^{\mu \nu a} / 8 \pi $. The constraint on the right-hand side comes from null measurements of the neutron EDM~\cite{Abel:2020pzs,Pospelov:1999mv,Liang:2023jfj}. Eq.~\eqref{eq:thetabar} defines the so-called strong CP problem\footnote{A cancellation between bare parameters in the Lagrangian may not be seen as a problem given that these parameters are nonphysical~\cite{Manohar:2018aog}. However, we expect the smallness of $\bar\theta_{\rm QCD}$ to be a consequence of a physical mechanism.}.

Among the different beyond the SM mechanisms proposed to solve the aforementioned issue, two well-known candidates stand out for their simplicity: the Peccei-Quinn (PQ) mechanism~\cite{Peccei:1977hh, Peccei:1977ur} and the Nelson-Barr (NB) mechanism~\cite{Nelson:1983zb,Barr:1984qx,Nelson:1984hg}. In the PQ mechanism a pseudo-Goldstone boson, called the axion~\cite{Wilczek:1977pj,Weinberg:1977ma,Kim:1979if,Shifman:1979if,Dine:1981rt,Zhitnitsky:1980tq}, associated to an anomalous global abelian symmetry (referred as PQ symmetry) dynamically sets $\bar \theta_{\rm QCD}=0$ when it relaxes to the minimum of its potential, generated non-perturbatively by QCD instantons. Nelson-Barr models, on the contrary, start by assuming that CP is an exact symmetry of the Lagrangian, spontaneously broken by the true vacuum of the theory. Both mechanisms involve new scalar degrees of freedom and, in the case of the KSVZ axion~\cite{Kim:1979if,Shifman:1979if}, new colored fermions as well. The QCD axion generally suffers from a 
severe quality problem~\cite{Georgi:1981pu}. Since gravitational effects are not expected to respect global symmetries, in order to preserve an acceptably small $\bar \theta_{\rm QCD}$, gauge invariant non-renormalizable operators violating the PQ symmetry of dimension equal to or higher than eleven must be forbidden by the symmetries and matter content of the ultraviolet (UV) theory~\cite{Kamionkowski:1992mf,Holman:1992us,Bonnefoy:2022vop}\footnote{This does not apply when the PQ symmetry emerges from a higher dimensional gauge field as in the string axiverse~\cite{Witten:1984dg,Wen:1985qj,Svrcek:2006yi}. In modern language, this is understood in terms of the generalized global symmetries they possess~\cite{Reece:2024wrn,Craig:2024dnl,Agrawal:2024ejr}.}. Nelson-Barr models are easier to protect against gravitational effects and their quality is straightforwardly preserved in theories where the NB mechanism is an automatic consequence of the gauge symmetry and matter content~\cite{Asadi:2022vys,Perez:2023zin}. We refer to this kind of models as {\it automatic NB models}. Last but not least, PQ models provide automatically a cold dark matter candidate~\cite{Preskill:1982cy,Abbott:1982af,Dine:1982ah} while NB models, in principle, do not\footnote{Ref.~\cite{Dine:2024bxv}, for instance, shows how ultralight dark matter can arise in NB models with a shallow scalar potential.}. Nevertheless, while PQ models are uncorrelated with the generation of a baryon asymmetry\footnote{Attempts were made by Kuzmin, Tkachev, and Shaposhnikov~\cite{Kuzmin:1992up}, leading to negative conclusions. These were followed up, for example, by Servant~\cite{Servant:2014bla}, considering supercooled phase transitions, or by Co and Harigaya~\cite{Co:2019wyp}, considering the kinetic energy of the rotation of the axion.}, baryogenesis is, to some extent, parasitic to Nelson-Barr models, as we aim to reinforce in this paper.

A necessary condition for baryogenesis is the existence of a CP violating phase~\cite{Sakharov:1967dj}, which must be large enough to account for the observed asymmetry. In models where CP is spontaneously broken, the Lagrangian is real and the only contribution to a non-zero CP phase comes from the vacuum of the theory. It is therefore crucial to examine the dynamical generation of a baryon asymmetry in the context of models with spontaneous CP violation (SCPV)~\cite{Lee:1973iz}. 

In NB models, the scale of SCPV is constrained from above by the need to preserve the quality of the mechanism, and therefore cannot be arbitrarily large. However, it cannot be arbitrarily low either, as inflation (or any alternative mechanism solving the horizon problem) must sufficiently dilute the domain walls associated with the spontaneous breaking of a gauged space-time symmetry~\cite{McNamara:2022lrw}. Consequently, baryogenesis must occur within a specific energy window.

In this work, we study the viability of generating the observed baryon asymmetry in the context of automatic Nelson-Barr models. In these models, the CP phase from the vacuum plays a two-fold role: it generates the CKM phase~\cite{ParticleDataGroup:2020ssz} and the baryon asymmetry via leptogenesis. 
We assume that the NB mechanism operates similarly in both the quark and lepton sectors. This assumption is further motivated by expected new physics in the lepton sector -- given the necessity of a mass generation mechanism for neutrinos -- as well as recent data from the T2K~\cite{T2K:2023smv}  and NO$\nu$A~\cite{NOvA:2021nfi,NOvA:2023iam} experiments, which suggest a nonzero CP phase in the PMNS matrix\footnote{The joint fit of T2K and NO$\nu$A shows a mild preference for the inverted ordering, in which case a null CP phase is excluded at $3\sigma$~\cite{Mikola:2024jnj}.}.

Several studies in the literature have explored the generation of a baryon asymmetry within the framework of Nelson-Barr models. Branco and Parada extended the minimal implementation of the NB mechanism~\cite{Bento:1991ez} to incorporate the lepton sector by using a discrete $\mathbb{Z}_4$ symmetry~\cite{Branco:2003rt}. However, as noted in Ref.~\cite{Asadi:2022vys}, such simplified models face tension with the thermal leptogenesis scenario. To address this, Ref.~\cite{Murai:2024alz} relaxed the quality bound on the SCPV scale by introducing, on top of the previous $\mathbb{Z}_4$, an extra approximate $\mathbb{Z}_4$. Ref.~\cite{Suematsu:2023jqa} explored a $\mathbb{Z}_4 \times \mathbb{Z}'_4$, supplemented by additional scalars beyond the vector-like fermions necessary for the mechanism, thereby enabling a significantly lower reheat temperature. Instead, we focus on automatic Nelson-Barr models, where the required texture of the fermion mass matrices naturally arises as a consequence of a gauge symmetry.

The paper is organized as follows. In Sec.~\ref{sec:NelsonBarr} we describe the NB mechanism in the context of the minimal (simplified) model that can host it. In Sec.~\ref{sec:upperboundCP}, the quality of the mechanism is discussed and different upper bounds on the SCPV scale are derived, depending on the nature of the theory. In Sec.~\ref{sec:leptogenesis}, we extend the Nelson-Barr mechanism to the lepton sector and study the viability of the model in terms of the quality of the mechanism and the ability to generate the observed baryon asymmetry. In Sec.~\ref{sec:results}, we perform a numerical scan to explore the feasible regions of parameter space and the correlations between the relevant parameters. In Sec.~\ref{sec:signatures}, we emphasize a connection between leptogenesis and $\bar{\theta}_{\rm QCD}$ that can offer a pathway to experimentally probe these theories in the near future. We draw our conclusions in Sec.~\ref{sec:conclusions}. Supplementary materials are provided in the appendices.

\section{The Nelson-Barr mechanism in a Nutshell}
\label{sec:NelsonBarr}
SCPV allows for an alternative solution to the strong CP-problem without invoking a pseudogoldstone boson at low energies - the axion. The key idea of SCPV models is to assume that CP is an exact (gauge) symmetry of nature: this implies that there exists a basis in which the UV Lagrangian can be written in terms of real-valued parameters and, consequently, $\bar{\theta}_{\rm QCD} = 0$ at high energies. In order to reproduce the observed order-one phase of the CKM matrix, complex scalars beyond the SM Higgs are then added to the theory whose complex vacuum expectation value (vev) breaks dynamically CP. A further mechanism - such as a discrete symmetry - has to be introduced, as it was first noted by Nelson~\cite{Nelson:1983zb} and generalized by Barr~\cite{Barr:1984qx}, to enforce that $\text{arg} \{ \text{Det} ({\cal M}_u {\cal M}_d)\}$ in Eq.~\eqref{eq:thetabar} vanishes, thus preventing these complex phases to spoil $\bar{\theta}_{\rm QCD}=0$ at tree-level.

To review how the Nelson-Barr mechanism works, we adopt the simplified model proposed by Bento, Branco and Parada (BBP) in Ref.~\cite{Bento:1991ez}. Within this model, NB criteria are achieved by considering an extended version of the SM, where the only additional fields are a vector-like down-type quark, $D_{L,R} \sim (3,1,-1/3)$, and a complex scalar singlet $X$, both odd under a further $\mathbb{Z}_2$ symmetry, under which the SM fields are even.
The most generic renormalizable Lagrangian for the down-type quark mass terms can be written as
\begin{equation}
\label{eq:genericNB}
-{\cal L} \supset  \mathsf{Y}_d^{ij} \, \bar Q_L^i H d_R^j + (\lambda_d^i X+ \lambda^{\prime i}_d X^*)\bar D_L d_R^i + m_D \bar D_L D_R + \text{h.c.},
\end{equation}
where $i,j=1,2,3$ are family indices; $H\sim(1,2,1/2)$, $Q_L\sim (3,2,1/6)$, and $d_R\sim(3,1,-1/3)$ are, respectively, the SM Higgs doublet, left-handed and right-handed quarks; $\mathsf{Y}_d$ and $\lambda_d\neq\lambda^{\prime}_d$ are Yukawa couplings encoded in a matrix and a vector, respectively, and $m_D$ is the allowed Dirac mass for the additional vector-like quark. Assuming CP-invariance, we can always find a basis in which all the above couplings are real.

For a range of parameters in the Higgs potential, the new scalar field can acquire a complex vev of the form $\langle X\rangle=v_{\rm CP} \, e^{\text{i}\alpha}/\sqrt{2}$, which breaks spontaneously CP and defines the scale of SCPV, $v_{\rm CP}$.
Proceeding further, below the electroweak scale, the Higgs acquires a vev\footnote{The complex phase associated to the SM Higgs vacuum can be rotated away by the residual electroweak symmetry in the scalar potential.}, $\langle H\rangle=v_{H}/\sqrt{2}$, leading to the following mass matrix for the down-type quarks:
\begin{equation}\label{eq:M44}
    \mathcal{M}_d^{AB}=
    \begin{pmatrix}
    \mathsf{M}_d^{ij} &  0 \\
     \mu_i &  m_D
    \end{pmatrix}~.
  \end{equation}  
Above, $\mathcal{M}_d^{AB}$ is written in block form, with the indices $A,B = 1,...,4$; $i,j=1,2,3$, and 
    \begin{eqnarray}
    \mathsf{M}_d^{ij} =&& \mathsf{Y}_d^{ij} v_H/\sqrt{2}, \\
    \mu_i=&&(\lambda_{d,i}\,  e^{\text{i}\alpha} + \lambda^{\prime}_{d,i}  e^{-\text{i}\alpha}) v_{\rm CP}/\sqrt{2}\,\nonumber.
    \end{eqnarray}
The crucial observation is that, due to the structure of the matrix, $\text{arg}\{ \text{Det} \, \mathcal{M}_d\}=0$, implying $\bar\theta_{\rm QCD}=0$ at tree-level.
Notice that, if $\lambda_d=\lambda^\prime_d$, all the entries in the mass matrix would become real, preventing the phase from surviving and getting propagated at low energies.

The mass matrix $\mathcal{M}_d$ can be diagonalized using a bi-unitary transformation such that $U_L^\dagger\mathcal{M}_d U_R=\text{diag}( \mathsf{\widehat  M}_d,\bar m_D)$, where $\mathsf{\widehat M}_d$ is the diagonal $3\times3$ matrix of the SM down-quark masses and $\bar m_D$ is the physical mass of the new vector-like quark, given by $\bar m_D^2 \simeq m_D^2+\sum_{i=1}^3 |\mu_i|^2$. This relation holds when the mass of the new vector-like down quark is much heavier than the standard model down-quarks, already guaranteed by the gap between the bottom mass and bounds from direct searches on new colored vector-like quarks~\cite{Alves:2023ufm}. 
Inverting $U_L^\dagger {\cal M}_d {\cal M}_d^\dagger U_L = \text{diag}( \mathsf{\widehat M}_d^2, \bar m_D^2)$, we can write 
\begin{equation}
\label{eq:masslight}
    (\mathsf{V}\mathsf{\widehat M}_d^2 \mathsf{V}^\dagger)_{ij} = \left(\mathsf{M}_d\right)_{ik}\left(\delta_{k\ell}+\frac{\mu^
*_k \mu_\ell}{\bar{m}_D^2}\right)\left(\mathsf{M}^T_d\right)_{\ell j}+\mathcal{O}([\mathsf{M}_d]^4[\mu]^2/\bar m_D^4)\,.
\end{equation}
where $\mathsf{V}$ is the $3\times 3$ upper-left block of $U_L$. $\mathsf{V}$ is nothing but the contribution from the down-quark sector to the CKM matrix, i.e. $\mathsf{V}\equiv\mathsf{V}_{\rm CKM}$ if the up-quark mass matrix is diagonal in the weak basis. Here and thereafter, we use the notation $[a]\equiv\mathcal{O}(|a_{i}|)$ and $[A]\equiv\mathcal{O}(|A_{ij}|)$ for any vector $a$ or matrix $A$.

We observe that the $3\times 3$ effective mass matrix for the SM quarks in the right-hand side of the above equation is approximately diagonalized by the matrix $\mathsf{V}$, which contains an order-one phase if $[\mu] \sim \bar m_D$. $\mathsf{V}$ is unitary up to ${\cal O}([\textsf{M}_d]^2 [\mu]^2/ \bar m_D^4) < 10^{-6}$. Furthermore, according to the analysis done in Ref.~\cite{Valenti:2021rdu}, $2 \lesssim [\mu] / m_D \ll 10^3$ is required in order to simultaneously reproduce the CKM phase and the SM quark masses.

\section{Quality of Nelson-Barr Solution}
\label{sec:upperboundCP}

One criticism of the Nelson-Barr mechanism is its sensitivity to the physics in the ultraviolet~\cite{Dine:2015jga} (which is known in the literature as quality issues), namely the contribution of higher-dimensional operators and radiative corrections to $\bar \theta_{\rm QCD}$ that can spoil the null tree-level result. We discuss them in what follows.

\subsection{Higher dimensional operators}

Higher-dimensional operators, ultimately suppressed by the Planck scale, are expected to induce shifts in the QCD vacuum angle.
The effect on $\bar \theta_{\rm QCD}$ from corrections to the tree-level mass matrix of Eq.\,\eqref{eq:M44}, $\delta {\cal M}_d$, can be parametrized as\footnote{See, for example, the appendix in Ref.\,\cite{FileviezPerez:2023rxn} for a derivation.}
\begin{equation}\label{eq:shiftThetaQCD}
\begin{split}
\Delta \bar \theta_{\rm QCD} &= \text{Im} \{ \text{Tr}\{ {\cal M}_d^{-1} \delta {\cal M}_d\} \}  + {\cal O}( \text{Tr}\{ {\cal M}_d^{-1}\delta {\cal M}_d \}^2) \\
& \simeq   \frac{ [\text{Im}\{\delta {\cal M}_d^{ij}\}]}{ [\mathsf{M}_d] } + \frac{[\text{Im}\{\delta {\cal M}_d^{44}\}]}{m_D} +\, \frac{[\mu]}{m_D}\frac{[\text{Im}\{e^{\text{i}\alpha}\delta{\cal M}_d^{i4}\}]}{[\mathsf{M}_d]}\,,
\end{split}
\end{equation}
where we assumed $\mu_i = e^{\text{i}\alpha}[\mu]$. It is worth noting that $\delta {\cal M}_d^{i4}$ could be real and still contribute to $\Delta \bar \theta_{\rm QCD}$ as indicated above, while $\delta {\cal M}_d^{4i}$ does not contribute at leading order. 
The terms $ \delta {\cal M}_d^{AB}$ in Eq.~\eqref{eq:shiftThetaQCD} contain phases coming from the vev $\langle X \rangle$. Since we are concerned about order of magnitude estimates of the quality bounds, we will take $[\text{Im}\{ \delta {\cal M}^{AB}\}] \sim [\delta {\cal M}^{AB}]$: this is motivated by knowing that ${\cal O}(1)$ phases are needed in order to reproduce the measured Jarlskog determinant~\cite{ParticleDataGroup:2020ssz}, by assuming no strong phase cancellations.

In the following, we start discussing the BBP simplified model~\cite{Bento:1991ez} (see Sec.~\ref{sec:NelsonBarr}). As we will show, the low quality of this solution motivates us to consider automatic Nelson-Barr models, where the role of the $\mathbb{Z}_2$ symmetry is taken on by an extra (spontaneously broken) gauge symmetry in the UV~\cite{Asadi:2022vys,FileviezPerez:2023rxn}.

\paragraph{Imposed $\mathbb{Z}_2$ symmetry.}
Consider the Lagrangian in Eq.~\eqref{eq:genericNB}, which represents the minimal implementation of the NB mechanism. In this context, the following dimension-five operators are allowed (among others),
\begin{equation}\label{eq:hdimop}
    \frac{1}{\Lambda} \bar Q_L H D_R ( \xi_1 X + \xi_2 X^*) + \frac{1}{\Lambda} \bar D_L D_R \big ( \xi_3 X^2 + \xi_4 (X^*)^2 \big)  + \text{h.c.},
\end{equation}
where $\Lambda$ is an ultraviolet scale. 
In the broken phase, the above operators contribute to the down-quark mass matrix, potentially compromising the tree-level prediction $\bar \theta_{\rm QCD} = 0$.
Using Eq.~\eqref{eq:shiftThetaQCD}, the shift in $\bar \theta_{\rm QCD}$ can be parametrized as
\begin{equation}
\Delta \bar \theta_{\rm QCD} \simeq [\xi]
\frac{v_{\rm CP}}{\Lambda}\left(\frac{[\mu]}{m_D} \,\frac{1}{ [\mathsf{Y}_d]} +\frac{v_{\rm CP}}{m_D}\right),
\end{equation}
where $[\mathsf{Y}_d]=[\mathsf{M}_d]/v_H$, and
$[\xi]$ quantifies the magnitude of the Wilson coefficients in Eq.~\eqref{eq:hdimop}. We take $[\xi] = 1$ in the following estimates.
As we argued in Sec.~\ref{sec:NelsonBarr}, to generate a quark mass spectrum consistent with experiment, $m_D \sim [\mu]$~\cite{Valenti:2021rdu}. Note that, as long as $m_D / v_{\rm CP} > [\mathsf{Y}_d]$, the dominant contribution to $\bar \theta_{\rm QCD}$ comes from the operators proportional to $\xi_{1,2}$ in Eq.~\eqref{eq:hdimop}. 
In the following, we will make the conservative choice\footnote{One could, for example, adopt the assumption of minimal flavor violation~\cite{Chivukula:1987py,DAmbrosio:2002vsn}. A spurion analysis would identify $[\xi] = [\mathsf{Y_d}]$, which would substantially relax the upper bound shown in Eq.~\eqref{eq:Z2bound}. We refer the reader to Ref.~\cite{Asadi:2022vys} for more details.} of taking $[\mathsf{Y}_d] = \sqrt{2} \, m_d / v_H = 3 \times 10^{-5}$. Then, $\Delta \bar \theta_{\rm QCD} < 10^{-10}$ implies that
\begin{equation}\label{eq:Z2bound}
    v_{\rm CP} \lesssim 40 \text{ TeV}\left(\frac{1}{[\xi] 
    } \right)\left(\frac{\Delta \bar \theta_{\rm QCD}}{10^{-10}}\right) \left(\frac{m_D / [\mu]}{1}\right)\left(\frac{\Lambda}{1.2 \times 10^{19} \text{ GeV}}\right),
\end{equation}
where we have considered the ultimate scenario where higher dimensional operators are suppressed by inverse powers of the Planck scale.

The spontaneous breaking of the CP symmetry generates domain walls~\cite{McNamara:2022lrw} that are not observed. One of the most popular mechanisms to dilute them is inflation. However, such a low upper bound is in tension with its simplest implementations (see Ref.~\cite{Asadi:2022vys} for a detailed discussion). One could consider particular scenarios where CP is not restored at high temperatures, which would prevent the formation of dangerous topological defects at the cost of a reduced parameter space in the scalar potential~\cite{Dvali:1995cc,Dvali:1996zr}. 
Alternatively, higher CP breaking scales compatible with the quality of the mechanism can be achieved if the role of the $\mathbb{Z}_2$ symmetry is replaced by a gauge symmetry in the UV. We discuss this scenario in the upcoming subsection.

 \paragraph{Automatic Nelson-Barr.}
 The tree-level zeroes in ${\cal M}_d$, see Eq.~\eqref{eq:M44}, play a crucial role in the implementation of the Nelson-Barr mechanism. Theories where these zeroes are an accident of the gauge symmetry group and matter content are referred to as {\it automatic Nelson-Barr models}.
 
 For simplicity, we extend the SM gauge group with an abelian $\text{U}(1)$\footnote{
 Non-abelian extensions could also enforce the required texture.}. The zeroes arise when ${\cal Q}_{D_R} \neq {\cal Q}_{d_R}$, being ${\cal Q}_f$ the gauge charge of the $f$ field under the new force. At least two new scalar fields are needed to generate a non-zero CP phase (see discussion in Refs.~\cite{Haber:2012np,Perez:2023zin}): we will use the subindex $a=1,2$ in $X_{a}$ to distinguish between them. The charge requirement ${\cal Q}_{X_a} = {\cal Q}_{D_L} - {\cal Q}_{d_R}$ must hold for the {\it bridge} term needed to implement the NB mechanism, $\bar D_L d_R X_a$, to be allowed. Since we work in a simplified model, anomaly cancellation in a UV completion may require additional charge constraints and/or extra fermions. 
 
 Generically, unless $|{\cal Q}_{X_a}| = |{\cal Q}_{D_R} - {\cal Q}_{d_R}|$, the non-renormalizable operators leading to the aggressive bound in Eq.~\eqref{eq:Z2bound} are forbidden. However, if $Q_{D_L} = Q_{D_R}$, dimension-five operators of the kind 
 \begin{equation}
   \frac{\xi_{ab}}{\Lambda}   \bar D_L D_R X_a^* X_b^{}\,,
 \end{equation}
 analogous to the operators
 weighted by $\xi_{3,4}$ in Eq.~\eqref{eq:hdimop}, still contribute to $\bar \theta_{\rm QCD}$. They lead to the following upper bound on the scale of CP violation
 \begin{equation}\label{eq:vCP10to9}
     v_{\rm CP} \lesssim 10^9 \text{ GeV}  \left(\frac{1}{[\xi]
     }\right)\left(\frac{\Delta \bar \theta_{\rm QCD}}{10^{-10}}\right) \left(\frac{m_D/v_{\rm CP}}{1}\right) \left(\frac{\Lambda}{1.2 \times 10^{19} \text{ GeV}}\right),
 \end{equation}
where we have taken the optimistic choice $m_D \sim v_{\rm CP}$, and $\Lambda$ to be the Planck scale.

 The bound above can be further relaxed if the SM vector-like new quarks are chiral under the new gauge symmetry (${\cal Q}_{D_L} \neq {\cal Q}_{D_R}$). In this case, $m_D$ can be associated to the breaking scale of the new gauge group, $v_S$. We assume $v_S \gtrsim v_{\rm CP}$. The leading corrections to the tree-level down-type quark mass matrix from non-renormalizable operators in this scenario occur at dimension-six, 
 \begin{equation}\label{eq:dim6}
     \frac{1}{\Lambda^2} X_a^* X_{b \neq a} \left( \xi_{1,ab} \bar D_L D_R S + \xi_{2,ab} \bar Q_L H d_R  \right) + \frac{\xi_{3,a}}{\Lambda^2} \bar Q_L H D_R X_a^* S + \text{h.c.},
 \end{equation}
 leading to 
 \begin{align}
 [\delta {\cal M}^{44}] &= \xi_{1,ab} \, v_{\rm CP}^2 \, v_S / \Lambda^2\, , \nonumber\\
 [\delta {\cal M}^{ij}] &= \xi_{2,ab}\,v_{\rm CP}^2  \, v_H / \Lambda^2\, , \nonumber\,\\
 [\delta {\cal M}^{i4}] &= \xi_{3,a} \,\,\,v_{\rm CP} \,  v_S \, v_H / \Lambda^2\,. \nonumber 
 \end{align}
 Here, $S$ is the scalar responsible for the breaking of the new gauge group. 
Remarkably, the last operator in Eq.~\eqref{eq:dim6} is always allowed in the presence of the bridge term, $\bar D_L d_R X_a$, due to the chiral nature of the new quarks under the new gauge symmetry.
The above contributions generate a shift to $\bar \theta_{\rm QCD}$, which according to Eq.~\eqref{eq:shiftThetaQCD} is given by
\begin{equation}\label{eq:shiftautomatic}
    \Delta \bar \theta_{\rm QCD} \simeq [\xi]  \frac{v_{\rm CP}^2}{\Lambda^2} \left[ \frac{1}{[\mathsf{Y}_d]} \left( 1 + \frac{[\mu]}{m_D} \frac{v_S}{v_{\rm CP}} \right) + \frac{v_S}{m_D} \right] .
\end{equation}
Since $v_S \gtrsim v_{\rm CP}$ and $[\mu] \sim m_D$, the second term between the round brackets is expected to be larger or equal than one. In order to derive an upper bound on the CP violating scale, let us assume that $v_{\rm CP} \sim v_{S}$, and $[\mu] / v_{\rm CP} < [\mathsf{Y}_d]$, which leads to 
\begin{equation} \label{eq:vCPbound}
v_{\rm CP} \lesssim 7 \times 10^{11} \text{ GeV} \left(\frac{\Delta \bar\theta_{\rm QCD}}{10^{-10}} \right)^{\tfrac{1}{2}}\left(\frac{1}{[\xi] } \right)^{\tfrac{1}{2}}  \left( \frac{v_{\rm CP}/v_S}{1}\right)^{\tfrac{1}{2}} \left( \frac{\Lambda}{1.2 \times 10^{19}\text{ GeV}}\right) .
\end{equation}
In general, the second term in the square brackets in Eq.~\eqref{eq:shiftautomatic} may dominate the shift in the angle, imposing a tighter bound. As we will show at the end of the next subsection, the finite naturalness criterion~\cite{Farina:2013mla,Perez:2023zin} requires the new down-quarks to be on the threshold of direct searches, $\bar m_D  \sim 2 \text{ TeV}$ (see Ref.~\cite{Alves:2023ufm} and references therein, and the discussion in the next subsection). The upper bound in this case reads
\begin{equation}
    v_{\rm CP} \lesssim 3 \times 10^{10} \text{ GeV} \left(\frac{\Delta \bar \theta_{\rm QCD}}{10^{-10}}\right)^{\tfrac{1}{3}} \left(\frac{1}{[\xi] }\right)^{\tfrac{1}{3}} \left(\frac{\Lambda}{1.2\times 10^{19} \text{ GeV}}\right)^{\tfrac{2}{3}} \left(\frac{m_D}{2 \text{ TeV}}\right)^{\tfrac{1}{3}}.
\end{equation}

\subsection{Radiative corrections}
\label{subsec:Radiative}

\begin{figure}[t]
\begin{equation*}
\begin{gathered}
\begin{tikzpicture}[line width=1.5 pt,node distance=1 cm and 1 cm]
\coordinate[label=left:$d_R$](dR);
\coordinate[right = 1 cm of dR](v1);
\coordinate[right = 2 cm of v1](v2);
\coordinate[right= 1 cm of v2,label=right:$d_L$](dL);
\coordinate[right= 1 cm of v1](vaux);
\coordinate[below =  0.75 cm of vaux,label=below:$\langle X_c \rangle$](Xk);
\coordinate[above = 1 cm of vaux](v4);
\coordinate[above left = 1 cm of v4,label=left:$\langle X_b \rangle$](Xi);
\coordinate[above right = 1 cm of v4,label=right:$\langle H \rangle$](Xj);
\coordinate[right = 0.5cm of v1,label=below:$D_L$](DLlabel);
\coordinate[right = 0.5cm of vaux,label=below:$d_R$](dRlabel);
\coordinate[above = 0.75 cm of v1,label=left:$X_a$](Xlabel);
\coordinate[above = 0.75 cm of v2,label=right:$H$](Xlabel);
\draw[scalarnoarrow] (v4)--(Xi);
\draw[scalarnoarrow] (v4)--(Xj);
\draw[fermion](dR)--(v1);
\draw[fermion](v1)--(vaux);
\draw[fermion](vaux)--(v2);
\draw[fermion](v2)--(dL);
\draw[scalarnoarrow](vaux)--(Xk);
\draw[fill=black] (v1) circle (.05cm);
\draw[fill=black] (v2) circle (.05cm);
\draw[fill=black] (vaux) circle (.05cm);
\draw[fill=black] (v4) circle (.05cm);
\semiloop[scalarloop]{v1}{v2}{0};
\semiloop[scalarloop1]{v1}{v2}{0};
\end{tikzpicture}
\end{gathered}
\qquad 
\begin{gathered}
\begin{tikzpicture}[line width=1.5 pt,node distance=1 cm and 1 cm]
\coordinate[label=left:$D_R$](DR);
\coordinate[right = 0.75 cm of DR](vS);
\coordinate[right = 0.5 cm of vS,label=below:$D_L$](labelDL);
\coordinate[above = 0.75 cm of vS,label=above:$\langle S \rangle$](S);
\coordinate[right = 1 cm of vS](v1);
\coordinate[right = 2 cm of v1](v2);
\coordinate[right= 0.75 cm of v2,label=right:$d_L$](dL);
\coordinate[right= 1.2 cm of v1,label=below:$d_{R}$](vaux);
\coordinate[below =  1 cm of vaux,label=below:$\text{}$](Xk);
\coordinate[above = 1 cm of vaux](v4);
\coordinate[above left = 1 cm of v4,label=left:$\langle X_b \rangle$](Xi);
\coordinate[above right = 1 cm of v4,label=right:$\langle H \rangle$](Xj);
\coordinate[above = 0.75 cm of v1,label=left:$X_a$](Xlabel);
\coordinate[above = 0.75 cm of v2,label=right:$H$](Xlabel);
\draw[scalarnoarrow] (v4)--(Xi);
\draw[scalarnoarrow] (v4)--(Xj);
\draw[fermion](dR)--(vS);
\draw[fermion](vS)--(v1);
\draw[fermion](v1)--(v2);
\draw[fermion](v2)--(dL);
\draw[scalarnoarrow](vS)--(S);
\draw[fill=black] (vS) circle (.05cm);
\draw[fill=black] (v1) circle (.05cm);
\draw[fill=black] (v2) circle (.05cm);
\draw[fill=black] (v4) circle (.05cm);
\semiloop[scalarloop]{v1}{v2}{0};
\semiloop[scalarloop1]{v1}{v2}{0};
\end{tikzpicture}
\end{gathered}
\end{equation*}
\begin{equation*}
\begin{gathered}
\begin{tikzpicture}[line width=1.5 pt,node distance=1 cm and 1 cm]
\coordinate[label=left:$H$] (i1);
\coordinate[below right = 1.5 cm of i1](v1);
\coordinate[below left = 1.5 cm of v1,label=left:$X_a$](i2);
\coordinate[above right = 1.5 cm of v1,label=right:$H$](i3);
\coordinate[below right = 1.5 cm of v1,label=right:$X_b$](i4);
\draw[scalar](i1)--(v1);
\draw[scalar](v1)--(i2);
\draw[scalar](v1)--(i3);
\draw[scalar](i4)--(v1);
\draw[fill=black] (v1) circle (.05cm);
\end{tikzpicture}
\end{gathered}  \ \  + \ \  
\begin{gathered}
\begin{tikzpicture}[line width=1.5 pt,node distance=1 cm and 1 cm]
\coordinate[label=left:$H$] (i1);
\coordinate[below left = 1.5 cm of v1,label=left:$X_a$](i2);
\coordinate[above right = 1.5 cm of v1,label=right:$H$](i3);
\coordinate[below right = 1.5 cm of v1,label=right:$X_b$](i4);
\coordinate[below right = 1.5 cm of i1](v1);
\coordinate[above = 0.40 cm of v1,label=$\,Q$](QL);
\coordinate[below = 0.90 cm of v1,label=$\,D$](UL);
\coordinate[below=0.2cm of v1](v1aux);
\coordinate[right = 0.6 cm of v1aux,label=$\,b$](tR);
\coordinate[left = 0.55 cm of v1aux,label=$\,b$](tR2);
\draw[scalar](0.35,-0.35)--(i1);
\draw[scalar](i2)--(0.35,-1.7);
\draw[scalar](1.85,-0.35)--(i3);
\draw[scalar](i4)--(1.85,-1.7);
\draw[fermion](1.1,-1) circle (1);
\draw[fermionp](1.1,-1) circle (1);
\draw[fermionpp](1.1,-1) circle (1);
\draw[fermionppp](1.1,-1) circle (1);
\draw[fill=black] (1.85,-1.7) circle (.05cm);
\draw[fill=black] (0.35,-1.7) circle (.05cm);
\draw[fill=black] (0.35,-0.35) circle (.05cm);
\draw[fill=black] (1.85,-0.35) circle (.05cm);
\end{tikzpicture}
\end{gathered}
 \left[ \ \ + \ \
\begin{gathered} 
\begin{tikzpicture}[line width=1.5 pt,node distance=1 cm and 1 cm]
\coordinate[label=left:$H$] (i1);
\coordinate[below left = 1.5 cm of v1,label=left:$X_a$](i2);
\coordinate[above right = 1.5 cm of v1,label=right:$H$](i3);
\coordinate[below right = 1.5 cm of v1,label=right:$X_b$](i4);
\coordinate[below right = 1.5 cm of i1](v1);
\coordinate[above = 0.40 cm of v1,label=$\,L$](QL);
\coordinate[below = 0.90 cm of v1,label=$\,N$](UL);
\coordinate[below=0.2cm of v1](v1aux);
\coordinate[right = 0.6 cm of v1aux,label=$\, \nu$](tR);
\coordinate[left = 0.55 cm of v1aux,label=$\, \nu$](tR2);
\draw[scalar](0.35,-0.35)--(i1);
\draw[scalar](i2)--(0.35,-1.7);
\draw[scalar](1.85,-0.35)--(i3);
\draw[scalar](i4)--(1.85,-1.7);
\draw[fermion](1.1,-1) circle (1);
\draw[fermionp](1.1,-1) circle (1);
\draw[fermionpp](1.1,-1) circle (1);
\draw[fermionppp](1.1,-1) circle (1);
\draw[fill=black] (1.85,-1.7) circle (.05cm);
\draw[fill=black] (0.35,-1.7) circle (.05cm);
\draw[fill=black] (0.35,-0.35) circle (.05cm);
\draw[fill=black] (1.85,-0.35) circle (.05cm);
\end{tikzpicture}
\end{gathered} \right]
\end{equation*}
\caption{Upper panel: Radiative corrections at 1-loop contributing to $\bar \theta_{\rm QCD}$ by inducing complex quark masses. Lower panel: Tree-level and 1-loop corrections to the quartic coupling $H^\dagger H X_a X_b^*$. In square brackets we add the would-be contribution from extending the Nelson-Barr mechanism to the neutral lepton sector (including right-handed neutrinos).}
\label{fig:deltaMqq}
\end{figure}
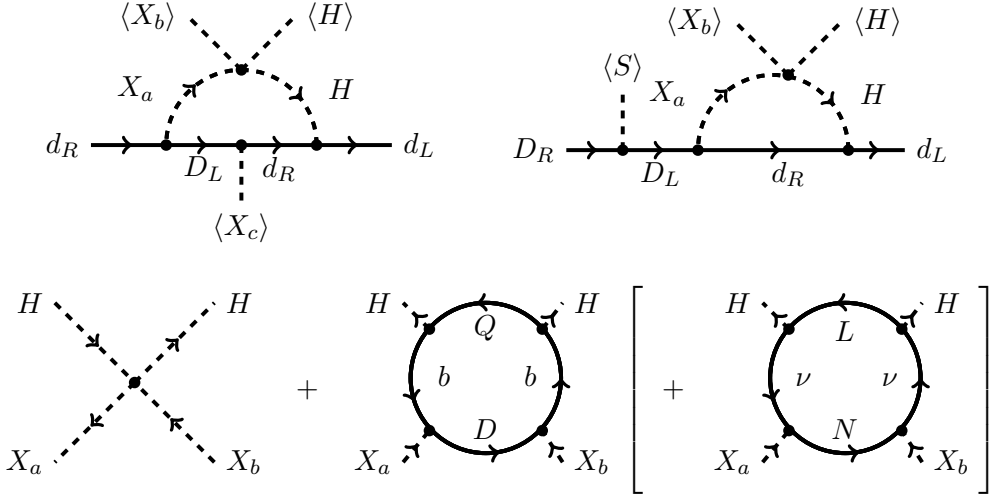
Quantum corrections start threatening the mechanism at one loop, via the diagrams displayed in the upper panel of Fig.~\ref{fig:deltaMqq}. These induce a shift in $\bar \theta_{\rm QCD}$ given by
\begin{equation}\label{eq:thetaQCDrad}
    \Delta \bar \theta_{\rm QCD} \sim \frac{\lambda_4}{16\pi^2} \frac{\bar m_D^2}{m_{X}^2}  \ln \left(\frac{m_{X}^2}{m_b^2} \right),
\end{equation}
where $\lambda_4$ is the quartic coupling weighting the $H^\dagger H X_a X_b^*$ ($a \neq b$) interaction. Here, we used $v_{\rm CP}$ as the scale for the mass of the $X$ fields, and the mass of the bottom quark as a subtraction scale.
$\Delta \bar \theta_{\rm QCD}  < 10^{-10}$ requires
\begin{equation}\label{eq:lambda4}
\lambda_4   \lesssim 2 \times 10^{-8} \left( \frac{v_{\rm CP}^2}{\bar m_D^2} \right) \frac{1}{\ln ( v_{\rm CP}^2 / m_b^2)}.
\end{equation}
After SCPV, $\lambda_4$ contributes to the Higgs mass, so that one expects that, at tree-level, $\lambda_4|_{\rm tree} < 10^{-12} (10^8 \text{ GeV} /v_{\rm CP})^2$.
 However, $\lambda_4$ cannot be arbitrarily small, as it will eventually be dominated by radiative corrections. The heavy down quarks contribute radiatively to the quartic coupling (see the bottom panel of Fig.~\ref{fig:deltaMqq}), inducing a shift of~\cite{Perez:2023zin} 
\begin{equation}
\label{eq:lam4oneloopnat}
    \delta \lambda_4 |_{\rm 1-loop} \simeq \frac{1}{16\pi^2}\frac{m_b^2}{v_H^2} \frac{\bar m_D^2}{v_{\rm CP}^2} \ln \left(\frac{\bar m_D^2}{m_b^2}\right).
\end{equation}
The finite naturalness criterion demands $\delta \lambda_4|_{\rm 1-loop}v_{\rm CP}^2  < m_H^2$, being $m_H$ the SM Higgs mass. As a consequence, $\bar m_D  \sim 2 \text{ TeV}$ as we already mentioned in the previous subsection. 
Eq.~\eqref{eq:lambda4} is still satisfied by the radiative quartic as long as $\bar m_D < 0.1 \, v_{\rm CP}$. One could read the latter condition as a lower bound on the CP violating scale, $v_{\rm CP} >20\text{ TeV}$.

\section{Leptogenesis}
\label{sec:leptogenesis}

In the previous sections we focused on the implementation of the NB mechanism in the quark sector of the SM. In general, one would expect that the UV completion hosting the NB mechanism acts \emph{democratically} on all fermions. In appendix~\ref{app:UV}, we show that leptons are naturally involved in the automatization of the NB mechanism. Moreover, recent data from the T2K~\cite{T2K:2023smv} and NO$\nu$A~\cite{NOvA:2021nfi,NOvA:2023iam} experiments hint towards a non-zero phase in the PMNS matrix, which motivates extending the NB mechanism to the lepton sector. In the following, we study whether the observed baryon asymmetry can be generated in the context of automatic Nelson-Barr models. We discuss the implications of successful leptogenesis in the available parameter space, considering that both quarks and leptons share the source of CP violation.

\subsection{A simplified model}
\label{sec:SimplifiedModel}
In the context of automatic Nelson-Barr models, the lagrangian in Eq.~\eqref{eq:genericNB} is promoted to
\begin{equation}
\label{eq:automaticNB}
-{\cal L} \supset   \bar Q_L \mathsf{Y}_d H d_R + \bar D_L (\lambda_{d1} X_1 + \lambda_{d2} X_2)d_R + \lambda_D S \bar D_L D_R + \text{h.c.}.
\end{equation}
Above, the fields are charged under a new $\text{U}(1)$ such that the zeroes in the down quark mass are guaranteed, which occurs when the new charges of the SM right-handed down quarks and the new $D_R$ are different, i.e. ${\cal Q}_{d_R} \neq {\cal Q}_{D_R}$. We note that the condition ${\cal Q}_X = {\cal Q}_{D_L}-{\cal Q}_{d_R}$ must also be required for the CKM matrix to inherit a complex phase. Furthermore, in order to relax the quality bound as discussed in Sec.~\ref{sec:upperboundCP}, we demand the new quarks to be chiral under the new $\text{U}(1)$; the charge condition ${\cal Q}_S = {\cal Q}_{D_L} - {\cal Q}_{D_R}$ allows them to get a mass proportional to $v_{S}$, the spontaneously symmetry-breaking scale of this new force.

In the broken phase, $\langle X_a \rangle = e^{i\alpha_a} v_{X_a} /\sqrt{2}$, $\langle S \rangle = v_S / \sqrt{2}$ and $\langle H \rangle = (0,v_H/\sqrt{2})^T$. The down-quark mass matrix is shown in Eq.~\eqref{eq:M44}, with $m_D = \lambda_D v_S / \sqrt{2}$, and $\mu_i$ now given by 
\begin{equation}
    \mu_i = (\lambda_{d1} e^{i\alpha_1} v_{X_1} + \lambda_{d2} e^{i \alpha_2} v_{X_2})/\sqrt{2}.
\end{equation}
Without loss of generality, we will assume that $v_{X_1} \sim v_{X_2} \sim v_{\rm CP}$. 
As in the BBP model presented in Sec.~\ref{sec:NelsonBarr}, here $\bar \theta = \text{arg} \{\det{\,\cal M}_d\}= 0$ at tree-level. In Sec.~\ref{sec:upperboundCP}, we extensively discussed the interplay between renormalizable operators and/or radiative corrections and the induced shifts in $\bar \theta$, examining under which conditions those are consistent with the current experimental bounds on the neutron electric dipole moment (nEDM).

In Appendix~\ref{app:UV}, we show how UV completions of the Lagrangian in Eq.~\eqref{eq:automaticNB} naturally involve leptons charged under the new force, and we discuss in detail some concrete charge assignments.

Mirroring the quark sector, in a bottom-up approach to the problem, we now identify the general requirements for the lepton sector. We therefore extend the SM with three right-handed neutrino singlets, $\nu_R$, and a single pair of neutral leptons, $N_L$ and $N_R$, vector-like under the SM. The relevant Lagrangian for the neutral leptons is given by
\begin{equation}\label{eq:Lagr}
\begin{split}
    - {\cal L}  &\supset  \bar L_L \mathsf{Y}_\nu \tilde H \nu_R + \bar L_L \lambda_\nu \tilde H N_R + \bar N_L  \mu \nu_R + m_N  \bar N_L N_R \\
    & \quad + \tfrac{1}{2} \overline{\nu_R^c} \,   \textsf{M}_R \nu_R  + \tfrac{1}{2} m_{L} \overline{N_L^c} N_L + \tfrac{1}{2}m_{R} \overline{N_R^c} N_R +   \tfrac{1}{2}\overline{N_R^c} m_{RR} \nu_R + {\rm h.c.},
    \end{split}
\end{equation}
where $\tilde H = i \sigma_2 H^*$. 
Assuming the same features of the quark sector, we require the charges of the fields under the new $\text{U}(1)$ gauge group to satisfy the following relations: 
\begin{itemize}
    \item[(i)] ${\cal Q}_{H}={\cal Q}_{\nu_R}-{\cal Q}_{L_L}$, so that SM neutrinos get a Dirac mass term with the three $\nu_R$; 
    \item[(ii)] ${\cal Q}_X= {\cal Q}_{N_L} - {\cal Q}_{\nu_R}$, to transfer the phase to the SM (complex PMNS matrix);
    \item[(iii)] ${\cal Q}_{N_R} \neq {\cal Q}_{\nu_R}$, to enforce the zeroes of the Nelson-Barr texture;
    \item[(iv)] $ {\cal Q}_{N_L} \neq {\cal Q}_{N_R}$, to avoid a vector-like mass term for $N_L$ and $N_R$.
\end{itemize}
Condition (ii) involves the scalar fields $X$ responsible for inducing a phase in the CKM matrix (quark sector). Regarding condition (iv), we will assume that the scalar responsible for the $\text{U}(1)$ breaking, $S$, singlet under the SM, has charge $|{\cal Q}_S| =| {\cal Q}_{N_L} - {\cal Q}_{N_R}|$ under the new symmetry, so that $N_L$ and $N_R$ get mass dynamically. 

Under (i)-(iv), in general, there are no Majorana mass terms allowed, and the previous Lagrangian, analogously to the quark sector, simply reads
\begin{equation}
\begin{split}
  -  {\cal L}  &\supset  \bar L_L \mathsf{Y}_\nu \tilde{H} \nu_R + \bar N_L   (\lambda_{\nu_1} X_1 + \lambda_{\nu_2} X_2) \nu_R+ \lambda_N S  \bar N_L N_R + {\rm h.c.}.
    \end{split}
\end{equation}
However, there are particular charge assignments leading to surviving Majorana masses. 
In the following, we assume that the $\nu_R$s get a dynamical Majorana mass from the spontaneous symmetry breaking of the new gauge group, while the rest of the Majorana mass terms are forbidden\footnote{Under particular choices of charges, other Majorana mass terms could also be allowed. For example, on top of $\mathsf{M}_R \neq \mathsf{0}$, $m_L \neq 0$ in Eq.~\eqref{eq:Lagr} if ${\cal Q}_S = - 2 {\cal Q}_{\nu_R}$ and ${\cal Q}_X = -2{\cal Q}_{N_L}$, while $m_R, \, m_{RR}\neq 0$ if ${\cal Q}_S = 2 {\cal Q}_{\nu_R}$ and ${\cal Q}_X = - 2 {\cal Q}_{N_R}$. In scenarios where the only Majorana mass is $m_L$, $m_R$ or $m_{RR}$, i.e. $|{\cal Q}_S| \neq 2 Q_{\nu_R}$, the rank of the heavy neutral lepton mass matrix is reduced, resulting in an increase in the number of light neutrinos. Further constraints need to be considered in such models to avoid spoiling the observed number of relativistic degrees of freedom, $N_{\rm eff}$~\cite{Planck:2018vyg}.}.
Coupling the SCPV fields $X_a$ as  above represents a conservative assumption in the sense that it does not introduce other complex parameters in the Lagrangian except the ones arising from the bridge term ($\bar N_L X_a \nu_R$). In fact, within this choice, $\mathsf{M}_R = \mathsf{Y}_R v_S$ is a real, symmetric matrix. The relevant Lagrangian for our study is then given by,
\begin{equation}\label{eq:LagrangianLepto}
\begin{split}
   - {\cal L}  &\supset  \bar L_L \mathsf{Y}_\nu \tilde H \nu_R + \bar N_L (\lambda_{\nu1} X_1 + \lambda_{\nu 2} X_2) \nu_R  + \lambda_N S  \bar N_L N_R + \tfrac{1}{2} S\,\overline{\nu_R^c} \,   \mathsf{Y}_R \nu_R  + {\rm h.c.},\\
  & \qquad \stackrel{\langle S \rangle, \langle X_{a}\rangle }{\rightarrow} \bar L_L \mathsf{Y}_\nu \tilde  H \nu_R +   \bar N_L \mu \, \nu_R  + m_N  \bar N_L N_R +  \tfrac{1}{2} \overline{\nu_R^c} \,   \mathsf{M}_R \nu_R  + {\rm h.c.},
    \end{split}
\end{equation}
where $m_N = \lambda_N \, v_S \in \mathbb{R}$, and $\mu = \lambda_{\nu 1} \langle X_1 \rangle + \lambda_{\nu 2} \langle X_2 \rangle \in \mathbb{C}$. From the perturbativity of the Yukawa couplings, the following inequalities hold $[\mathsf{M}_R], m_N \lesssim v_S$, and $[\mu] \lesssim v_{\rm CP} \leq v_S$. Eq.~\eqref{eq:LagrangianLepto} involves two heavy scales $v_S$ and $v_{\rm CP}$. Should condition (iv) be relaxed, an additional scale $m_N$, independent of $v_S$, would emerge.

One concrete example of a UV completion leading to the Lagrangians in Eqs.~\eqref{eq:automaticNB} and~\eqref{eq:LagrangianLepto} is the model discussed in Ref.~\cite{Perez:2023zin}, with $|{\cal Q}_S| = 2 |{\cal Q}_{\nu_R}|$.
In Appendix~\ref{app:UV}, we discuss this particular example, as well as other  explicit examples of UV theories that satisfy the  conditions (i)-(iv) in both the quark and lepton sectors.

The simplified model described above contains the ingredients for a type-I see-saw mechanism~\cite{Minkowski:1977sc,Yanagida:1979as,Gell-Mann:1979vob,Mohapatra:1979ia} and calls for thermal leptogenesis. In the following, we will address whether the observed baryon asymmetry can arise within this framework, consistently with the quality constraints discussed in Sec.\,\ref{sec:upperboundCP}.

\subsection{High-scale leptogenesis}
High-scale leptogenesis via the decay of heavy Majorana neutrinos has been extensively studied in the literature. For comprehensive reviews on this topic, see for example Refs.~\cite{Davidson:2008bu,Buchmuller:2004nz}. In the vanilla leptogenesis scenario~\cite{Fukugita:1986hr}, the lightest heavy neutrino, $N_1$, is responsible for leptogenesis. The Yukawa couplings parametrizing the interactions of the heavy neutrinos with the SM lepton doublets and the Higgs boson, in general c-numbers, can satisfy the required ingredients to generate a lepton asymmetry~\cite{Sakharov:1967dj,Fukugita:1986hr}: (i) they transfer the total lepton number violation coming from the Majorana mass term to the SM, (ii) they can source the needed CP violation via the interference between 1-loop and tree-level decay of the heavy neutrinos, (iii) their smallness can help the departure from thermal equilibrium. Subsequently, the lepton asymmetry is transferred to the baryon sector via the sphaleron processes, which violate baryon ($B$) and lepton ($L$) numbers in three units while preserving $B-L$~\cite{Kuzmin:1985mm}. The mass of the lightest heavy neutrino, $M_{N_1}$, is related to the active neutrino masses via the type-I seesaw mechanism. Since there is a minimum amount of CP asymmetry required to generate the observed baryon asymmetry, see Eq.~\eqref{eq:etaB}, the upper bound on the active neutrino masses translates into a lower bound on the lightest heavy neutrino mass, $M_{N_1} \gtrsim 10^8$ GeV, commonly referred to as the Davidson-Ibarra (DI) bound~\cite{Davidson:2002qv}. 

The Nelson-Barr model in Eq.~\eqref{eq:LagrangianLepto}, however, has two major differences with respect to the vanilla scenario studied in the literature:
\begin{enumerate}
    \item[(i)] The starting inputs are real Yukawa couplings and Majorana masses; the only CP phase comes from the vacuum of the theory, $\langle X_a \rangle$. Notice that such a phase is shared between the quark and the lepton sectors, and is inherited at low energies by both the CKM and the PMNS matrices.
    \item[(ii)] There are in total five heavy neutral leptons, whose mass parameters come, up to perturbative Yukawa couplings, from two scales: the CP breaking scale, $v_{\rm CP}$, and the new gauge symmetry breaking scale, $v_S \gtrsim v_{\rm CP}$. 
\end{enumerate}
Difference (i) implies that leptogenesis must occur below the SCPV scale. Yet, $v_{\rm CP}$ cannot be arbitrarily high, as it is bounded from above to preserve the quality of the NB mechanism, as we discussed in Sec.~\ref{sec:upperboundCP}. 
Whatever mechanism is responsible for diluting the topological defects arising from the breaking of CP~\cite{McNamara:2022lrw} should occur below $v_{\rm CP}$ and above the scale of leptogenesis; otherwise, the lepton asymmetry would also be erased. 

We then require the first temperature of the universe, $T_{\rm RH}$, to be above the lowest starting temperature for successful leptogenesis, dubbed $T_i$. Hence, we expect
\begin{equation}\label{eq:window}
T_i \leq T_{\rm RH} \leq v_{\rm CP}, \qquad \text{ or, schematically, }\qquad \begin{gathered}
\begin{tikzpicture}[line width=1.5 pt,node distance=1 cm and 1.5 cm]
\filldraw[fill=Neptuno, draw=none] (0,0) rectangle (4,0.5);
\filldraw[fill=Flamingo, draw=none] (0,0.9) rectangle (4,1.5);
\node (A) at (0, 0) {};
\node (B) at (0, 2.1) {${\rm E \, \text{[GeV]} }$};
\draw[] (0,0.5) -- (4,0.5);
\draw[] (0,0.9) -- (4,0.9);
\node (Z) at (-0.75, 0.7) { $T_i$};
\node (D) at (2.2, 1.2) { $\Delta \theta_{\rm QCD} > 10^{-10}$};
\node (E) at (2.2, 0.2) { $m_\nu > 0.1 \text{ eV}$};
\draw[->, to path={-| (\tikztotarget)}]
  (A) edge (B);
\end{tikzpicture}
\end{gathered}\quad .
\end{equation}

This hierarchy delineates the window of scales where leptogenesis within automatic Nelson-Barr models is viable. The upper boundary of this window is set by the quality of the mechanism (the red region is disfavored by unacceptably large shifts to $\bar \theta_{\rm QCD}$ from higher dimensional operators). This reflects on different upper bounds on $v_{\rm CP}$, depending on the specific NB model considered. 
The lower boundary arises from a minimal allowed temperature for successful leptogenesis, $T_i$, which is typically of the order $T_{i} \sim M_{N_1}$. As we discuss in the upcoming subsections, $M_{N_1}$ is bounded from below by the Davidson-Ibarra bound as a consequence of a maximal allowed value for the heaviest active neutrino (the turquoise region is ruled out by active neutrino masses that exceed observational limits). 

Our aim in this work is to study the interplay between the two boundaries and show whether such a window exists and how it maps onto the parameter space of the theory.

\subsubsection{Majorana neutrino masses}
\label{subsec:Majorana}

In the broken phase, Eq.~\eqref{eq:LagrangianLepto} generates the following Majorana masses for the eight neutral leptons of the theory,
\begin{equation}\label{eq:heavyN}
   - {\cal L} \supset  \frac{1}{2} \, f_N^TC \, \mathsf{M}^{8 \times 8}  \, f_N + \text{h.c.}, \qquad \text{with} \quad \mathsf{M}^{8 \times 8} = \begin{pmatrix} 0^{3\times 3} & \mathsf{M}_D^{3\times 5} \\ & &\\
(\mathsf{M}_D^T)^{5\times 3} \quad & \mathsf{M}_H^{5 \times 5} \end{pmatrix},
\end{equation}
where $f_N^T = (\nu_{L_1}^c, \nu_{L_2}^c , \nu_{L_3}^c, N_L^c , \nu_{R_1}, \nu_{R_2}, \nu_{R_3}, N_R)$, and 
\begin{equation}\label{eq:matrices}
\mathsf{M}_D = \begin{pmatrix} 0 &  &  &  & 0 \\
                      0 &  & \mathsf{m}_{D}^{3\times 3} &  & 0 \\
                      0 &  &  &  & 0 \end{pmatrix}, \quad \text{and} \quad
\mathsf{M}_H = \begin{pmatrix} 0 & \mu_1 & \mu_2 & \mu_3 & m_N \\
                        \mu_1 & m_{R_1} & 0 & 0 & 0 \\
                        \mu_2 & 0 & m_{R_2} & 0 & 0 \\
                        \mu_3 & 0 & 0 & m_{R_3} & 0 \\
                        m_N & 0 & 0 & 0 & 0 \end{pmatrix}.
\end{equation}
Above, $\mathsf{m}_D = (\mathsf{Y}_\nu \mathsf{O}_R) v_H/\sqrt{2}$, where $\mathsf{O}_R$ is the orthogonal matrix that diagonalizes $\mathsf{M}_R$ after rotating the $\nu_R$s, that is, $\mathsf{O}_R^T \mathsf{M}_R \mathsf{O}_R = \text{diag}(m_{R_1}, m_{R_2}, m_{R_3})$. Without loss of generality, we choose $\mathsf{O}_R$ such that $m_{R_1} \leq m_{R_2} \leq m_{R_3}$. We also note that the $\mu_i$ masses in $\mathsf{M}_H$ correspond to those in Eq.~\eqref{eq:M44}, up to a reshuffling induced by $\mathsf{O}_R$.

Applying the seesaw block diagonalization~\cite{Grimus:2000vj}, the three light and five heavy masses in the physical basis are given by 
\begin{equation}\label{eq:seesaw}
\begin{split}
    \mathsf{\widehat M}_L &= - \mathsf{U}_L^T \mathsf{M}_D \mathsf{M}_H^{-1} \mathsf{M}_D^T \mathsf{U}_L + {\cal O}([\mathsf{m}_D]^3 / [\mathsf{M}_R]^2),  \\
     \mathsf{\widehat M}_H &= \textsf{U}_H^T \textsf{M}_H \textsf{U}_H + {\cal O}([\mathsf{m}_D]^2/[\mathsf{M}_R]),
\end{split}
\end{equation}
where $\textsf{U}_{L,H}$ are the unitary matrices that diagonalize, \emph{\`a la Takagi}~\cite{Autonne1915,TAKAGI}, the light and heavy mass matrices.

We note that, considering the structure of the mass matrices outlined above, the masses of the active neutrinos depend only on the light Dirac masses $\text{m}_D^{ij}$, and $m_{R_i}$. Consequently, $\mu_i$ and $m_N$ are not affected by the heaviest active neutrino mass bound, $m_\nu \lesssim \, 0.1 \text{ eV}$~\cite{KATRIN:2021uub,Planck:2018vyg,DESI:2024mwx}. A significant implication is that, unlike the vanilla leptogenesis scenario widely studied in the literature, $M_{N_1}$ cannot be generically identified as the lightest eigenvalue of $\textsf{M}_R$, $m_{R_1}$. We find that, for a mass matrix texture as given in Eq.~\ref{eq:heavyN}, the lightest heavy neutrino mass is related to the mass scales in $\textsf{M}_H$ as
\begin{equation}\label{eq:SamBound}
    M_{N_1} \simeq \text{Min} \! \left\{ m_N, m_{R_1}, m_{R_i} \left( \frac{m_N^2}{m_N^2+ |\mu_i|^2} \right) \right\}.
\end{equation}
We refer the reader to Appendix~\ref{app:Sambound} for its analytical derivation in the scenario of three scales (corresponding to degenerate entries of $\mathsf{M}_R$).

Eq.~\eqref{eq:matrices} allows us to locate the complex phases in the heavy neutrino mass matrix. In the limit $[\mu] \ll [\mathsf{M}_R] \text{ or }m_N$, $\mathsf{M}_H$ becomes symmetric and real up to ${\cal O}([\mu]/ \text{Max}\{[\mathsf{M_R}],m_N\})$. This limit is disfavored by leptogenesis as the complex phases needed to generate the CP asymmetry would be suppressed. For example, if all Yukawa couplings in the new sector were ${\cal O}(1)$, heavy masses would be associated with the vevs generating them: $[\mathsf{M}_R] \sim m_N \sim v_S$ and $[\mu] \sim v_{\rm CP}$. If this is the case, we expect the vevs to be correlated, i.e. $v_{\rm CP} \sim v_S$.

\begin{figure}[t]
\begin{equation*}
\hspace{-0.5cm}
\begin{gathered}
 \resizebox{.2\textwidth}{!}{
\begin{tikzpicture}[line width=1.5 pt,node distance=1 cm and 1.5 cm]
\node (A) at (0, 0) {};
\node (B) at (0, 3) {${\rm Mass}$};
\node (C) at (-0.75, 0.7) { $[\mathsf{M}_R]$};
\node (D) at (1.2,3.5) {$\boxed{A}$};
\node (E) at (-0.75,2.05) {$m_N$};
\draw[] (0.15,0.5) -- (2,0.5);
\draw[] (0.15,0.7) -- (2,0.7);
\draw[] (0.15,0.9) -- (2,0.9);
\draw[] (0.15,2)--(2,2);
\draw[] (0.15,2.1)--(2,2.1);
\draw[->, to path={-| (\tikztotarget)}]
  (A) edge (B);
\end{tikzpicture}}
\end{gathered} \qquad
 \begin{gathered}
 \resizebox{.2\textwidth}{!}{%
\begin{tikzpicture}[line width=1.5 pt,node distance=1 cm and 1.5 cm]
\node (A) at (0, 0) {};
\node (B) at (0, 3) {${\rm Mass}$};
\node (C) at (-0.75, 0.7) { $m_N$};
\node (D) at (1.2,3.5) {$\boxed{B}$};
\node (E) at (-0.75,2.05) {$m_R$};
\draw[] (0.15,0.6) -- (2,0.6);
\draw[] (0.15,0.7) -- (2,0.7);
\draw[] (0.15,1.8)--(2,1.8);
\draw[] (0.15,2)--(2,2);
\draw[] (0.15,2.2)--(2,2.2);
\draw[->, to path={-| (\tikztotarget)}]
  (A) edge (B);
\end{tikzpicture}}
\end{gathered} \qquad
 \begin{gathered}
 \resizebox{.2\textwidth}{!}{%
\begin{tikzpicture}[line width=1.5 pt,node distance=1 cm and 1.5 cm]
\node (A) at (0, 0) {};
\node (B) at (0, 3) {${\rm Mass}$};
\node (C) at (-0.75, 1.1) { $m_R$};
\node (D) at (1.2,3.5) {$\boxed{C}$};
\node(E) at (-0.75,1.1) {$m_R$};
\node(G) at (-0.75,2.05) {$|\mu|$};
\draw[] (0.15,0.4) -- (2,0.4);
\draw[] (0.15,1) -- (2,1);
\draw[] (0.15,1.2) -- (2,1.2);
\draw[] (0.15,2)--(2,2);
\draw[] (0.15,2.1)--(2,2.1);
\draw[->, to path={-| (\tikztotarget)}]
  (A) edge (B);
\end{tikzpicture}}
\end{gathered} \qquad
  \begin{gathered}
 \resizebox{.2\textwidth}{!}{%
\begin{tikzpicture}[line width=1.5 pt,node distance=1 cm and 1.5 cm]
\node (A) at (0, 0) {};
\node (B) at (0, 3) {${\rm Mass}$};
\node (C) at (-0.75, 2) { $m_R$};
\node (D) at (1.2,3.5) {$\boxed{D}$};
\draw[] (0.15,0.4) -- (2,0.4);
\draw[] (0.15,1) -- (2,1);
\draw[] (0.15,1.8)--(2,1.8);
\draw[] (0.15,2)--(2,2);
\draw[] (0.15,2.2)--(2,2.2);
\draw[->, to path={-| (\tikztotarget)}]
  (A) edge (B);
\end{tikzpicture}}
\end{gathered}
\end{equation*}
\caption{Possible heavy neutrino physical mass spectra for different hierarchies. We follow the notation from the main text. Panel A corresponds to the hierarchies $m_N \gg [\mathsf{M}_R] \gg |\mu_i|$ and $m_N \gg |\mu_i| \gg [\mathsf{M}_R]$. In both cases $[\mu]$ does not impact the heavy neutrino masses. Panel $B$ represents the mass spectrum when $[\mathsf{M}_R] \gg m_N \gg |\mu_i|$.  Panel C involves the hierarchies $|\mu_i| \gg m_N \gg [\mathsf{M}_R]$, where the lightest heavy neutrino mass is either $m_{R_1}$ or $m_{R_i} m_N^2 / |\mu_i|^2$, and $|\mu_i| \gg [\mathsf{M}_R] \gg m_N$, with the lightest heavy neutrino mass always being given by $m_{R_i} m_N^2/|\mu_i|^2$. We note that in the latter case, $m_N$ does not impact the heavy neutrino masses. Panel D corresponds to the hierarchy $[\mathsf{M}_R] \gg |\mu_i| \gg m_N$ (except for very hierarchical $\mathsf{M}_R$, which lead to the spectrum from Panel B instead). Although we assumed three decoupled scales to reproduce the panels above, similar hierarchies can be found relaxing this condition.}
\label{fig:cartoon}
\end{figure}
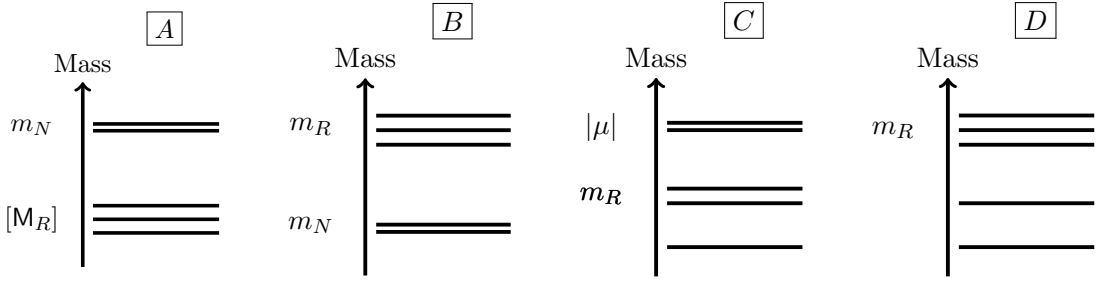

Fig.~\ref{fig:cartoon} summarizes different hierarchies that the heavy neutrino mass spectrum can exhibit, depending on the relationships among the scales participating in $\mathsf{M}_H$. Taking into account that scenarios with $[\mu] \gtrsim m_N,[\mathsf{M}_R]$ are preferred by leptogenesis, 
\begin{itemize}
    \item[(i)] If $[\mu]$ is the heaviest scale (panel C), $M_{N_1} \simeq \text{Min}\left\{m_{R_1}, m_{R_i}m^2_N/|\mu_i|^2\right\}$. Generically there will be a mass gap between the lightest, $N_1$, and the second lightest, $N_{2}$, heavy neutrinos. Hence, vanilla leptogenesis is expected to occur. 
    \item[(ii)] If $ [\mu] \sim m_N \gg [\mathsf{M}_R]$ (panel A),  whether vanilla leptogenesis or resonant leptogenesis~\cite{Pilaftsis:2003gt} takes place depends on the hierarchy of the Yukawa couplings in $\mathsf{M}_R$.
    \item[(iii)] If $[\mu] \sim [\mathsf{M}_R] \gg m_N$ (panel C or panel D), $M_{N_1}\simeq m_{R_i} (m_N/| \mu_i|)^2$, and vanilla leptogenesis is expected.
    \item[(iv)] If $m_N \sim [\mathsf{M}_R] \sim [\mu]$, all heavy neutrino masses will be roughly at the same scale, and whether vanilla or resonant leptogenesis occur will depend on the particular values of the Yukawa couplings entering in $\mathsf{M}_H$.
\end{itemize}

We proceed under the assumption of hierarchical heavy neutrinos, whereby only the lightest heavy neutrino plays a significant role in leptogenesis. This corresponds to scenarios A, C and D in Fig.~\ref{fig:cartoon}. In our numerical analysis in Sec.~\ref{sec:results}, this is enforced by imposing $M_{N_{j \neq 1}} > 5 M_{N_1}$ as a rule of thumb~\cite{Blanchet:2006dq}.  

\subsubsection{The CP asymmetry}
\label{sec:CPasymm}

In the physical basis, the interactions between the heavy neutrinos and the leptonic $\text{SU}(2)_L$ doublets are given by
\begin{equation}
    -{\cal L} \supset \bar L_{Li} \mathsf{h}^{i\ell} \tilde H  N_\ell + {\rm h.c.} , \qquad \text{where} \quad \mathsf{h}^{i \ell} \equiv (\mathsf{Y}_\nu \mathsf{O}_R)^{ik} \mathsf{U}_H^{*k \ell}.
\end{equation}
In our construction of the $5 \times 5$ matrix $\mathsf{U}_H$, $k = 2,3,4$ correspond to the three flavours of $\nu_R$s. Meanwhile, the index $\ell$ runs over the five heavy flavors. In particular, $\ell=1$ corresponds to the lightest heavy neutrino.
\begin{figure}[t]
\begin{equation*}
\begin{split}
&\begin{gathered}
\begin{tikzpicture}[line width=1.5 pt,node distance=1 cm and 1 cm]
\coordinate[label=left:$N_1$] (X);
\coordinate[right = 0.75 cm of X](v1);
\coordinate[above right = 1 cm of v1,label=right:$H$](H);
\coordinate[below right = 1 cm of v1,label=right:$\ell_{L\alpha}$](ell);
\draw[fill=black](v1) circle (.05cm);
\draw[fermionnoarrow](X)--(v1);
\draw[scalar](v1)--(H);
\draw[fermion](v1)--(ell);
\end{tikzpicture}
\end{gathered} \quad  + \quad 
\begin{gathered}
\begin{tikzpicture}[line width=1.5 pt,node distance=1 cm and 1 cm]
\coordinate[label=left:$N_1$] (X);
\coordinate[right = 0.75 cm of X](v1);
\coordinate[above right = 1 cm of v1](H);
\coordinate[below right = 1 cm of v1](D);
\draw[fill=black](v1) circle (.05cm);
\draw[fermionnoarrow](v1)--(X);
\draw[fermion](H)--(v1);
\draw[scalar](D)--(v1);
\draw[fermionnoarrow](H)--(D);
\coordinate[right = 0.75 cm of H,label=right:$H$](HF);
\coordinate[right = 0.75 cm of D,label=right:$\ell_{L\alpha}$](ellF);
\draw[scalar](H)--(HF);
\draw[fermion](D)--(ellF);
\draw[fill=black](H) circle (.05cm);
\draw[fill=black](D) circle (.05cm);
\coordinate[right = 0.7 cm of v1, label=right:$N_{j \neq 1}$](Ni);
\end{tikzpicture}
\end{gathered} 
\quad + \quad
    \begin{gathered}
\resizebox{.32\textwidth}{!}{%
\begin{tikzpicture}[line width=1.5 pt,node distance=1 cm and 1.5 cm]
\coordinate[label = left: $N_1$] (p1);
\coordinate[right = 0.8cm  of p1](p1a);
\coordinate[right = of p1a](p1b);
\coordinate[right = 1 cm of p1b](p2);
\coordinate[above right = 1 cm of p2, label=right: $H$] (v1);
\coordinate[below right =  1 cm of p2,label=right: $\ell_{L\alpha}$] (v2);
\coordinate[above right = 1cm of p1a,label=$\ell_{L\alpha}$](vaux1);
\coordinate[below right = 1cm of p1a,label=$H$](vaux2);
\coordinate[right=0.5cm of p1b,label=above:$N_{j \neq 1}$](vaux3);
\draw[fermionnoarrow] (p1) -- (p1a);
\draw[scalar] (p2) -- (v1);
\draw[fermion] (p2) -- (v2);
\draw[fermionnoarrow] (p1b)--(p2);
\semiloop[fermionnoarrow]{p1a}{p1b}{0};
\hello[scalarnoarrow]{p1a}{p1b}{0};
\draw[fill=black] (p2) circle (.05cm);
\draw[fill=black] (p1a) circle (.05cm);
\draw[fill=black] (p1b) circle (.05cm);
\end{tikzpicture}}
\end{gathered}
\end{split}
\end{equation*}
\caption{Tree-level and one-loop level diagrams for the decays of the lightest heavy neutrino $N_1$ that can contribute to the CP asymmetry.}
\label{fig:diagrams}
\end{figure}
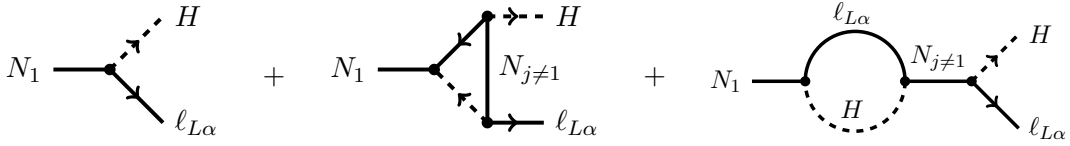
The CP asymmetry, generated by the interference of the vertex correction and self-energies shown in Fig.~\ref{fig:diagrams}, is defined as (see for example~\cite{Davidson:2008bu,Blanchet:2012bk})
\begin{equation}\label{eq:epsilonCPaa}
    \epsilon_{\rm CP} \equiv \frac{ \sum_\alpha \textstyle  \Gamma_{N_1 \to H \ell_\alpha}-\Gamma_{N_1 \to H^\dagger \bar \ell_\alpha}}{ \Gamma_D}=  \sum_{j\neq 1} \frac{\text{Im}\{ [\mathsf{h}^\dagger \mathsf{h}]_{1 j}^2 \}}{8\pi [\mathsf{h}^\dagger  \mathsf{h}]_{11}} f_{\rm loop} \left[ \frac{M_{N_j}^2}{M_{N_1}^2} \right],
\end{equation}
where $\Gamma_D$ is the total decay rate of $N_1$, given by\footnote{We are assuming that, in whatever theory that UV-completes our simplified model, the branching ratio $\text{Br}(N_1 \to H \ell) \sim 1$.}
\begin{equation}\label{eq:GammaD}
   \Gamma_D \equiv \sum_\alpha \Gamma_{N_1 \to H \ell_\alpha} + \Gamma_{N_1 \to H^\dagger \bar \ell_\alpha}  = \frac{[\mathsf{h}^\dagger \mathsf{h}]_{11}}{8\pi}M_{N_1}.
\end{equation}
For hierarchical heavy neutrinos, $ r \equiv  M_{N_{j\neq 1}}^2 /  M_{N_1}^2 \gg 1$, the loop function is approximately given by
\begin{equation}\label{eq:floop}
    f_{\rm loop} [r] = \sqrt{r} \left[ 1 - (1+r) \log \left( \frac{1+r}{r}\right) + \frac{1}{1-r} \right] \simeq - \frac{3}{2 \sqrt{r}}.
\end{equation}

As shown in Ref.~\cite{Davidson:2002qv}, for a given $M_{N_1}$ there is a maximum value of $\epsilon_{\rm CP}$ reachable. The proof of this bound, referred to as the Davidson–Ibarra (DI) bound, relies on the assumption that the numbers of light and heavy neutrinos are equal. Although our mass matrix in Eq.~\eqref{eq:matrices} does not satisfy the latter criterion, a very similar bound can be obtained in our case:
\begin{equation}\label{eq:DI}
|\epsilon_{\rm CP}| \leq \frac{3}{8\pi} \frac{M_{N_1}}{v_H^2} \, m_{\nu_3},
\end{equation}
where $m_{\nu_3}$ is the heaviest light neutrino.
We refer the reader to Appendix~\ref{app:DIbound} for the proof.  

As anticipated earlier, the existence of a minimum required $\epsilon_{\rm CP}$ to generate $\eta_B^{\rm obs}$ combined with the upper bound on the heaviest active neutrino mass, leads Eq.~\eqref{eq:DI} to impose a lower bound on the mass of the lightest heavy neutrino.

\subsubsection{Dynamics}\label{eq:dynamics}

The dynamics of $N_1$ and the generated $B-L$ asymmetry are dictated by the Boltzmann equations: 
\begin{eqnarray}\label{eq:Boltzmann1}
\frac{dN_{N_1}}{dz} &=& -(D +  S) (N_{N_1} -N_{N_1}^{\rm eq}), \\
\frac{dN_{\rm B-L}}{dz} &=& - \epsilon_{\rm CP} D (N_{N_1} - N_{N_1}^{\rm eq}) -  W  \, N_{\rm B-L},\label{eq:Boltzmann2}
\end{eqnarray}
where $z \equiv M_{N_1} / T$,  $\epsilon_{\rm CP}$ is defined in Eq.~\eqref{eq:epsilonCPaa}, $N_{N_1}$ ($N_{\rm B-L}$) is the number density of $N_1$ (the total $B-L$ asymmetry) normalized by the number density of the photons at $z \to 0$. 
$D$ accounts for the relative strength of the thermal averaged decay and inverse decay rates with respect to the expansion rate of the universe,
\begin{equation}
D(z) = K  z\frac{K_1(z)}{K_2(z)}\,.
\end{equation}
Here, $K_1(z)$ and $K_2(z)$ are the first and second modified Bessel functions, respectively. The \emph{washout parameter} $K$, defined as 
\begin{equation}
\label{eq:Kwash}
 K \equiv \frac{\Gamma_D}{H(z=1)} =  \frac{\tilde m_1}{m_\star} \frac{m_{R_1}}{M_{N_1}}\,, 
\end{equation}
quantifies the strength of the interactions responsible for either generating (through inverse decays) or depleting (through decays) the $N_1$ population.
Above, $m_\star \equiv 4\pi v_H^2 \sqrt{8 \pi^3 g_\star / 90} / M_{\rm Pl}$ is a constant (up to variations in the relativistic degrees of freedom $g_\star$~\cite{Borsanyi:2016ksw}), with $M_{\rm Pl} = 1.2 \times 10^{19} \text{ GeV}$, and the total decay rate in the vacuum, $\Gamma_D$, given in Eq.~\eqref{eq:GammaD}. Following the literature, we define the effective neutrino mass $\tilde m_1 \equiv [ \mathsf{h}^\dagger \mathsf{h}]_{11} v_H^2 / (2\,m_{R_1})$ as a proxy for the scale of active neutrino masses. We note that $K$ differs from the usual definition in the literature by the ratio $m_{R_1}/M_{N_1}$ since, in our case, the lightest heavy neutrino mass $M_{N_1}$ does not always correspond to $m_{R_1}$, as shown in Eq.~\eqref{eq:SamBound}.

$\Delta L = 1$ scatterings may affect both the population of heavy neutrinos via $S$ in Eq.~\eqref{eq:Boltzmann1}, as well as the washout processes encoded in $W$ in Eq.~\eqref{eq:Boltzmann2}. The latter also receives contributions from inverse decays, as well as non-resonant $\Delta L = 2$ scatterings mediated by heavy neutrinos~\cite{Giudice:2003jh}.

We will consider the two scenarios where the abundance of $N_1$ is either dynamically generated (starting from a null thermal abundance at high temperatures) or thermally generated. While the former is more conservative, the latter may occur easily in the context of automatic Nelson-Barr models due to the scattering processes mediated by the new gauge boson, $N_1 N_1 \to (Z')^* \to \bar f f$, with $f$ being both new and/or standard model fermions, depending on the force. See appendix~\ref{app:UV} and Refs.~\cite{Plumacher:1996kc,FileviezPerez:2021hbc} for more details. 

In our analysis, we make the following assumptions, which simplify considerably the dynamics:
\begin{itemize}
\item Flavor effects become relevant at temperatures below $10^{11}$ GeV~\cite{Barbieri:1999ma,Abada:2006ea,  Nardi:2006fx}, which covers the regime for automatic Nelson-Barr models, given the upper bound on the SCPV scale coming from demanding quality to the mechanism, see Eq.~\eqref{eq:vCPbound}. Flavor effects generically help achieve successful leptogenesis, but do not impact the lower bound for $M_{N_1}$~\cite{Davidson:2007xu,Blanchet:2006be,Josse-Michaux:2007alz}, which is our main interest. Hence, adopting a conservative approach, we neglect flavor effects and trace over the final lepton flavors, as we did in Eq.~\eqref{eq:Boltzmann2}.

\item $\Delta L =1$ scatterings play a relevant role in the weak washout regime ($ K < 1$). We find that, in order to be consistent with neutrino data, most of the selected points in our numerical analysis are in the strong washout regime. Therefore, we neglect the effect of $\Delta L =1$ scatterings in our analysis, although we have checked numerically that they do not impact our conclusions. 
 Including them, the washout term is dominated by the inverse decays, and given by
\begin{equation}
    W_{\rm ID} =\frac{1}{2} \frac{N_{N_1}^{\rm eq}(z)}{N_\ell^{\rm eq}}D = \frac{1}{4} K \, z^3K_1(z) \,,
\end{equation}
where $N_{N_1}^{\rm eq} = 3 z^2 K_2(z) / 8$ and $N_\ell^{\rm eq} =3/4$.
\item Non-resonant $\Delta L = 2$ scatterings are neglected, as these are suppressed by two insertions of the Dirac Yukawa couplings. 
\item Thermal and spectator effects are also neglected. The former can give corrections to the masses and couplings at high temperatures. See Ref.~\cite{Giudice:2003jh} for a detailed study on their impact in leptogenesis. For a thermal population of $N_1$, the lower bound on its mass stays roughly the same\footnote{In the scenario of a dominant population of $N_1$, the bound could be weakened by an order of magnitude~\cite{Giudice:2003jh}.}. In the strong washout regime, thermal effects are practically negligible~\cite{Davidson:2008bu}. Spectator processes~\cite{Buchmuller:2001sr} are interactions in the plasma that can alter the chemical potentials of particles in thermal equilibrium and affect the dynamics of leptogenesis. Their proper inclusion in the dynamics have at most $20\%-40\%$ effects~\cite{Nardi:2005hs}.
\end{itemize}
Eqs.~\eqref{eq:Boltzmann1} and \eqref{eq:Boltzmann2} clearly show the three ingredients needed for leptogenesis: (i) a non-zero CP violation, see Eq.~\eqref{eq:epsilonCPaa}, (ii) lepton number violation induced by the $\Delta L = 1$ decay (and scatterings) of the lightest heavy Majorana neutrino $N_1$, and (iii) departure of thermal equilibrium, $N_{N_1} \neq N^{\rm eq}_{N_1}$, in order to achieve a non-zero leptonic asymmetry.   
Sphaleron processes will then transfer the asymmetry generated in the lepton sector to the baryon sector, respecting $B-L$. The latter conversion depends on the $\text{SU}(2)_L$-charged degrees of freedom (the same in our model as in the SM), which gives $a_{\rm sph} = 28/79$~\cite{Harvey:1990qw}. The baryon asymmetry today is then given by
\begin{equation}\label{eq:etaBlepto}
\eta_{B} = a_{\rm sph}\left(\frac{3}{4}   \epsilon_{\rm CP} \, \kappa_f \right) \frac{n_\gamma(T_i)}{n_\gamma(T_{\rm rec})},
\end{equation}
where $N_{\rm B-L} (T_i)$ is displayed in between brackets, $T_i$ was introduced above as the starting temperature for leptogenesis, and $n_\gamma(T_{\rm rec})$ ($n_\gamma(T_i)$) is the number density of photons evaluated at the recombination temperature $T_{\rm rec}$ (at $T_i$).

The efficiency factor is defined as
\begin{equation}
\kappa_f = - \frac{4}{3}\int_{z_{\rm RH}}^\infty d z' \frac{dN_{N_1}}{dz'} e^{\int_{z'}^\infty dz'' W_{ID} (K,z'')},
\end{equation}
where $z_{\rm RH}$ ideally corresponds to the first temperature of the universe, i.e. $z_{\rm RH} = M_{N_1}/T_{\rm RH}$.
For the efficiency factor in the two scenarios studied, we use the analytical approximations from Ref.~\cite{Buchmuller:2004nz}
\begin{equation}
\label{eq:keff}
\kappa_f ( K)= \begin{cases} \displaystyle \frac{2}{z_B( K)  K} \left( 1 - e^{-\frac{1}{2} z_B( K)  K}\right) & \text{if thermal}, \\
                    \kappa_f^+( K) + \kappa_f^-( K)  & \text{if dynamical}, \end{cases}
\end{equation}
where $\kappa_f^\pm$ are given by
\begin{equation*}
 \kappa_f^-(K) =  2e^{-\frac{3}{8} \pi K} \left(1- e^{\frac{2}{3} \bar N (K)}\right), \quad \text{and} \quad
  \kappa_f^+(K) = \displaystyle \frac{2}{z_B( K) K} \left( 1 - e^{-\frac{2}{3} z_B(K) K \bar N (K)}\right),
\end{equation*}
with 
\begin{align}
\bar N ( K) &= \frac{9\pi K}{16}\left( 1 + \sqrt{\frac{9\pi K/16}{ N_{N_1}^{\rm eq}(z=0)}}\right)^{-2}, \\ \label{eq:zb}z_B(K) &\simeq 1+  \frac{1}{2}\ln \left[1 + \frac{\pi K^2}{1024}\left( \ln \left[ \frac{5^5 \pi K^2}{1024}\right]\right)^5 \right].
\end{align}

When does leptogenesis occur? This is a key question regarding the viability of leptogenesis within Nelson-Barr models. Following Ref.~\cite{Buchmuller:2004nz}, we estimate the lowest possible initial temperature for leptogenesis as 
\begin{equation}
\label{eq:Ti}
    T_i\equiv\frac{M_{N_1}}{z_i}\simeq\frac{M_{N_1}}{z_B - 2 e^{-3/K}}\,.
\end{equation}
Inflation, or whatever mechanism that solves the horizon problem, must take place after SCPV in order to erase the domain walls. To ensure this, we require the reheating temperature to be lower than $v_{\rm CP}$. 
We can then interpret $T_i$ as a lower bound on the first temperature of the universe. In the weak washout regime ($K\lesssim 1$), $T_i\sim M_{N_1}$, while in the strong washout regime $T_i$ is up to one order of magnitude smaller than $M_{N_1}$. The former is understood from the small $K$ expansion of $z_B = 1 + {\cal O}(K^2)$. The latter comes from the upper bound on $K \lesssim 10^3$, which suppresses $T_i$ by a factor of $\sim 10$ with respect to $M_{N_1}$.

\section{Methods and Results}\label{sec:results}

In this section, we perform a numerical study of leptogenesis for the simplified model parametrized in Eq.~\eqref{eq:LagrangianLepto}, which reproduces the features of the Nelson-Barr mechanism in the lepton sector (see discussion in Sec.~\ref{sec:leptogenesis}).

\subsection{Sampling}
To efficiently explore the parameter space, we first allow the two heavy vevs, $v_{\rm S}$ and $v_{\rm CP}$, to vary logarithmically within the interval $[10^7, 10^{14}]\,{\rm GeV}$. These boundaries are justified as follows: numerical scans show that there are no viable parameter points for the heavy vevs below $10^7$ GeV. On the upper end, $v_{\rm CP}$ is constrained by the quality bound, while $v_{\rm S}$ is limited by the canonical seesaw bound. We then pick specific mass matrices by randomly sampling the 13 real Yukawa couplings $(\mathsf{Y_\nu}\mathsf{O}_R)^{ij},(\mathsf{O}_R^T\mathsf{Y}_R\mathsf{O}_R)^{ii}$, and $\lambda_N$ on a uniform logarithmic scale. We similarly sample the magnitude of the 3 complex parameters $|\mu^i|/v_{\rm CP}$, and we uniformly sample their phases in the interval $[0,2\pi)$. For each point in a $200\times 200$ grid $(v_{\rm CP},v_{\rm S})$ such that $v_{\rm CP}<v_{\rm S}$, we randomly sampled the 19 parameters described above $10^3$ times.

We fix the overall scale of $(\mathsf{Y_\nu}\mathsf{O}_R)$ in order to obtain a mass for the heaviest active neutrino that saturates the bound $m_{\nu_3} \le0.1\,{\rm eV}$.  Invoking perturbativity, we bound the magnitude of all Yukawa couplings to be less than one\footnote{\label{foot:Yupbnd}We have chosen 1 as an upper bound for the Yukawa couplings (instead of $\sqrt{4\pi}$) to stay safe from RGE effects. Although we do not expect relevant contributions from the new sector to the running as these are neutral fields (under the SM forces), new charged fermions may be present in specific UV completions (see appendix~\ref{app:UV}). We thank Riccardo Rattazzi for pointing this out to us.}, allowing them to vary along the span of 1, 2, and 5 orders of magnitude. In the plots, we indicate the different spreads in the Yukawa couplings with different colors: blue, orange, and green points are log-uniformly sampled from the intervals $[10^{-5},1)$, $[10^{-2},1)$, and $[10^{-1},1)$ respectively. 

Once all the mass matrices are generated, we numerically diagonalize \emph{\`a la Takagi} to find the masses in the physical basis and we analytically compute - using the formulae listed in Sec.~\ref{sec:leptogenesis}  - the unflavored CP asymmetry $\epsilon_{\rm CP }$ and the washout parameter $K$ for each one of our specific realizations, inferring the predicted $\eta_B$ using Eq.~\eqref{eq:etaBlepto}.

\subsection{Successful Leptogenesis}
\label{subsec:successufullept}
\begin{figure}[tbp]
\includegraphics[width=0.475\linewidth]{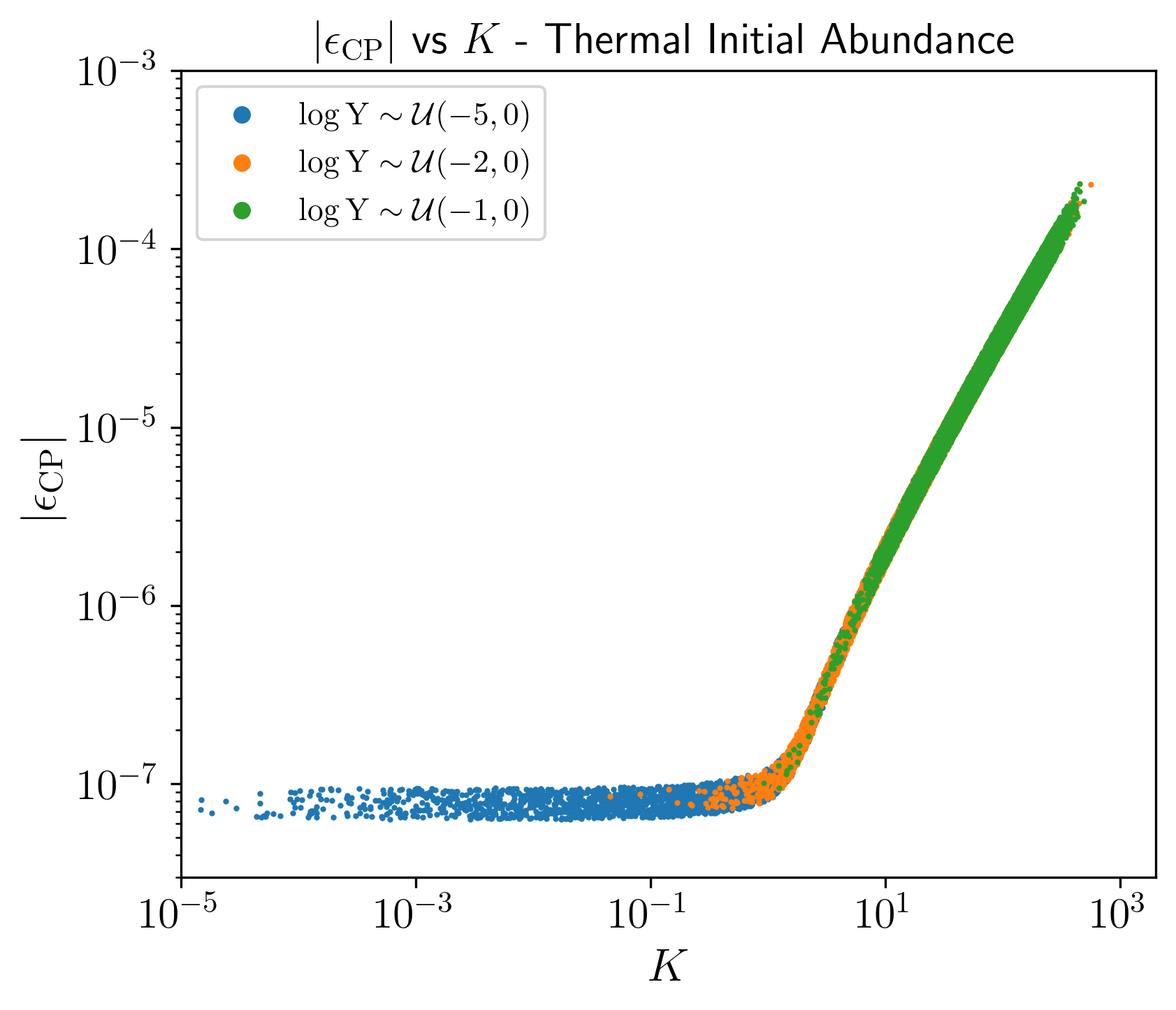}	
\includegraphics[width=0.49\linewidth]{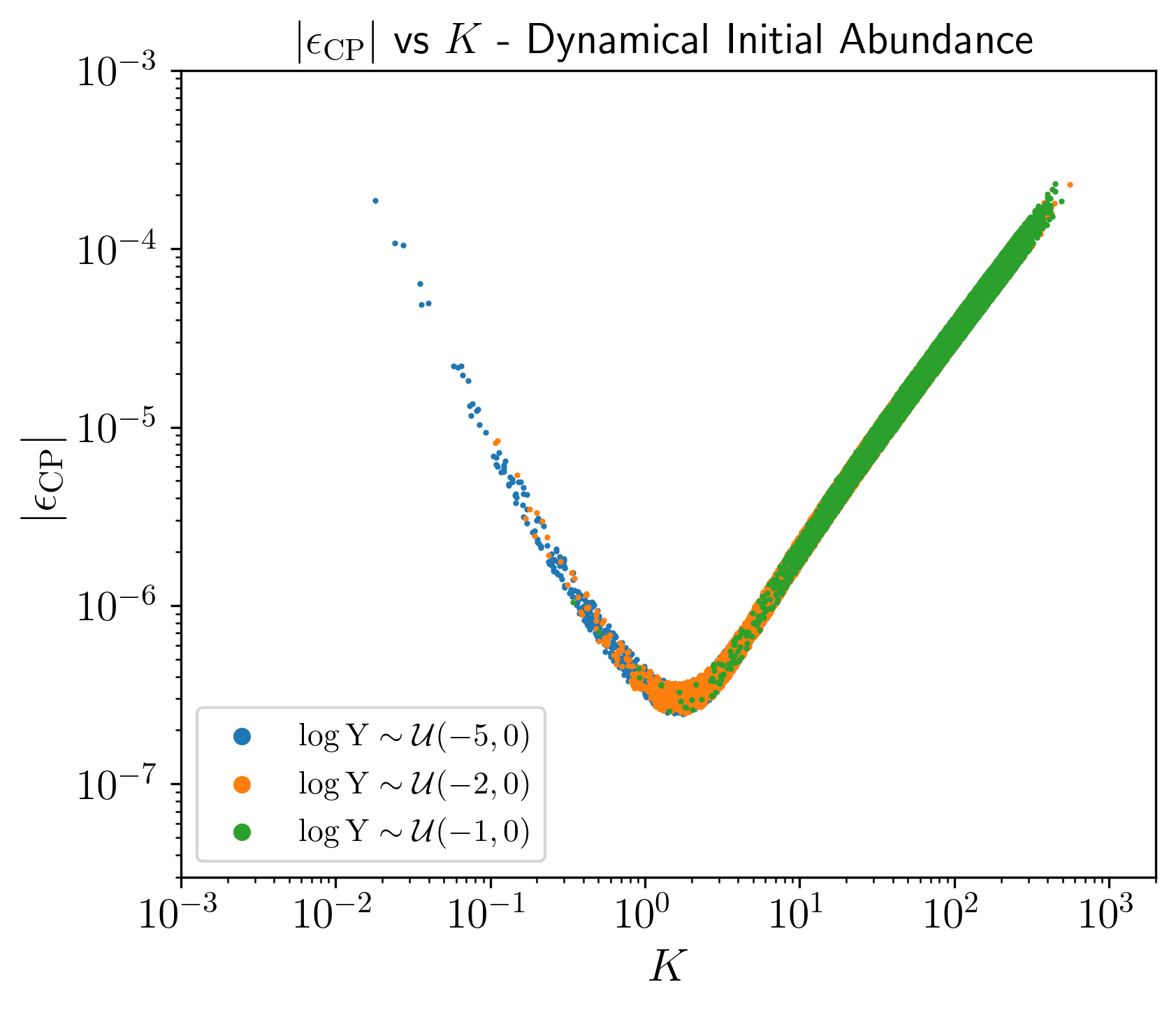}\\
\includegraphics[width=0.49\linewidth]{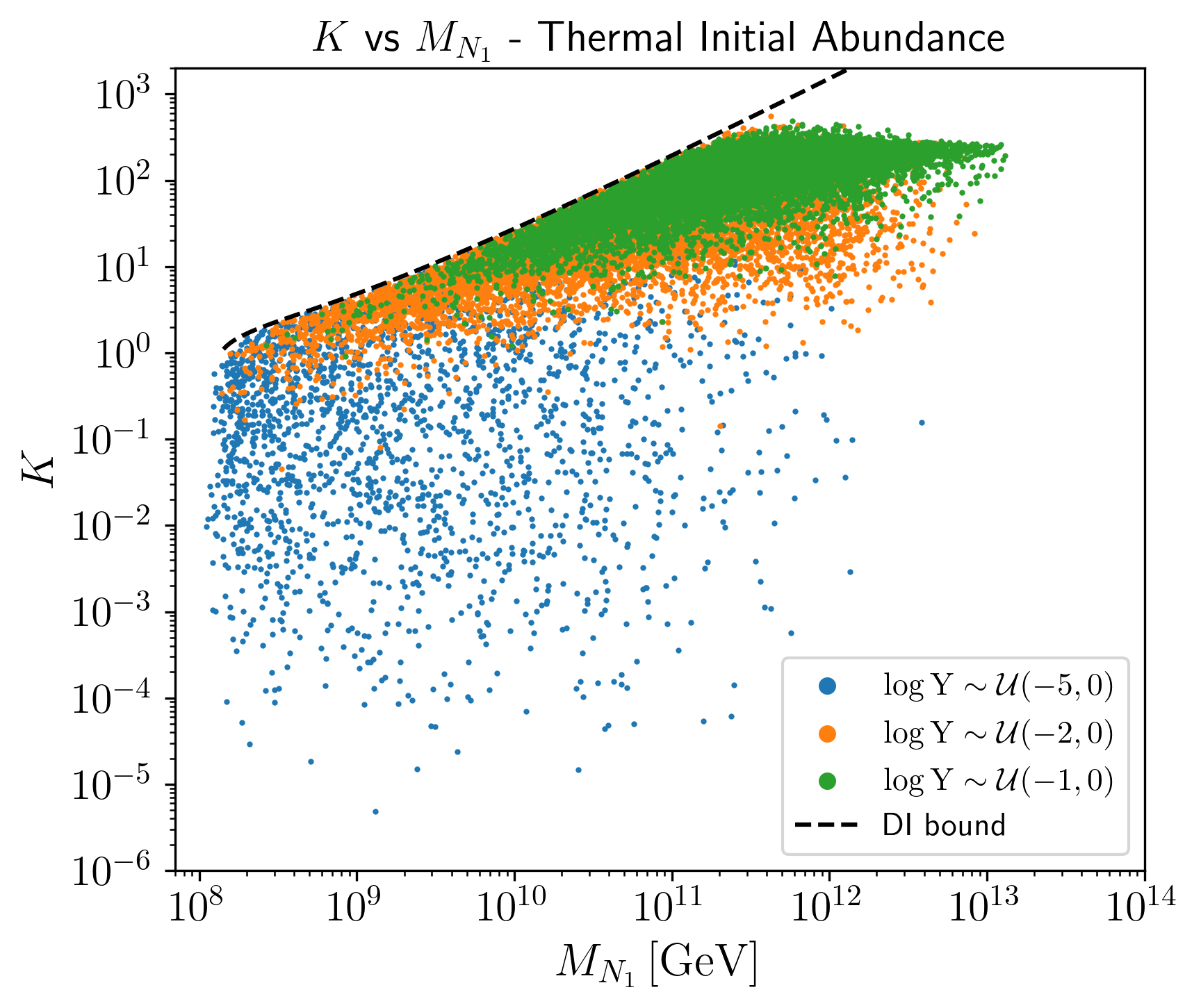}
\includegraphics[width=0.49\linewidth]{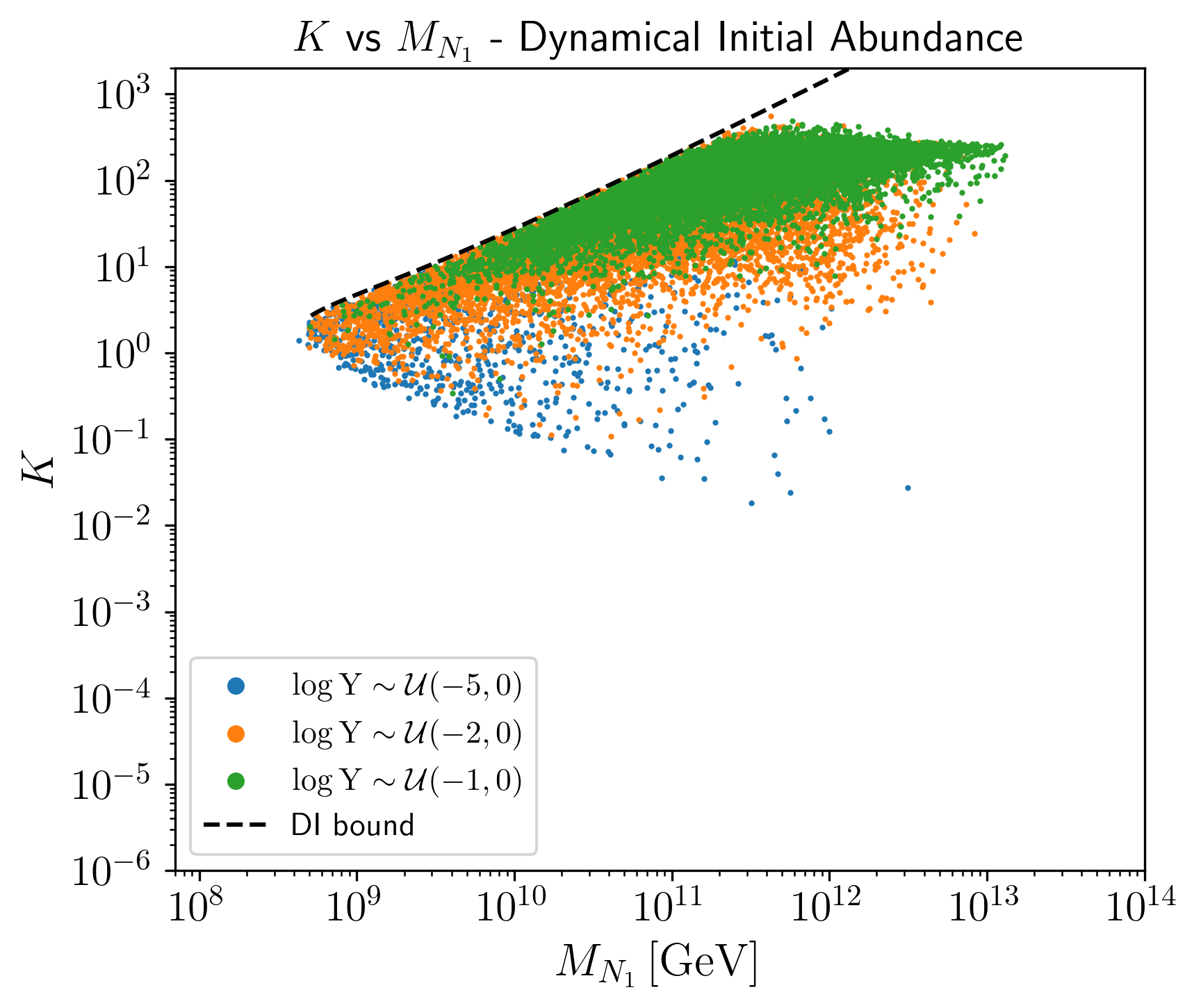}\\
\includegraphics[width=0.49\linewidth]{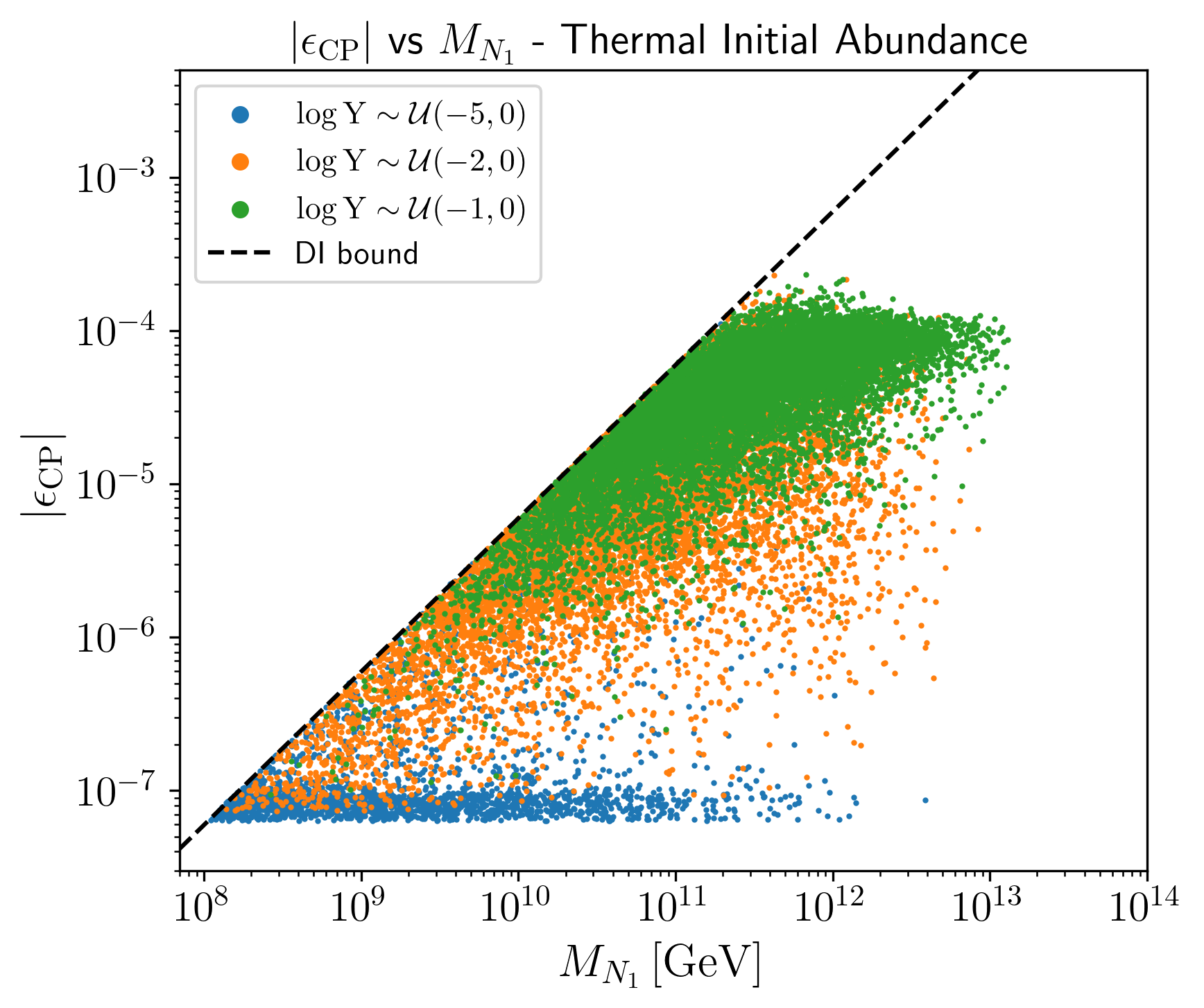}
\includegraphics[width=0.49\linewidth]{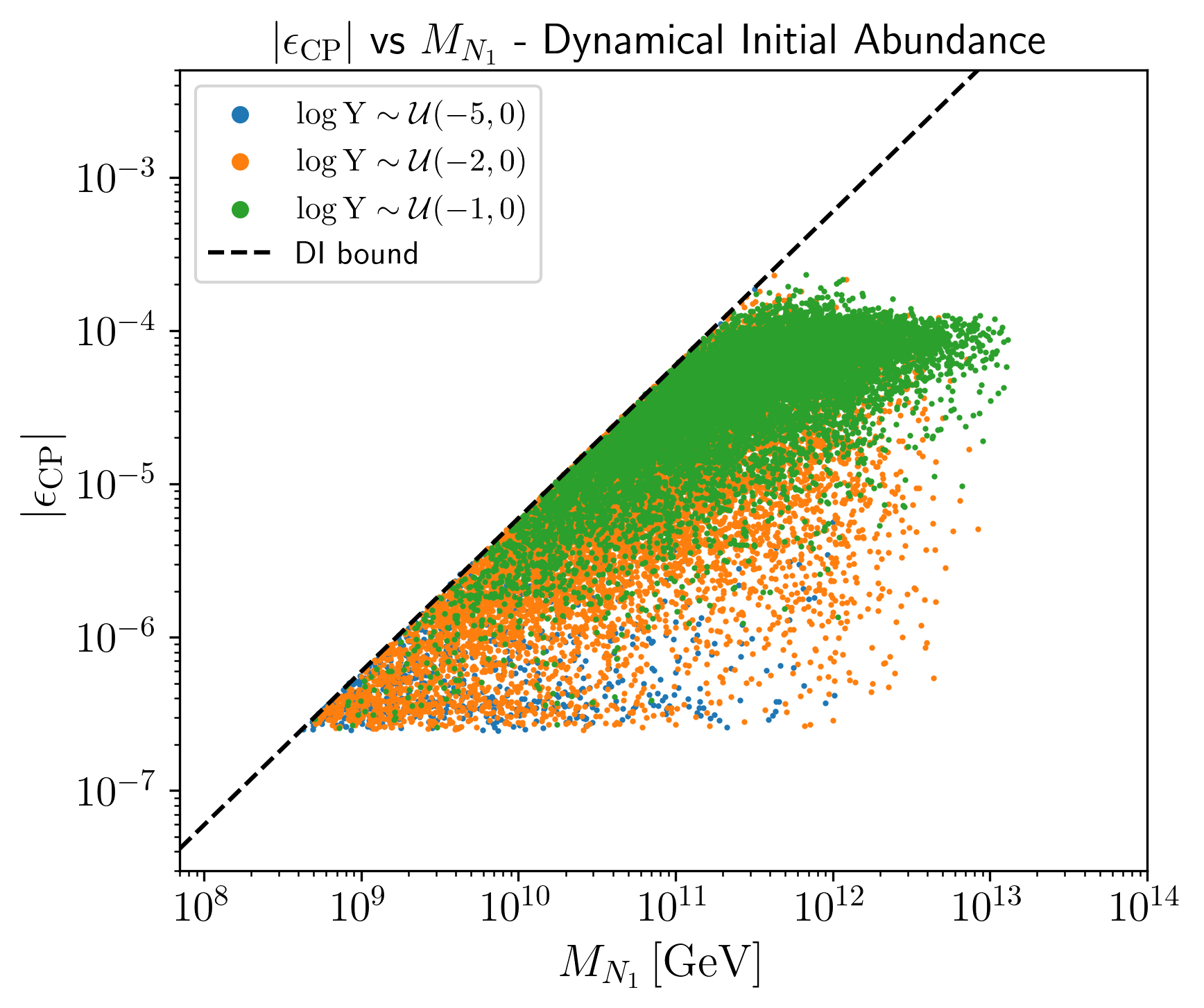}
\caption{
    Scatter plots showing the parameter space that lead to successful leptogenesis, obtained by randomly sampling the magnitude of the Yukawa couplings, Y, from log-uniform distributions with varying ranges. Specifically, the log values of Y are drawn from a uniform distribution $ \log \mathsf{Y} \sim \mathcal{U}(n, 0)$, where \( n = -1 \), \( -2 \), or \( -5 \) sets the lower bound of the distribution. This corresponds to logarithmically sampling $\mathsf{Y}$ over one (green points), two (orange points), or five (blue points) orders of magnitude below the upper bound, which is fixed at $ \mathsf{Y}=1 $ (see discussion in footnote \ref{foot:Yupbnd}).  The points satisfy $\eta_B^{\rm obs}<|\eta_B| < 1.5 \,\eta_B^{\rm obs}$, $v_{\rm CP}>T_i$, and $M_{N_{j \neq 1}} > 5 M_{N_1}$.}
    \label{fig:scatters}
\end{figure}
We consider successful samplings the ones that meet the following three conditions:
\begin{itemize}
    \item  They must generate enough baryon asymmetry to match the observed value $\eta_B^{\rm obs}$ given in Eq.~\eqref{eq:etaB}. Thus, we require $\eta_B^{\rm obs}< |\eta_B|< 1.5 \, \times \eta_B^{\rm obs}$.
    \item The initial temperature for leptogenesis $T_i$, computed according to Eq.~\eqref{eq:Ti}, must be lower than $v_{\rm CP}$ to avoid the restoration of the CP symmetry. Hence, we impose $T_i<v_{\rm CP}$.
    \item The mass spectrum of the heavy neutrinos must be hierarchical so that only the asymmetry produced by the $N_1$s population matters. We conventionally keep only the points with $M_{N_{j \neq 1}} > 5 M_{N_1}$ as a rule of thumb~\cite{Blanchet:2006be}.
\end{itemize}
We work within two possible scenarios regarding the initial population of $N_1$s: 
\begin{itemize}
    \item[(i)] TIA: thermal initial abundance (left panels), where the lightest heavy neutrino $N_1$ is taken to be in thermal equilibrium prior to leptogenesis, i.e. $N_{N_1}(0) = 3/4$.
    \item[(ii)] DIA: dynamical initial abundance (right panels), with $N_{N_1}(0) = 0$, i.e. assuming that the $N_1$s are produced exclusively by inverse decays. 
\end{itemize} 
For the two scenarios above, the efficiency factors are computed according to Eq.~\eqref{eq:keff}.

In Fig.~\ref{fig:scatters}, we plot all the successful points in our scan in order to show the parameter space consistent with leptogenesis from different perspectives.

In the first row of Fig.~\ref{fig:scatters}, $|\epsilon_{\rm CP}|$ is plotted against $K$. The efficiency curve for both scenarios is easily recognized. This is a consequence of the efficiency factor, $\kappa_f$, scaling inversely proportional to $\epsilon_{\rm CP}$ for a fixed baryon-to-photon ratio, as shown in Eq.\,\eqref{eq:etaBlepto}. The thickness of the band is given by our selection rule for the $\eta_{\rm B}$ of the successful points. The reader may immediately identify the lower bound on $|\epsilon_{\rm CP}|$, which corresponds to the maximum efficiency achievable: $K\lesssim 1$ ($K\sim 1$) for the TIA (DIA) case.
Note that, as expected~\cite{Buchmuller:2004nz,Buchmuller:2003gz}, the strong washout regime, $K>1$, is preferred if Yukawa couplings are ${\cal O}(1)$ (green points). This is because $\tilde m_1 \sim m_{\nu_3}$\footnote{For $[\lambda_N], [\mathsf{O}_R^T\mathsf{Y}_R \mathsf{O}_R], [\mu]  = {\cal O}(1)$, $m_{R_1} \sim m_N \sim \mu$ and so, according to Eq.~\eqref{eq:SamBound}, $m_{R_1} \sim M_{N_1}$. In this case, the hierarchy between $\tilde m_1$ and $m_\star$ determines whether we are in the strong ($\tilde m_1 > m_\star$) or weak ($\tilde m_1 < m_\star$) wash-out regimes.\label{fn:mRmN1}}. Wider ranges of variation (on a logarithmic scale) of the Yukawa couplings allow for a larger dissociation between $\tilde m_{1}$ and $m_{\nu_3}$, which in turn leads to a broader range of $K$ values (see the orange and blue points). Since in the strong washout regime the efficiency curves for both TIA and DIA are the same, the two plots for $K\gtrsim 10$ are perfectly coincident.
In the DIA plot, we can identify an upper bound for $\epsilon_{\rm CP}$ for each washout regime. In the weak washout regime, the curve stops because the CP asymmetry is saturated\footnote{We have checked that with more statistics -- for $(\mathsf{O}_R^T \mathsf{Y}_R\mathsf{O}_R)^{ii}, \lambda_N, [\mu]/v_{\rm CP}  \sim {\cal O}(1)$ and the Yukawa couplings relevant for the light neutrinos varying along the span of 5 orders of magnitude -- $\epsilon_{\rm CP} > 10^{-4}$ can be achieved for $K \ll 1$, approaching $\epsilon_{\rm CP}\sim10^{-2}$ (loop effect).}.  
In the strong washout regime, the common upper bound for both TIA and DIA comes from $K$ reaching its maximum value. As we discuss in Appendix~\ref{app:DIbound}, the upper bound $K < m_{\nu_3}/m_\star$~\cite{Buchmuller:2003gz} derived in the vanilla leptogenesis scenario (3 heavy neutrinos) does not hold in our case -- although we do not expect a significant deviation. 
Indeed, our results include points with $10^2<K<10^3$. 
The plots in the second row show an alternative correlation in the parameter space of the theory, this time $K$ vs $M_{N_1}$. For a given $K$, there is a certain $\epsilon_{\rm CP}$ that gives the right asymmetry, which in turn fixes the lightest $M_{N_1}$ allowed by saturating the DI bound (see Eq.~\eqref{eq:DI}, represented with a dashed line in these figures). We note that, as $M_{N_1}$ grows, the washout parameter tends to $K \sim 10^2$ in both TIA and DIA. 
This behavior reflects the seesaw mechanism: as $M_{N_1}$ approaches the canonical seesaw limit, the spread in Yukawa coupling narrows, leading to $\tilde m_1 \sim m_{\nu_3}$. Conversely, for smaller values of $M_{N_1}$, the spread widens (for a fixed range, i.e., a given color), resulting in a stronger decoupling between $\tilde{m}_1$ and $m_{\nu_3}$, and thus a broader variation in $K$. 

The third-row plots explicitly depict the DI bound. The upper bound on $\epsilon_{\rm CP}$ for both regimes has already been discussed. The lower bound on $\epsilon_{\rm CP}$ corresponds to the maximum value of the efficiency factor $\kappa_f$.
The blue points are statistically favored where the efficiency is maximized, that is, $K\lesssim 1$ ($K\sim 1$) for TIA (DIA). This can be clearly seen from the accumulation of points at the minimum $|\epsilon_{\rm CP}|$ allowed.
The lowest value of $M_{N_1}$ leading to successful leptogenesis is easily identified: $M_{N_1}\gtrsim 10^8$ GeV ($M_{N_1} \gtrsim 4 \times 10^{8}$ GeV) for the TIA (DIA) scenario.
Such a lower bound sets the lowest possible reheating temperature via Eq.~\eqref{eq:Ti}.

A comment on the mass scale of the new fermions regarding finite naturalness is in order. As we discussed in subsection~\ref{subsec:Radiative}, finite naturalness constraints on the radiatively generated quartic coupling require the new quarks to lie near the TeV scale~\cite{Perez:2023zin}. However, this bound is significantly relaxed in the case of the new neutral leptons. As shown in the one-loop diagram (square brackets) in Fig.\ref{fig:deltaMqq}, their contribution is suppressed by two insertions of the Dirac Yukawa coupling and by the hierarchy between the mass of the neutral fermions and $v_{\rm CP}$ (width of the viable window for leptogenesis). Using the analog of Eq.~\eqref{eq:lam4oneloopnat} for the lepton, $\delta \lambda_4|_{\rm 1-loop}v_{\rm CP}^2  \lesssim m_H^2$ reads as 
\begin{equation}
    \frac{1}{16\pi^2}\frac{2 m_{\nu_3}}{v^2_{H}}M^3_{N_1}\ln\left(\frac{M_{N_1}^2}{m_h^2}\right)\lesssim m_H^2 \,\, \Rightarrow  \, M_{N_1}\lesssim 0.4\times10^7,\,
\end{equation}
which agrees with the order of magnitude of the constraint derived in Ref.~\cite{Farina:2013mla} by demanding finite naturalness in Type-I seesaw models.

\subsection{Quality bounds}
\label{subsec:resultsQB}
%
\begin{figure}[tbp]
     \centering
\includegraphics[width=0.49\linewidth]{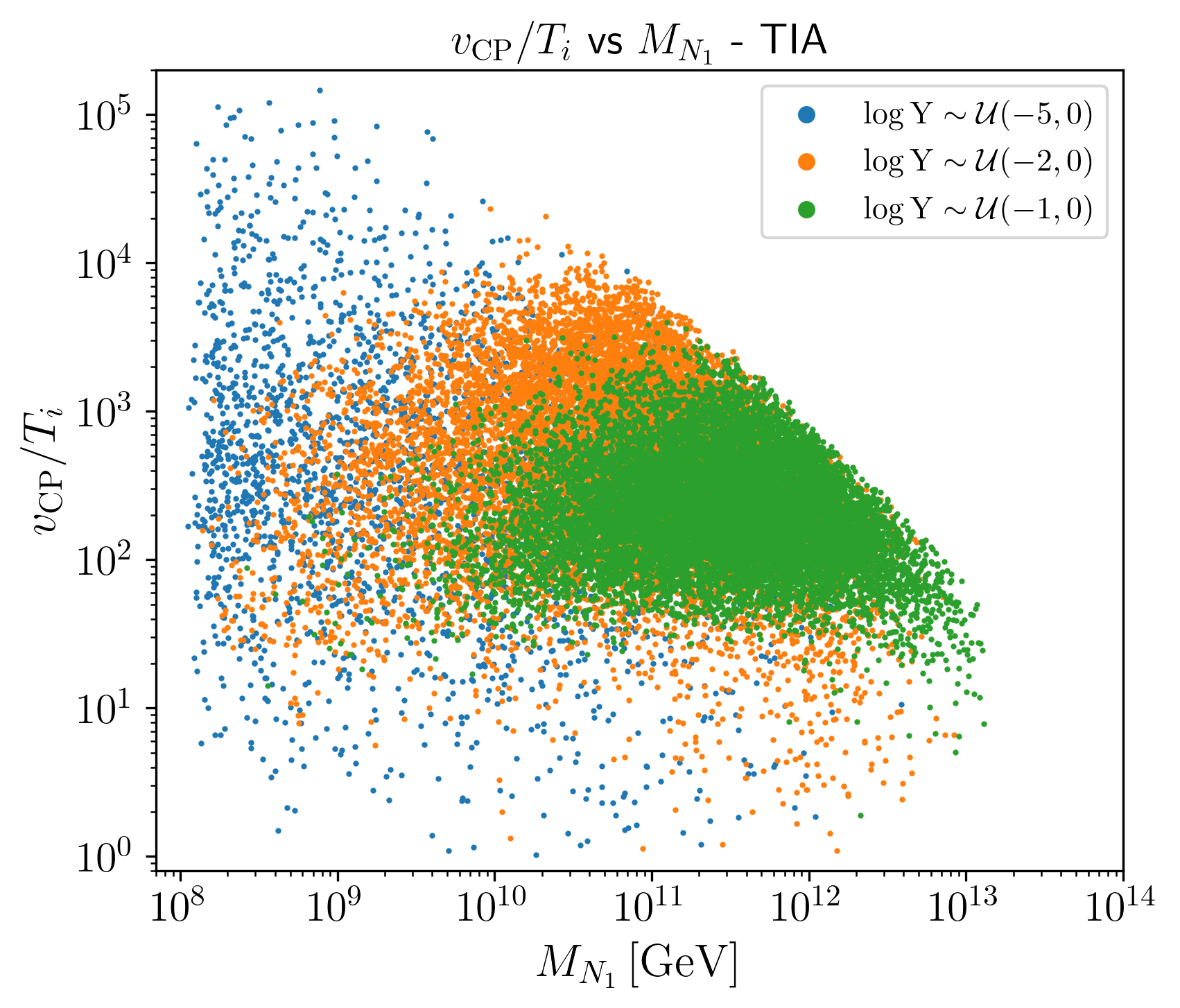}
\includegraphics[width=0.49\linewidth]{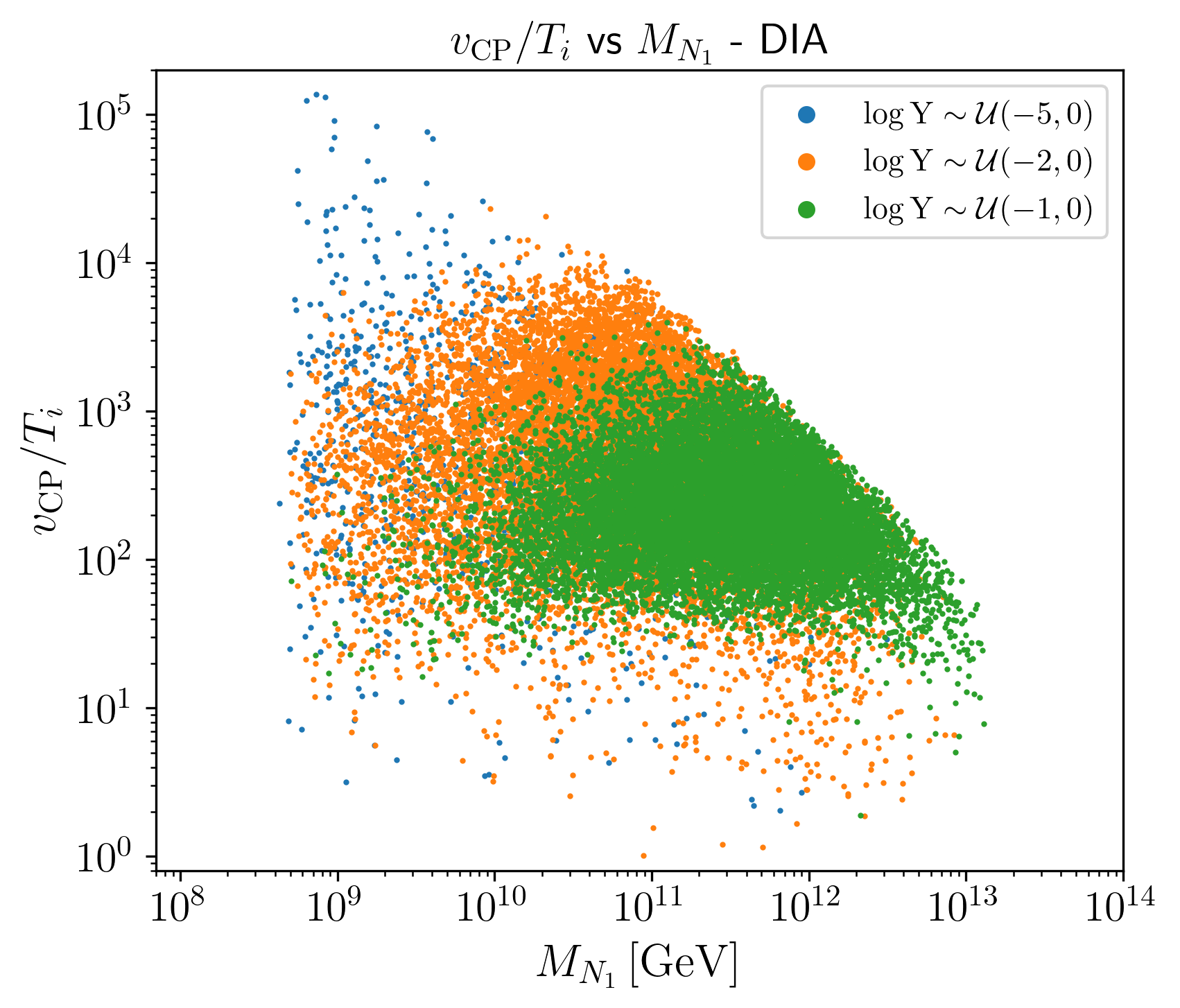}\\
\includegraphics[width=0.49\linewidth]{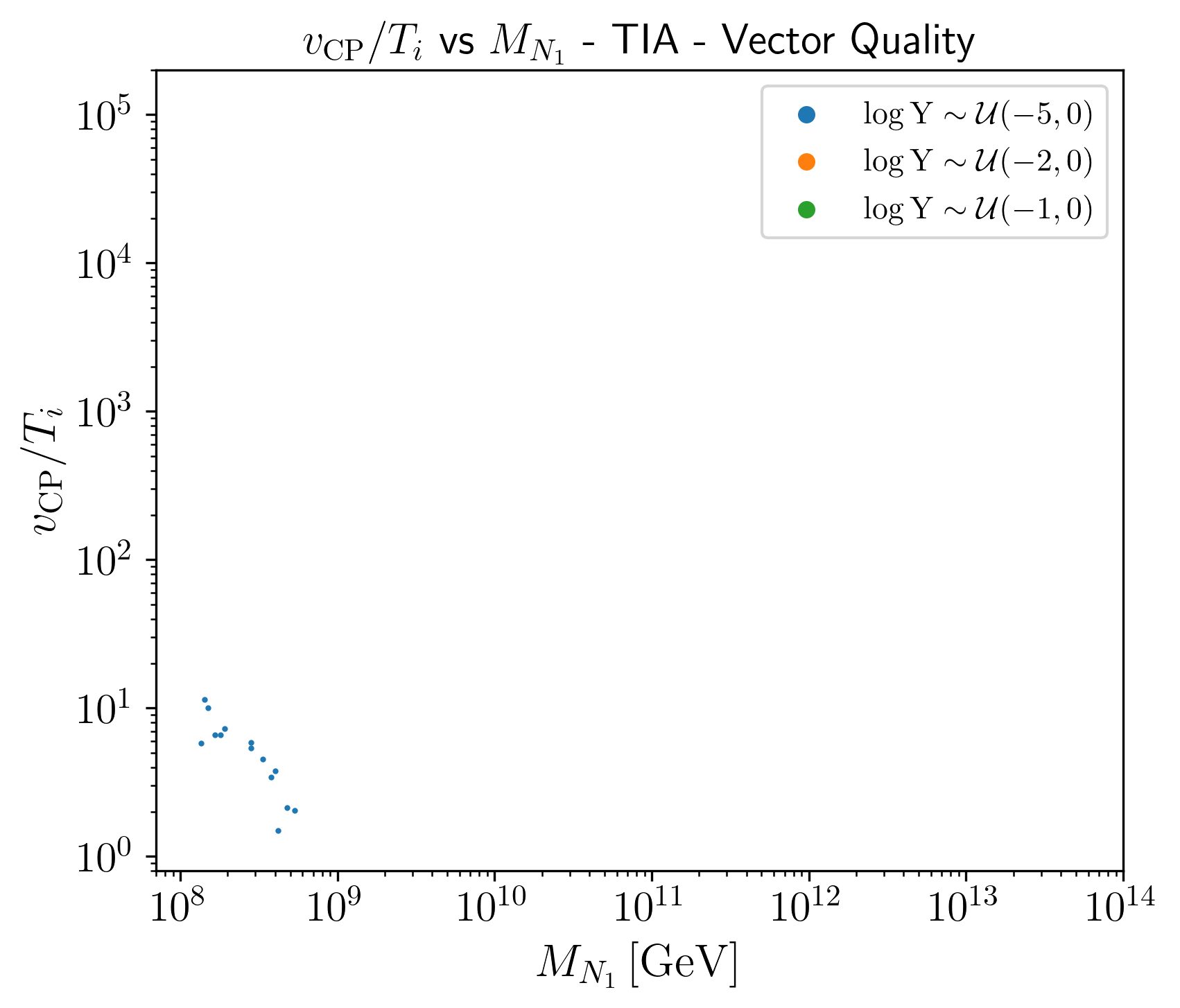}
\includegraphics[width=0.49\linewidth]{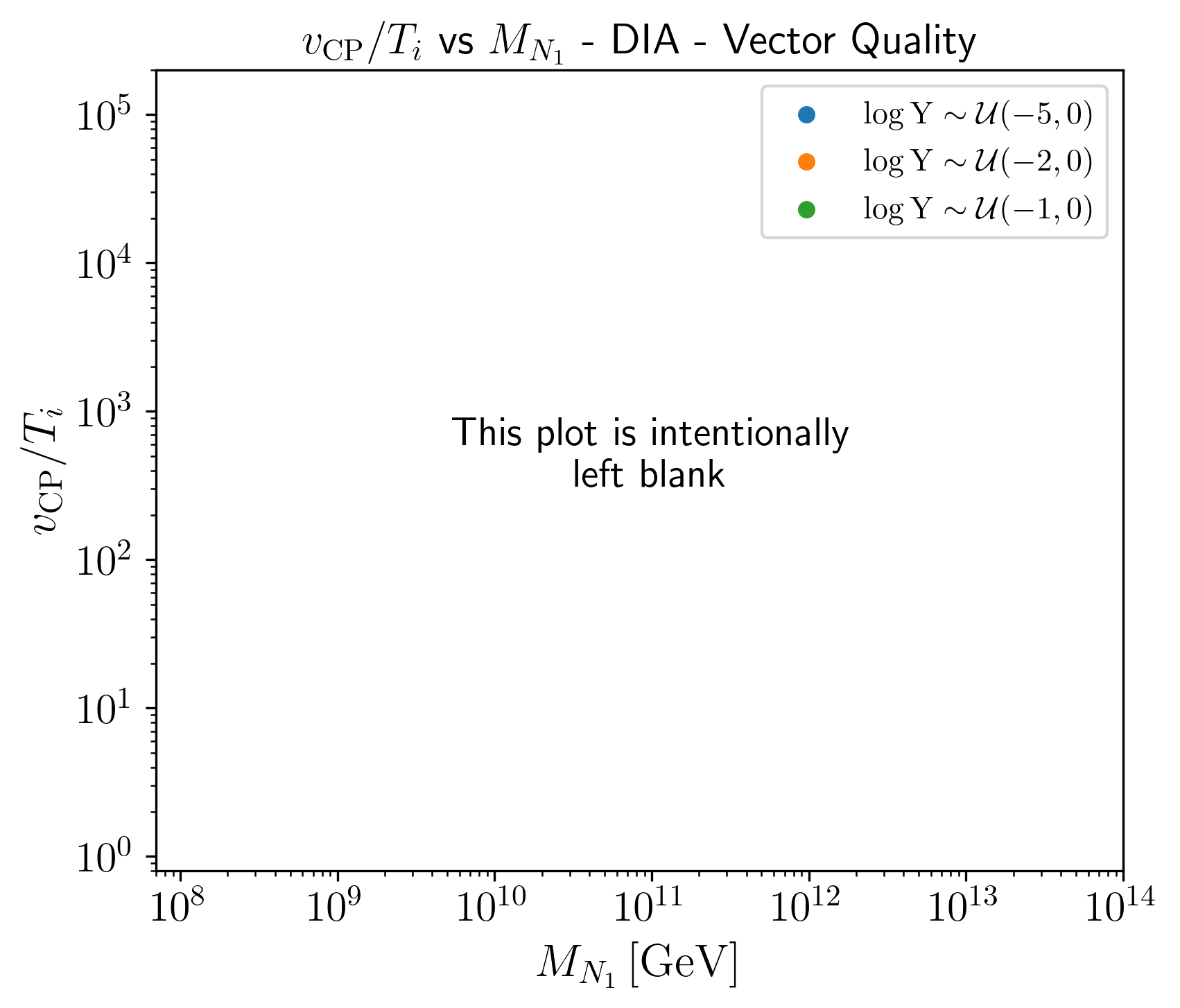}\\
\includegraphics[width=0.49\linewidth]{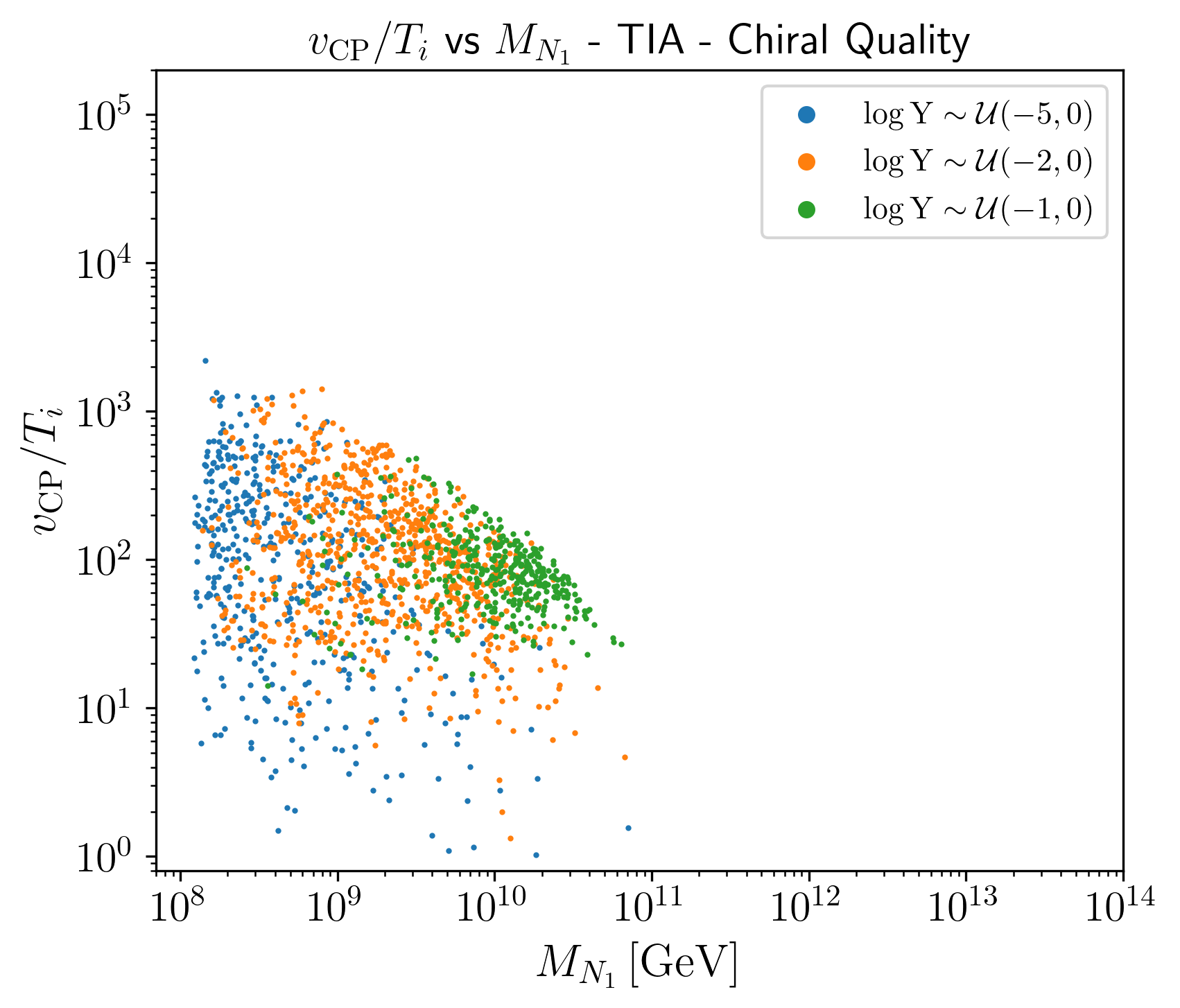}
\includegraphics[width=0.49\linewidth]{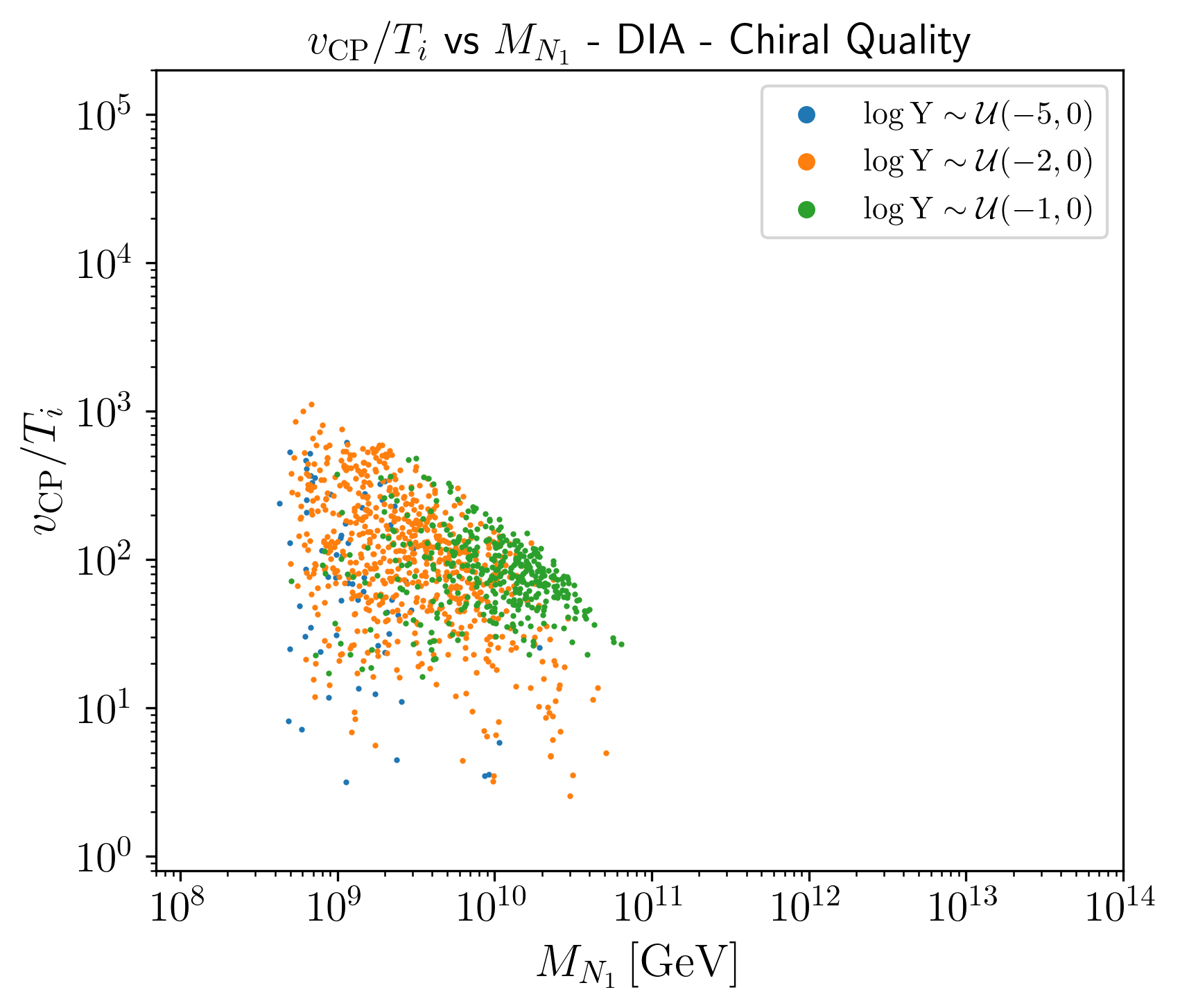}

     \caption{Scatter plots of the successful points for leptogenesis as a function of $M_{N_1}$ and the ratio $v_{\rm CP}/T_i$ (required to be greater than one by our selection procedure), for a TIA (left) and DIA (right) of $N_1$s. In the first row, we show all successful points from our numerical analysis. In the second and third row, we show which of these succesful points are compatible with the quality bounds for automatic Nelson-Barr models with new vector-like and chiral quarks, respectively. Different colors of the points indicated different spreads taken for the Yukawa couplings, according to the legends.}
     \label{fig:vcpTwindow} 
 \end{figure}

 We are now in a position to address whether leptogenesis is viable within automatic Nelson-Barr models. Each point in our sampled parameter space has an associated temperature at which leptogenesis begins, see Eq.~\eqref{eq:Ti}, and a given $\Delta \bar\theta_{\rm QCD}$. In this subsection, we apply the quality bounds discussed in Sec.~\ref{sec:upperboundCP} to the successful points in order to determine the width of the viable window in energy scale.
 
In Fig.\,\ref{fig:vcpTwindow}, we plot the ratio between the scale of SCPV, $v_{\rm CP}$, and the lowest temperature at which leptogenesis becomes successful, $T_i$. We take this dimensionless quantity as a proxy for a the width of the viable energy window for leptogenesis.

In the first row, the plots display the successful points from the parameter scan used in our numerical analysis. As indicated by the different colors, a wider spread in Yukawa couplings allows for a broader window. The points forming the sharp boundary on the upper right of the plots maximize the ratio $v_{\rm CP}/T_i$, achieved when $v_{\rm CP} \sim 10^{14}$~GeV and $T_i \sim 0.1\, M_{N_1}$ (strong washout regime). This cutoff is a consequence of our imposed upper limit on $v_{\rm CP}$ in the scan. However, as we will see in the plots below, this cutoff is purely chosen for convenience since the points consistent with the mechanism’s quality remain unaffected by it.

Green points (for which Yukawa couplings are order one) favor the strong washout regime and therefore $T_i\sim 0.1\, M_{N_1}$, and $v_{\rm CP}\lesssim v_{S}$ (see subsection~\ref{subsec:Majorana}). These points cluster around $v_{\rm CP}/T_i\sim 10^2$, bounded by two horizontal cuts. The upper cut is given by the maximum ratio $v_{\rm CP}/T_i \leq v_S / (0.1 M_{N_1})$ allowed. The minimum $M_{N_1}$ accessible within an order of magnitude spread in the Yukawa couplings is $M_{N_1} \sim 10^{-3} \, v_S$ (see panel C in Fig.~\ref{fig:cartoon}). 
Conversely, the emergent lower cut arises when minimizing the ratio. The largest $M_{N_1}$ possible for our numerics is $v_S/5$, due to the imposed hierarchy $M_{N_1} < M_{N_{j\neq 1}}/5$, which yields a minimum ratio of $v_{\rm CP} / T_i \sim 50$. 
These horizontal cuts relax for the orange and blue points, with the blue ones spanning all five orders of magnitude, as expected. These are statistically favored in the range of highest efficiency,  i.e., $K \lesssim 1$ ($K \sim 1$) for TIA (DIA). For most of these points $T_i\sim M_{N_1}$, and the spread in the plots reflects the spread in the magnitude of the Yukawa couplings.

In the second row, we show the successful points for leptogenesis that are compatible with the quality bound, $\Delta \bar \theta_{\rm QCD} < 10^{-10}$, for automatic Nelson-Barr models with new vector-like quarks. See Sec.~\ref{sec:upperboundCP}, Eq.~\eqref{eq:vCP10to9}.  The absence of points shows that this scenario is incompatible with high-scale leptogenesis starting from a DIA; and only a few points survive for the TIA case. We notice that relaxing condition (iv) in subsection \ref{sec:SimplifiedModel}, i.e.\,considering vector-like neutral leptons, includes a third scale $m_N$ (uncorrelated with $v_{\rm CP}$ and $v_S$). We explicitly verified by more simulations that the results of the scan do not change the available parameter space shown in Fig.~\ref{fig:vcpTwindow}.  The statement that high-scale leptogenesis requires chiral new fermions, already anticipated in Ref.~\cite{Asadi:2022vys}, is strongly based on the assumptions we made. We note some of them could be relaxed (e.g.\,allowing for degenerate neutrinos enabling resonant leptogenesis, or including flavor effects). However, we also note that the bound in Eq.~\eqref{eq:vCP10to9} assumes a particularly optimistic framework. For instance, when naturalness considerations are included and the new quarks are assumed to reside near the TeV scale, the quality bound becomes more stringent, driving $v_{\rm CP}$ down to the TeV range, and thereby creating clear conflict with high-scale leptogenesis.

Plots from the third row show the successful points for leptogenesis consistent with quality bounds for new chiral fermions (vector-like under the SM). See Sec.~\ref{sec:upperboundCP}, Eq.~\eqref{eq:vCPbound}. 
As illustrated by the points, a broader spread in the Yukawa couplings leads to smaller values of $M_{N_1}$ and a wider viable window.

We conclude that automatic Nelson-Barr models with new chiral fermions (vector-like under the SM group) are compatible with high-scale leptogenesis and with the simplest mechanisms for diluting unobserved topological defects. In the next section, we focus on this type of automatic NB models.

\section{Potential signatures}
\label{sec:signatures}
%
High-scale leptogenesis is notoriously inaccessible to experimental verification. However, in the context of automatic Nelson-Barr models, the quark and lepton sectors become correlated due to their shared origin of the CP-violating phase. In this framework, the scale of spontaneous CP violation, $v_{\rm CP}$, provides a connection between leptogenesis and $\bar \theta_{\rm QCD}$.  

In the previous section we showed that automatic NB models with new chiral fermions are compatible with high scale leptogenesis. As extensively discussed in Sec.~\ref{sec:upperboundCP}, higher-dimensional operators can induce a shift in $\bar \theta_{\rm QCD}$, given by Eq.~\eqref{eq:shiftautomatic}. 
Assuming that higher-dimensional operators are suppressed by negative powers of the Planck scale, and both the CP phase and the magnitude of the Wilson coefficients, $[\xi]$, to be order one, Eq.~\eqref{eq:shiftautomatic} becomes
\begin{equation}
\label{eq:deltathetaQCD}
\Delta \bar \theta_{\rm QCD}   \simeq 3\times 10^{-18}\left(\dfrac{v_{\rm CP}}{10^{8} \;{\rm GeV}}\right)^2
\left(\frac{3\times 10^{-5}}{[\mathsf{Y}_d]}\right)\left(1+\frac{v_S}{v_{\rm CP}}\right).
\end{equation}
In the above, we picked the new quark to be down-like, $[\mathsf{Y}_d]= 3 \times 10^{-5}$, but a similar result is expected for up-type quarks. We also have assumed that $m_D / v_{\rm CP} \gg [\mathsf{Y}_d]$. Relaxing this assumption would shift the points towards more testable shifts in $\bar \theta_{\rm QCD}$.

\begin{figure}[]
\includegraphics[width=0.49\linewidth]{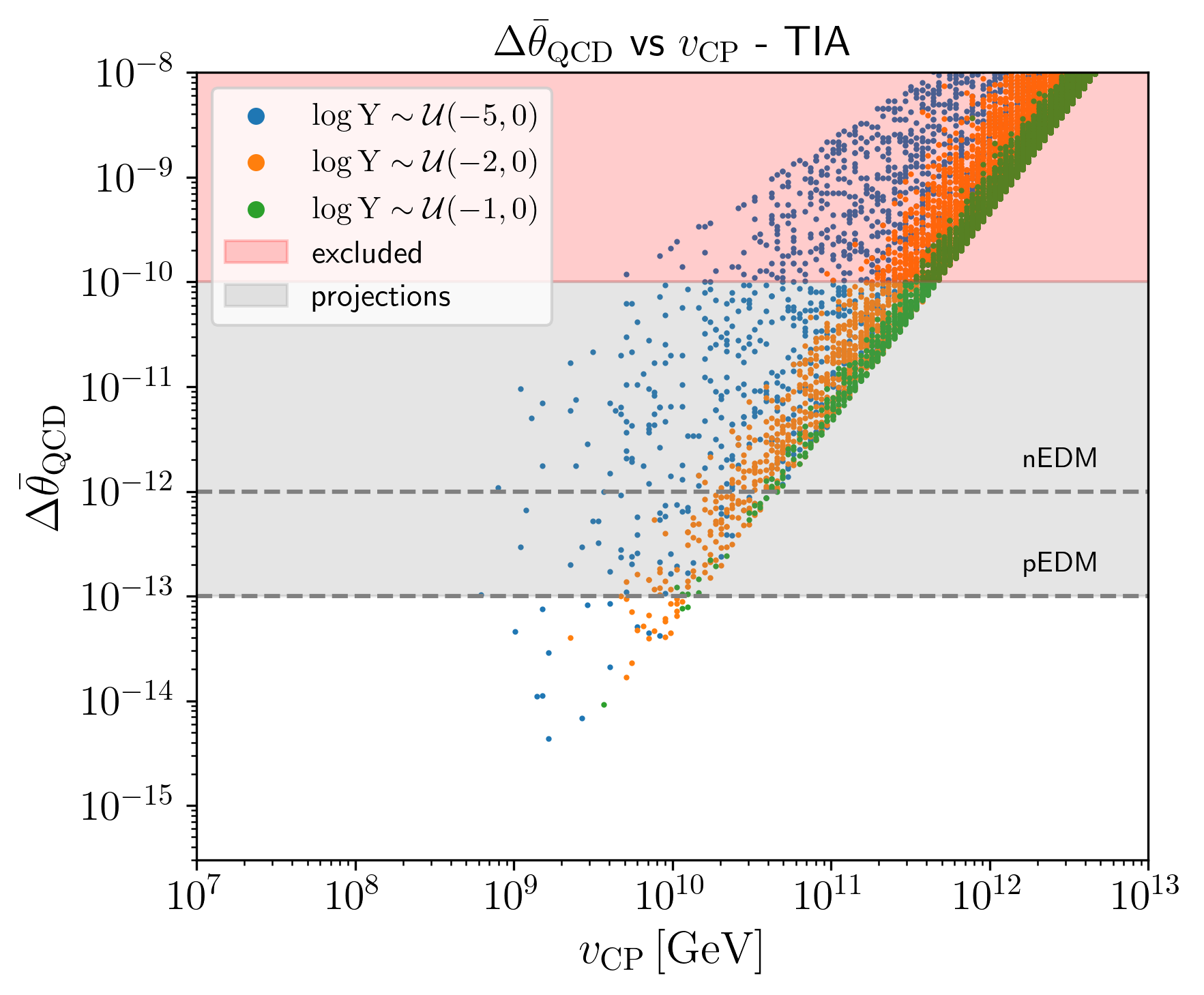}
\includegraphics[width=0.49\linewidth]{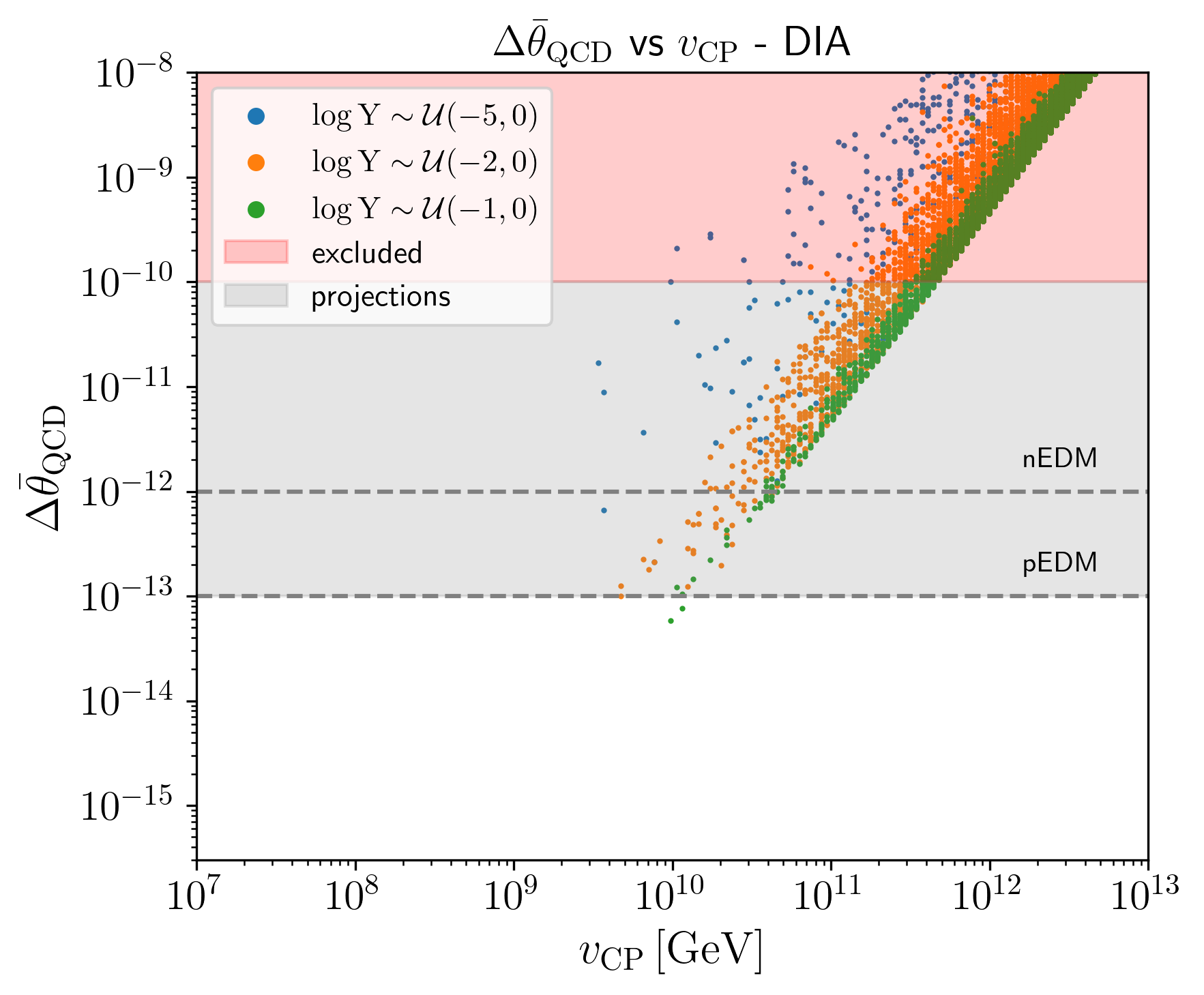}\\
\includegraphics[width=0.49\linewidth]{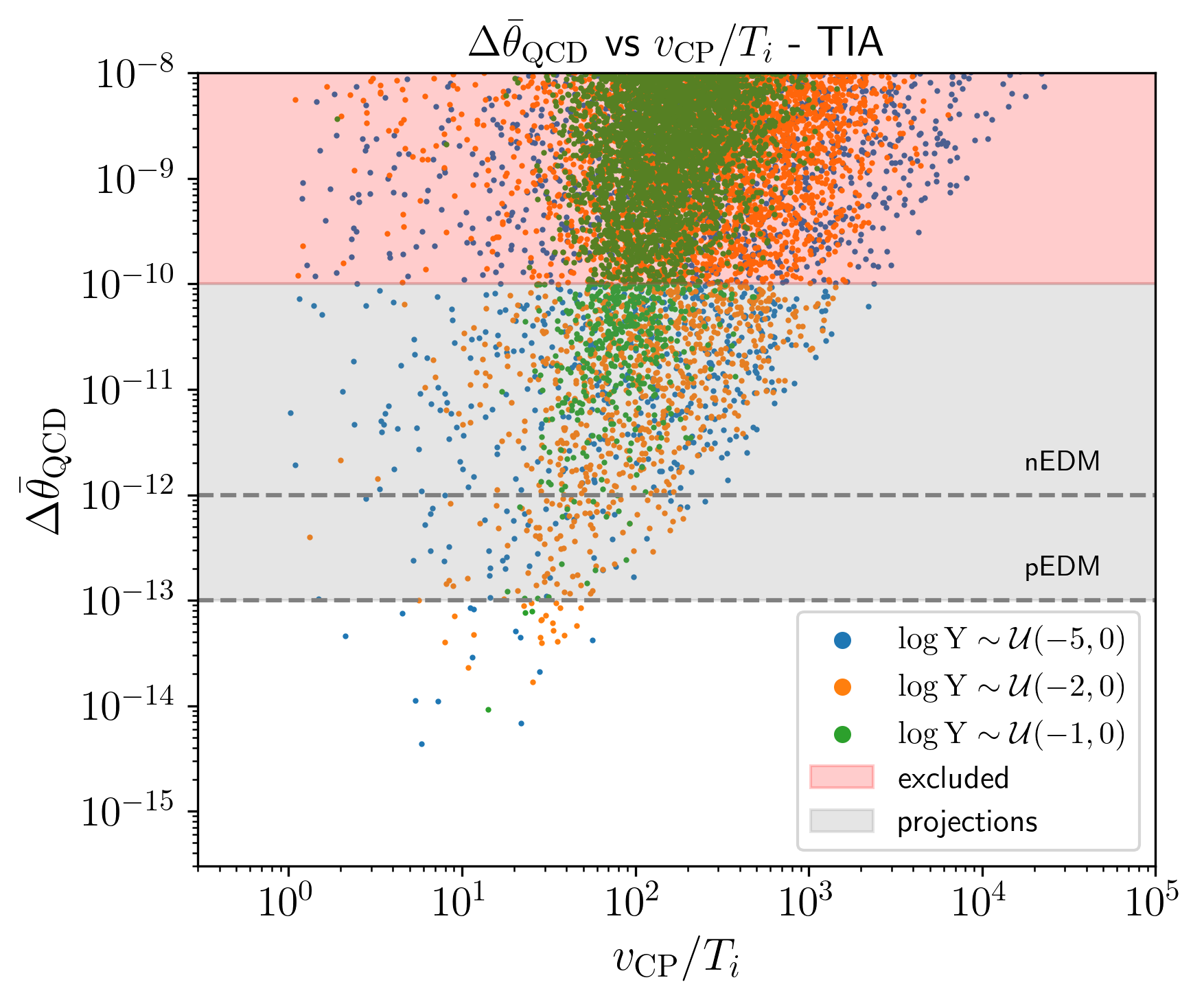}
\includegraphics[width=0.49\linewidth]{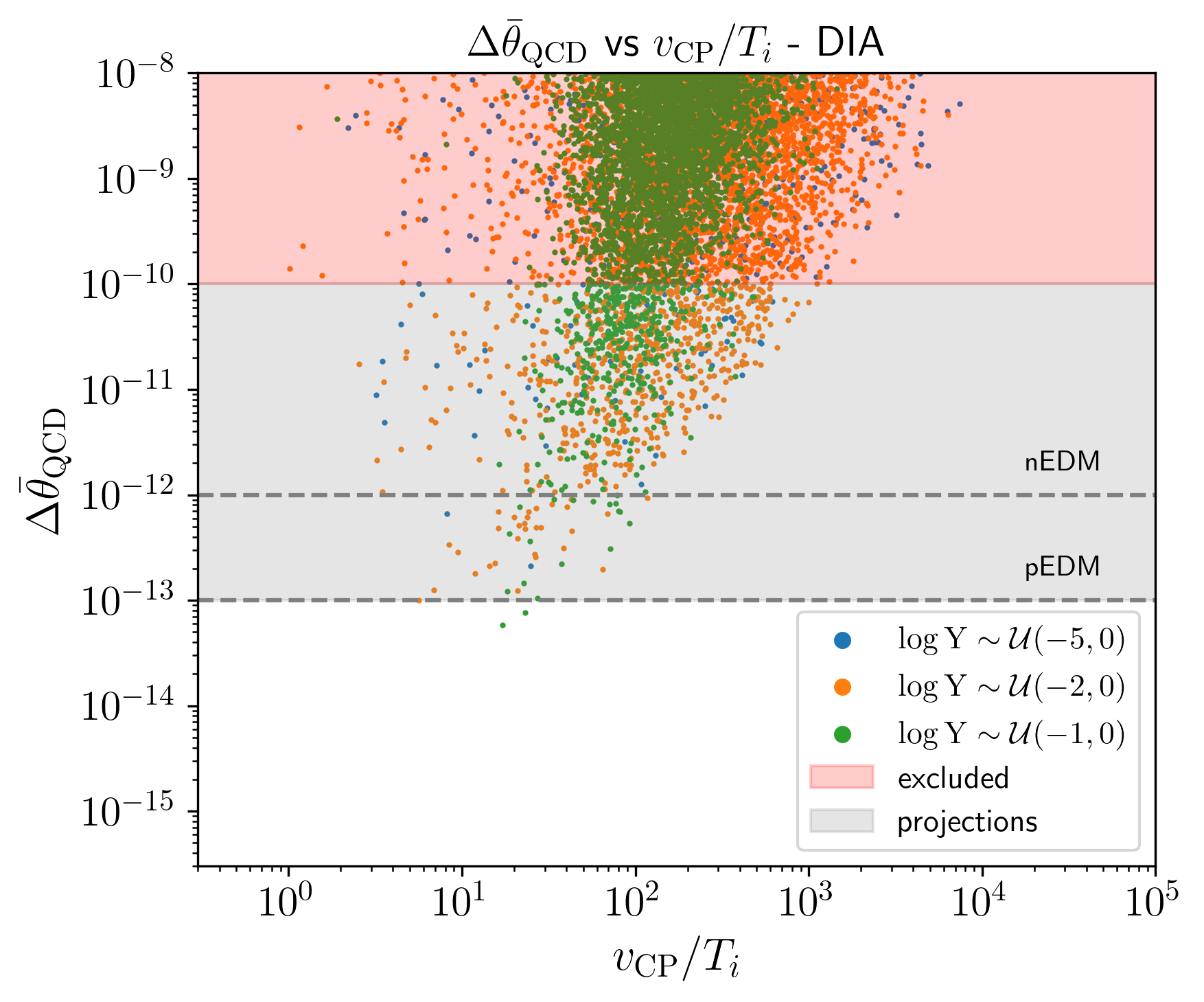}\\
\includegraphics[width=0.49\linewidth]{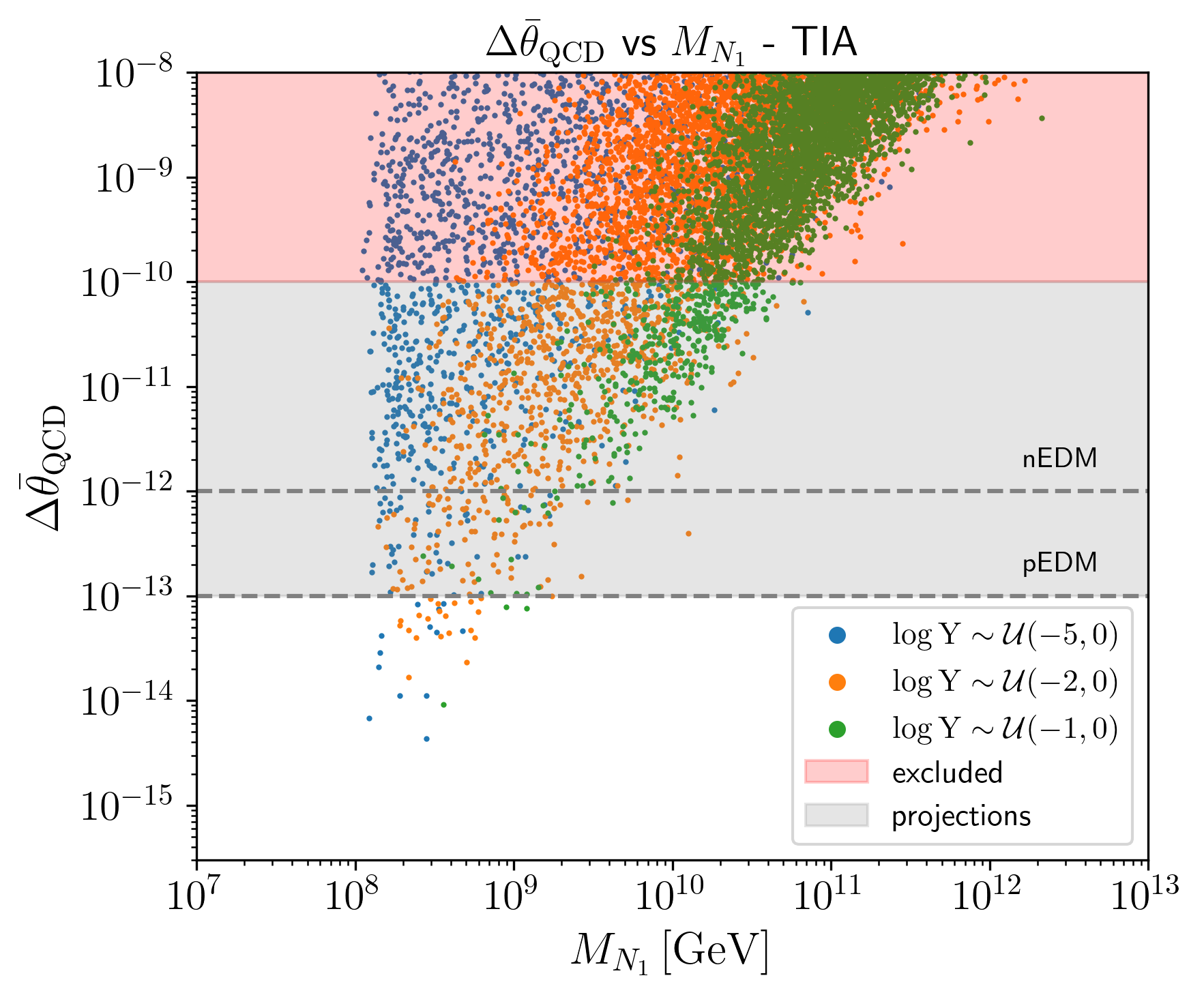}
\includegraphics[width=0.49\linewidth]{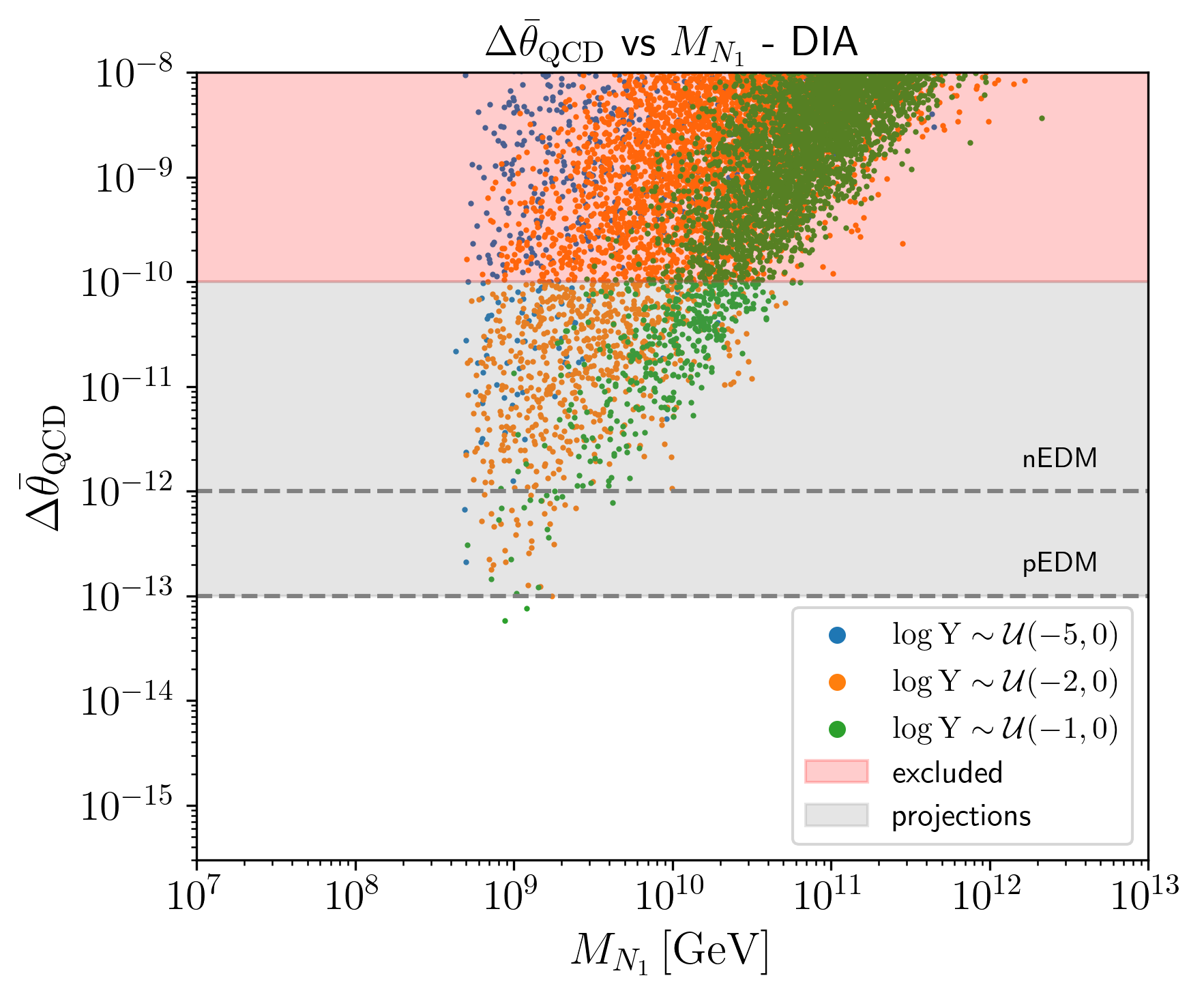}

\caption{Scatter plots of the successful leptogenesis points and their induced shift in $\bar\theta_{QCD}$ computed with Eq.\,\eqref{eq:deltathetaQCD} as a function of $v_{\rm CP}$ (first row), $v_{\rm CP}/T_i$ (second row), and $M_{N_1}$ (third row), for a TIA (left) and DIA (right) of $N_1$s. In red, the zone currently excluded from electric dipole moment experiments.
In gray, delimitated by dashed lines, we show the projective bounds from future experiments on the nEDM~\cite{n2EDM:2021yah,EuropwanEDMprojects:2025okn,nEDM:2019qgk,TUCAN:2022koi}, and the proton electric dipole moment proton electric dipole moment (p-EDM)~\cite{Alexander:2022rmq}.}
\label{fig:DeltaTheta}
\end{figure}

In Fig.~\ref{fig:DeltaTheta}, we show the scatter plots of the points which achieve successful leptogenesis, according to our previous three criteria (see subsection~\ref{subsec:successufullept}). On the y-axis we compute the shift $\Delta\bar\theta _{\rm QCD}$ according to Eq.\,\eqref{eq:deltathetaQCD}. The red shaded area is excluded by the current leading upper bound on the nEDM, $|d_n| < 2.2 \times 10^{-26} \ e\,\text{cm}$~\cite{Abel:2020pzs}, measured at PSI. Such a bound translates into an upper bound on $\bar \theta_{\rm QCD}$ via~\cite{PhysRevD.19.2227,Crewther:1979pi}
\begin{equation}
    |d_N| \sim |\bar \theta_{\rm QCD}| \frac{m_q^*}{\Lambda_{\rm QCD}}\frac{e}{2m_N} \sim |\bar \theta_{\rm QCD}| \times 10^{-16} \ e\, \text{cm},
\end{equation}
where $m_q^* = m_u m_d m_s / (m_u m_d + m_s m_u + m_s m_d)$ is the reduced quark mass, $m_N \sim 1$ GeV is the mass of the nucleon, and $\Lambda_{\rm QCD} \sim 200$ MeV. We also show, in gray, potential constraints  coming from future nucleon EDM measurements. Experiments of ultra-cold neutrons conducted directly within superfluid helium allow higher densities of neutrons and electric fields, enabling improved sensitivities~\cite{EuropwanEDMprojects:2025okn}. For example, the n2EDM experiment, which is currently under construction, is designed to measure the nEDM with sensitivity $|d_n| < 1 \times 10^{-27} \ e\, \text{cm}$, with further possibility to go below the $10^{-27} \ e \,\text{cm}$ regime (e.g. $|d_n| < 5 \times 10^{-28} \ e \,\text{cm}$ in n2EDM Magic)~\cite{n2EDM:2021yah}. Other collaborations~\cite{nEDM:2019qgk,TUCAN:2022koi} aim for similar sensitivities. These experiments will allow to probe $|\bar \theta_{\rm QCD}| > 10^{-12}$, shown on the plots with a dashed line labeled by nEDM. 

Alternatively, projections on the proton EDM (pEDM) are very promising. The pEDM experiment~\cite{pEDM:2022ytu,CPEDM:2019nwp} proposes to search for a pEDM by measuring the vertical rotation of the polarization of a stored proton beam in a storage ring. Phase-1 aims for $|d_p| < 10^{-29} \ e \, \text{cm}$, which improves the sensitivity to $\bar \theta_{\rm QCD}$ by three orders of magnitude. We show this projection on the plots with a dashed line labeled by pEDM. Phase-2 aims to be sensitive to the SM prediction of $10^{-31}\ e \, \text{cm}.$ 

In the first row of Fig.~\ref{fig:DeltaTheta}, $v_{CP}$ is shown on the x-axis. As is particularly evident in the TIA scenario, the successful points lie within two diagonal bands with different slopes. The lower band, where most of the green points are located, corresponds to the regime where $v_S\sim v_{\rm CP} $, leading to $\Delta\bar\theta _{\rm QCD}\propto v_{\rm CP}^2$. The upper band, reached only by the blue points, arises when a larger spread in the Yukawa couplings allows for $v_{\rm CP}\ll v_S $ while still maintaining a sufficiently large CP-violating phase. In this case, $\Delta\bar\theta _{\rm QCD}\propto v_{\rm CP}$. 

In the second row, the shift in $\bar\theta_{\rm QCD}$ is plotted against the available window for leptogenesis, parametrized by $v_{\rm CP}/T_i$. As already shown in Fig.~\ref{fig:vcpTwindow}, smaller predictions for $\Delta \bar \theta_{\rm QCD}$ correspond to narrower windows in which leptogenesis can successfully occur.

In the third row, $\Delta\bar\theta_{\rm QCD}$ is plotted against the mass of the lightest heavy neutrino, $M_{N_1}$. 
There is an upper bound on $M_{N_1}$ consistent with maintaining the quality of the mechanism, approximately $M^{\rm max}_{N_1}\sim 8\times10^{10} \text{ GeV}$. Lower neutrino masses lead to a smaller shift in $\bar \theta_{\rm QCD}$.  This trend reflects a general feature: improved experimental sensitivity to $\Delta \theta_{\rm QCD}$ favors both a lower scale for the new heavy neutral fermions and lower scale for SCPV, as shown in the first-row plots.

In both TIA and DIA scenarios, we can appreciate that the vast majority of the successful points are within reach of future EDM experiments~\cite{EuropwanEDMprojects:2025okn}.

\section{Conclusions}
\label{sec:conclusions}

In this study, we have numerically shown that the observed baryon asymmetry can be reproduced in automatic Nelson-Barr models through leptogenesis while maintaining a high-quality solution to the strong CP problem. In these models, where all sources of CP violation are tied to a single symmetry-breaking scale, $v_{\rm CP}$, CP violation must account for the nonzero phase in the quark sector (Jarlskog determinant), but also for the observed baryon asymmetry.  We consequently studied the resulting correlations among the observed CP-violating phenomena.

In automatic NB models, the required quark mass texture to achieve a vanishing tree-level $\bar \theta_{\rm QCD}$ is a consequence of a gauge symmetry (in our scenario, an abelian force). This framework allows for a higher spontaneous CP violation (SCPV) scale facilitating the implementation of mechanisms to dilute the domain walls associated with its breaking. 
On the model building side, we mirrored the quark sector by introducing five heavy neutral leptons, following the same structure in the couplings to the SCPV scalar fields. In addition, we let the three right-handed neutrinos to get a Majorana mass term from the vev of the field that breaks the extra gauge symmetry, $v_{\rm S}$, keeping the number of relevant scales and phases to a minimum.
To be maximally conservative, we adopted intentionally pessimistic assumptions to test NB leptogenesis: we modeled leptogenesis dynamics using only decays and inverse decays, leaving out other processes (such as  scatterings, flavor, thermal effects,\ldots), that usually enhance the generated asymmetry. We further imposed a hierarchical spectrum for the heavy leptons so that the Davidson–Ibarra bound holds, ruling out resonant enhancements.

A numerical scan over the two scales at play, $v_{\rm CP}$ and $v_{\rm S}$, and a randomized sampling of Yukawa couplings over several orders of magnitude manifest a robust window of two to three orders of magnitude between $v_{\rm CP}$ and the leptogenesis temperature $T_i$, assuming the new quarks are chiral under the new gauge group. This gap is wide enough to accommodate any mechanism needed to dilute SCPV‐induced domain walls and still reheats the universe above $T_i$. For vector-like quarks, the quality bound tightens and high-scale leptogenesis faces clear tension -- though our conservative assumptions suggest that more complex models or a fuller treatment of leptogenesis dynamics could reopen viable parameter space.

High-scale leptogenesis is notoriously difficult to experimentally constrain. However, NB models implementing it quantitatively correlate -- due to to the common SCPV scale -- the induced shift in the QCD vacuum angle, $\Delta\bar\theta_{\rm QCD}$, coming from higher-dimensional operators, with the baryon asymmetry predicted, $\eta_B$. From our numerical analysis, a clear correlation emerges between $\Delta\bar\theta_{\rm QCD}$ and the available leptogenesis window $v_{\rm CP}/T_i$; improved sensitivity to $\bar\theta_{\rm QCD}$ corresponds to narrower windows, thereby constraining high-scale leptogenesis.
Another interesting insight we can deduce from our numerics is that enforcing the NB quality bound caps the lightest heavy–neutrino mass at $M_{N_1}\lesssim8\times10^{10}\,\mathrm{GeV}$.
Remarkably, the vast majority of the parameter space available induces a shift in $\bar\theta_{\rm QCD}$ that lies within the reach of current and future nucleon EDM experiments and well above the SM prediction of $\bar \theta_{\rm QCD} \sim 10^{-15}$. Although we cannot exclude Nelson-Barr models if such a shift is not observed, further tuning may be needed to naturally accommodate high-scale leptogenesis while still having a high-quality solution to the strong-CP problem.

So far, we have shown that automatic NB models solve a ten order of magnitude discrepancy between the observed upper bound on $\bar\theta_{\rm QCD}<10^{-10}$ and the naive theoretical expectation for it to be $\mathcal{O}(1)$ via a high-quality mechanism that is robust under UV corrections and, under certain assumptions, compatible with high-scale leptogenesis. 
However, the introduction of new heavy fermions generically adds dangerous radiative corrections to the effective scalar potential, 
and a similar fine tuning to the one that
NB models succeed to solve may arise between different contributions to a physical
quantity. Demanding finite naturalness, the lowest mass solutions found $M_{N_1} \sim 10^8 \text{ GeV}$, require three-four orders of magnitude in fine tuning between the squares of the bare mass of the Higgs and its radiative contributions, manifesting the usual tension with high-scale leptogenesis. A possibility to relax the tuning in this case is to consider enhancements of the produced asymmetry in order to evade the Davidson-Ibarra bound (e.g.\,population dynamics, resonant spectrum, ARS leptogenesis\ldots). 
A second way would be to explore a low scale baryogenesis scenario in which the out-of-equilibrium decays of the complex fields responsible for SCPV produce an asymmetry in the primordial universe. However, quick thermalization with the SM bath due to the hypercharges as well as mixings between all neutral leptons strongly challenge the mechanism. We leave studies in these directions for future work.

In summary, automatic Nelson-Barr models with chiral fermions offer a high-quality solution to the strong CP problem and can naturally accommodate high-scale leptogenesis, successfully accounting for the observed matter-antimatter asymmetry.

\acknowledgments
We thank Pavel Fileviez Pérez and Mark Wise for numerous discussions on Nelson-Barr models. We also thank Federico Cima, Jim Cline, Miguel Escudero, Michele Papucci, Ryan Plestid, and Riccardo Rattazzi.
S.P. thanks the TH department at CERN for its hospitality during the completion of this work. This material is based upon work supported by the U.S. Department of Energy, Office of Science, Office of High Energy Physics, under Award Number DE-SC0011632.

\appendix
\section{Automatic Nelson-Barr UV completions}
\label{app:UV}
As discussed in the main text, we aim to gauge the $\mathbb{Z}_2$ (or any discrete symmetry imposing the zeroes in the mass matrix) usually introduced in simplified NB models such as the minimal theory proposed by BBP~\cite{Bento:1991ez}. For simplicity, we assume that it is a consequence of an abelian symmetry, $\text{U}(1)$.

The simplest attempt would be to assume that the new vector-like down (or up) quarks are vector-like under the entire symmetry group $\SU(3) \otimes \SU(2)_L\otimes \text{U}(1)_Y \otimes \text{U}(1)$:
$$D_L \sim (3,1,-1/3,{\cal Q}_D), \qquad \text{and} \qquad D_R \sim (3,1,-1/3,{\cal Q}_D),$$
and that the complex scalars responsible for SCPV have quantum numbers $X_a \sim (1,1,0,{\cal Q}_D)$.
Although the $\mathbb{Z}_2$ symmetry has been gauged, the vector-like nature of the new fermions leads to contributions from dimension five operators to $\Delta \theta_{\rm QCD}$, with a quality bound that is in tension with thermal leptogenesis (see Eq.~\ref{eq:vCP10to9} and the discussion in subsection~\ref{subsec:resultsQB}, around Fig.~\ref{fig:vcpTwindow}).

In order to make the new quarks chiral under the new symmetry (still vector-like under the SM), we assume they carry different $\text{U}(1)$ charges. In this case, however, the triangle anomalies ${\cal A}[\SU(3)^2\otimes \text{U}(1)]$, ${\cal A}[\text{U}(1)_Y\otimes \text{U}(1)^2]$, ${\cal A}[\text{U}(1)_Y^2 \otimes \text{U}(1)]$, and ${\cal A}[\text{U}(1)^3]$ do not cancel\footnote{For the charges being equal in magnitude but opposite sign, ${\cal A}[\text{U}(1)_Y\otimes\text{U}(1)^2]=0$, but the other two still survive.}. In particular, ${\cal A}[\SU(3)^2 \otimes \text{U}(1)]$ requires at least an extra field charged under QCD and the new force. By limiting ourselves to SM ingredients, one could try charging the quark doublets under $\text{U}(1)$, but that would spoil the anomalies involving $\SU(2)_L$ unless lepton doublets are also $\text{U}(1)$-charged. Alternatively, one could try charging the right-handed quarks under the new symmetry. We encourage the reader to explore all possible charge combinations and see for themselves why this approach fails.

Therefore, we are forced to add either exotic fields (on top of $D_L$ and $D_R$) or extend the new force to the lepton sector. In the context of Nelson-Barr models, the latter is encouraged by the experimental evidence of flavour number violation and the suggestion of CP violation in the PMNS matrix.

A simple way to generate the textures required by the Nelson-Barr mechanism is to consider a $\text{U}(1)$ symmetry under which the SM is anomaly-free, plus the following new vector-like (under the SM gauge group) fermions:~\cite{Perez:2023zin}
\begin{align*}
&U_R \sim (3,1,2/3,-{\cal Q}_R), \quad && U_L \sim (3,1,2/3,{\cal Q}_L)\\
&D_R \sim (3,1,-1/3,{\cal Q}_R), \quad && D_L \sim (3,1,-1/3,-{\cal Q}_L),\\
&E_R \sim (1,1,-1,{\cal Q}_R), \quad && E_L \sim (1,1,-1,-{\cal Q}_L),\\
&N_R \sim (1,1,0,-{\cal Q}_R), \quad && N_L \sim (1,1,0,{\cal Q}_L).
\end{align*}

 Motivated abelian (anomaly-free) extensions of the standard model are $\text{U}(1)_{R}$ and $\text{U}(1)_{B-L}$ with three right-handed neutrinos $\nu_R$. The automatization of the NB mechanism in the quark sector in these models was studied in Ref.~\cite{Perez:2023zin}. We note that this UV completion satisfies the four requirements listed in subsection~\ref{sec:SimplifiedModel} for the lepton sector.

 To implement high-scale leptogenesis in these models, Majorana neutrino masses are needed. In the $\text{U}(1)_R$ scenario, under the charge relation $|{\cal Q}_L+{\cal Q}_R|=2|{\cal Q}_{\nu_R}|$, the right-handed neutrinos $\nu_R$ get a Majorana mass from the spontaneous symmetry breaking of $\text{U}(1)_R$. In the context of $\text{U}(1)_{\rm B-L}$, the new scalars $X_a$ cannot couple simultaneously to quarks and leptons as they have different $B-L$ charges. In order to implement NB in both quark and lepton sector, and avoid heavy charged relics (assuming that their mass lies below the reheat temperature of the universe),  ${\cal Q}_L = 1/3$ and $X \sim (1,1,0,2/3)$~\cite{Perez:2023zin}. With ${\cal Q}_R  = 5/3$, Majorana masses for the right-handed neutrinos would be allowed. 
 
We close this appendix by including another possibility where both lepton and baryon numbers are local, i.e. $\text{U}(1)_B$ and $\text{U}(1)_L$ are gauged. The fermion content of this theory was proposed in Ref.~\cite{FileviezPerez:2011pt}. Gauge anomalies are cancelled by adding a family of vector-like fermions. Their quantum numbers under $\text{SU}(3) \otimes \SU(2)_L\otimes \text{U}(1)_Y \otimes \text{U}(1)_B \otimes \text{U}(1)_L$ are listed in Table~\ref{tab:QN}. 
   \begin{table}[t]
   \centering
  \begin{tabular}{| c | c |  c  | c | c | c|   }
  \hline
  Field & $\SU(3)$ & $\SU(2)_L$ & $\text{U}(1)_Y$ & $\text{U}(1)_B$ & $\text{U}(1)_L$\\
  \hline
  \hline
  $Q_L ~(\times 3)$ & 3 & 2 & 1/6& 1/3 & 0 \\
  $L_L ~ (\times 3)$ & 1 & 2 & -1/2 &  0 & 1\\
  $u_R ~(\times 3)$ & 3 & 1 & 2/3 & 1/3 & 0 \\
  $d_R ~(\times 3)$ & 3 & 1 & -1/3 & 1/3 & 0 \\
  $e_R ~(\times 3)$ & 1 & 1 & -1 & 0 & 1 \\
  $\nu_R ~ (\times 3)$ & 1 & 1 & 0 & 0 & 1\\
  \hline
  \hline
  $q_L$ & 3 & 2 & 1/6 & $B-1$ & 0 \\
  $\ell_L$ & 1 & 2 & -1/2 & 0 & $L-3$\\
  $U_R $ & 3 & 1 & 2/3 & $B-1$ & 0\\
  $D_R$ & 3 & 1 & -1/3 & $B-1$ & 0\\
  $E_R$ & 1 & 1 & -1 & 0 & $L-3$\\
  $N_R$ & 1 & 1 & 0 & 0 & $L-3$\\
  \hline
  $q_R$ & 3 & 2 & 1/6 & $B$ & 0 \\
  $\ell_R$ & 1 & 2 & -1/2 & 0 & $L$  \\
  $U_L$ & 3 & 1 & 2/3 & $B$  & 0\\
  $D_L$ & 3 & 1 & -1/3 & $B$ & 0 \\
  $E_L$ & 1 & 1 & -1 & 0 & $L$ \\
  $N_L$ & 1 & 1 & 0 & 0 & $L$\\
  \hline 
  \end{tabular}
  \caption{Quantum numbers of the field content under the  $\text{SU}(3) \otimes \SU(2)_L\otimes \text{U}(1)_Y \otimes \text{U}(1)_B \otimes \text{U}(1)_L$ gauge group. The first box contains just the SM fields. $B$ and $L$ are unconstrained baryon and lepton charges, respectively.}
  \label{tab:QN}
  \end{table}
  As long as $B \neq 4/3$, and $L \neq 4$, condition (iii) is satisfied in both the quark and lepton sectors. We assume that the local baryon and lepton numbers are broken spontaneously by the vevs of $S_B \sim (1,1,0,1,0)$ and $S_L\sim (1,1,0,0,3)$, respectively, generating heavy masses for the new fermions\footnote{In Ref.~\cite{FileviezPerez:2011pt}, where this model was proposed, $S_L\sim (1,1,0,2)$ generates Majorana masses for the neutrinos via the type-I seesaw, while the new leptons get mass via the electroweak symmetry breaking. The latter is in tension with the branching fraction of the SM Higgs to two photons, as the new Yukawa couplings need to be ${\cal O}(1)$.}.
  
  To implement the  NB mechanism, we need the new scalars $X_a$ responsible for SCPV to bridge both sectors. This fixes the charge of $X_a$ to be $|{\cal Q}_{X_a}|= |B-1/3|$. To extend the mechanism to the lepton sector, condition (ii) must be fulfilled, i.e. $|{\cal Q}_{X_a} | =| L-1 |$.  As an automatic consequence of anomaly cancellation, conditions (i) and (iv) are satisfied. 
  
  Regarding high-scale leptogenesis in this scenario; for example, by fixing $L= 3$, Majorana masses for $\nu_R$ ($\mathsf{M}_R$ dynamically generated by $\langle X_a \rangle$) and for $N_R$ (bare mass $m_R$) are allowed. Instead, fixing $L=-1$, on top of the dynamical Majorana masses for the $\nu_R$s, the couplings $N_R^T C \nu_R S_L$ and $N_L^T C N_L X^*$ generate dynamically a $m_{RR} \neq 0$ and $m_L \neq 0$, see Eq.~\eqref{eq:LagrangianLepto}, respectively. 
  
  We note that, within this UV completion, baryon number is broken in one unit. As noted in Ref.~\cite{Duerr:2013dza}, dimension-nine operators such as 
  \begin{equation}
\frac{\xi}{\Lambda^5} \, Q_L Q_L Q_L L_L S_B^* S_L^* X_a + \cdots
  \end{equation}
can mediate proton decay. Here, $\Lambda$ is the scale that UV-completes the theory presented. In spite of the severe constraints on the lifetime of the proton, $\tau_p \lesssim 10^{34}$ years~\cite{Super-Kamiokande:2020wjk}, any of the scales $\langle X \rangle$, $\langle S_L \rangle$ and $\langle S_B \rangle$ are consistent with experiment if the cutoff of the theory is at the Planck scale. 

It is worth noting that the new gauge symmetry allowing for the Nelson-Barr mechanism also generates interactions amongst the new and the SM sectors. In the aforementioned examples, the lightest neutral lepton $N_1$ could easily thermalize at early times via the following interactions:
\begin{equation}\label{eq:N1scattering}
    \begin{gathered}
\begin{tikzpicture}[line width=1.5 pt,node distance=1 cm and 1.5 cm]
\coordinate[label =left: $N_1$] (i1);
\coordinate[below right= 1cm of i1](v1);
\coordinate[ right= 0.5cm of v1,label= below:$Z'$](vaux);
\coordinate[below left= 1cm of v1, label= left:$N_1$](i2);
\coordinate[right = 1 cm of v1](v2);
\coordinate[above right = 1 cm of v2, label=right: $f_{\rm SM}\text{, }F$] (f1);
\coordinate[below right =  1 cm of v2,label=right: $ f_{\rm SM}\text{, }F$] (f2);
\draw[fermionnoarrow] (i1) -- (v1);
\draw[fermionnoarrow] (i2) -- (v1);
\draw[vector] (v1) -- (v2);
\draw[fermion] (v2) -- (f1);
\draw[fermion] (f2) -- (v2);
\draw[fill=gray] (v1) circle (.1cm);
\draw[fill=gray] (v2) circle (.1cm);
\end{tikzpicture}
\end{gathered} :\quad 
\begin{gathered}
    \sigma_{N_1 N_1 \to (Z')^* \to \bar f f} = \frac{(g')^4 \sqrt{s}}{8\pi} \frac{\sqrt{s-4M_{N_1}^2}}{(s-M_{Z'}^2)^2+\Gamma_{Z'}^2M_{Z'}^2}.
\end{gathered}
\end{equation}
The strength of the gauge coupling (or, alternatively, the mass of the new gauge boson\footnote{The spontaneous breaking scale of the new symmetry $\text{U}(1)$, $v_S$, will determine the new gauge boson mass up to the gauge coupling, $M_{Z'}= {\cal O}(g' v_S)$.}) will determine which scenario of leptogenesis takes place: thermal or dynamical. A study on how these interactions affect the initial thermal population of heavy leptons, as well as the interplay with the out-of-equilibrium condition required at the time of leptogenesis, can be found in Refs.~\cite{Plumacher:1996kc,FileviezPerez:2021hbc}. 

\section{The lightest heavy neutrino mass}
\label{app:Sambound}
In this appendix, we will analytically derive the relation of Eq.\,\eqref{eq:SamBound} for a $3\times3$ matrix with the same features of the heavy mass matrix for the simplified model in Eq.\,\eqref{eq:LagrangianLepto}.

The real positive singular values of the following mass matrix (complex symmetric) 
\begin{equation}
    \mathcal{M}=
    \begin{pmatrix}
        0&\mu&m_N\\
        \mu&m_R&0\\
        m_N&0&0
    \end{pmatrix}\,,
\end{equation}
where $m_N, m_R\in \mathbb{R}$ and $\mu \in \mathbb{C}$, can be expressed as the solutions of the following third-order algebraic equations
\begin{equation}
\label{eq:3ordeqs}
    x^3\pm m_Rx^2-\bar{m}^2_N x\mp m_Rm_N^2=0\,,
\end{equation}
where $\bar{m}^2_N=m^2_N+|\mu|^2$.

The discriminant is the same for both equations since it is bilinear in the coefficients of the $x^2$ and $x^0$ terms, and it reads as
\begin{equation}
    \Delta=4\bar{m}^6_N-27m_N^4m_R^2+18\bar{m}^2_Nm_N^2m^2_R+m_R^2\bar{m}_N^4+4m_N^2m_R^4\,.
\end{equation}
Using that $m_R,m_N>0$, and $\bar{m}_N>m_N$ by definition, we have that $\Delta>0$. Hence, the two equations above will have three real solutions each. In order to correctly identify the three mass eigenvalues, we want to isolate the three positive solutions.

For clarity, we can rewrite the two equations in \eqref{eq:3ordeqs} as 
\begin{equation}
    m_R =\pm \frac{x(x^2-\bar{m}^2_N)}{x^2-m_N^2}\,.
\end{equation}
Using again the property that all the known coefficients are real and positive, the three mass eigenstates can be identified as follows.
\begin{itemize}
\item Picking the minus sign, $x>0$ iff
\begin{equation}
    {\rm (a):} \;\begin{cases}
    x>m_N\\
    \bar{m}_N>\sqrt{x^2+x-\frac{m_N^2}{x^2}}
    \end{cases}
\end{equation}
\item Picking the plus sign, $x>0$ iff
\end{itemize}
\begin{equation}
    {\rm (b):} \;\begin{cases}
    x>m_N\\
    \bar{m}_N<\sqrt{x^2+x-\frac{m_N^2}{x^2}}
    \end{cases}\qquad {\rm or} \qquad {\rm (c):} \;
    x<m_N
\end{equation}
Now, we want to determine which root is the lightest. We can have two mass hierarchies (assuming all non-degenerate mass parameters).
\begin{enumerate}
    \item $\mathbf{\underline{m_R<m_N}:}$ In this case, conditions in case (a) imply that $m_N<x<\bar{m}_N$. Conditions for case (b) give $x>\bar{m}_N$. Hence, case (c) corresponds to the lightest eigenvalues for $x$.
    \item $\mathbf{\underline{m_R>m_N}:}$ Case (c) trivially gives the lightest eigenvalue since $x$ is smaller than the lightest scale.
\end{enumerate}

As we proved above, the lightest mass eigenstate is always expressed by the solution $x_c$ to the following equation
\begin{equation}
    m_R = \frac{x(\bar{m}^2_N-x^2)}{m_N^2-x^2}\,\quad {\rm with }\quad x<m_N\,.
\end{equation}
We can find another upper-bound on the solution, rewriting the above equation as
\begin{equation}
\frac{m_R}{x_c}=\frac{(\bar{m}^2_N-x_c^2)}{m_N^2-x_c^2}=1+\frac{|\mu|^2}{m_N^2-x_c^2}>1+\frac{|\mu|^2}{m_N^2}=\left(\frac{\bar{m}_N}{m_N}\right)^2\Rightarrow x_c<m_R\left(\frac{m_N}{\bar{m}_N}\right)^2\,.
\end{equation}
Hence, we conclude that
\begin{equation}
    M_{N_1}\equiv x_c <{\rm Min}\left\{m_N,m_R\left(\frac{m_N}{\bar{m}_N}\right)^2\right\}\,.
\end{equation}

\section{The Davidson-Ibarra bound for more than three heavy neutrinos}
\label{app:DIbound}
The CP asymmetry summing over the lepton flavors and expanding the loop function around  $M_{N_1}^2 / M_{N_{j \neq 1}}^2$, see Eq.~\eqref{eq:floop}, is given by
\begin{equation}\label{eq:B1}
    \epsilon_{\rm CP} = -\frac{3M_{N_1}}{2} \frac{\text{Im}\{\sum_{j} [\mathsf{ h}^\dagger \mathsf{h}]_{1 j}^2 M_{N_j}^{-1}\}}{8\pi [\mathsf{h}^\dagger  \mathsf{h}]_{11}},
\end{equation}
where $\mathsf{h}^{i \ell} \equiv (\mathsf{Y}_\nu \mathsf{O}_R)^{ik} \mathsf{U}_H^{*k \ell}$ with $k=2,3,4$. This implies that $\mathsf{h}^\dagger \mathsf{h} = \mathsf{U}_H^T M_D^T M_D \mathsf{U}_H^* (2/v_H^2)$, see Eqs.~\eqref{eq:matrices}.
Then, we can rewrite $\text{Im}\{ \cdots \}$ in  Eq.~\eqref{eq:B1} as
\begin{equation}
\begin{split}
 \text{Im}\{ [\mathsf{h}^\dagger \mathsf{h} \, \mathsf{\widehat M}_H^{-1} \mathsf{h}^T \mathsf{h}^* ]_{11}\} &=(2/v_H^2)^2 \text{Im}\{ [\mathsf{U}_H^T \mathsf{M}_D^T \mathsf{M}_D \mathsf{U}_H^* \, \mathsf{\widehat M}_H^{-1} \mathsf{U}_H^\dagger \mathsf{M}_D^T \mathsf{M}_D \mathsf{U}_H]_{11} \}\\
 & = - (2/v_H^2)^2 \text{Im}\{ [\mathsf{U}_H^\dagger \mathsf{M}_D^T \mathsf{M}_D \mathsf{U}_H \, \mathsf{\widehat M}_H^{-1} \mathsf{U}_H^T \mathsf{M}_D^T \mathsf{M}_D \mathsf{U}_H^*]_{11} \},
 \end{split}
\end{equation}
where in the last step we used that $\text{Im}\{z \} = -\text{Im}\{ z^* \}$. By using the identities in Eq.~\eqref{eq:seesaw}, i.e. $\mathsf{U}_H \, \mathsf{\widehat M}_H^{-1} \mathsf{U}_H^T = \mathsf{M}_H^{-1}$, and $\mathsf{M}_L = -\mathsf{M}_D \, \mathsf{M}_H^{-1} \mathsf{M}_D^T$, we can rewrite the above expression as
\begin{equation}\label{eq:equation}
  \text{Im}\{ [\mathsf{h}^\dagger \mathsf{h} \, \mathsf{\widehat M}_H^{-1} \mathsf{h}^T \mathsf{h}^* ]_{11}\}  =  (2/v_H^2)^2 \text{ Im}\{ [\mathsf{U}_H^\dagger \mathsf{M}_D^T \mathsf{M}_L \mathsf{M}_D \mathsf{U}_H^*]_{11} \}.
\end{equation}

In what follows, we will make use of the Casas-Ibarra parametrization~\cite{Casas:2001sr}. From the seesaw relation in Eq.~\eqref{eq:seesaw} we can write 
\begin{equation}\label{eq:IRRT}
   \mathbb{I}_{3 \times 3} = \left (e^{i \pi/2} \sqrt{\mathsf{\widehat M}_L^{-1}} \mathsf{U}_L^T \mathsf{M}_D \mathsf{U}_H \, \sqrt{\mathsf{\widehat M}_H^{-1}} \right )  \left (e^{i \pi/2}\sqrt{\mathsf{\widehat M}_L^{-1}} \mathsf{U}^T_L \mathsf{M}_D \mathsf{U}_H \sqrt{\mathsf{\widehat M}_H^{-1}} \right )^T \equiv \mathsf{R}^T \mathsf{R}.
\end{equation}
Above, $\sqrt{\mathsf{\widehat M}} = \text{diag}(\sqrt{\mathsf{\widehat M}_{11}}, \cdots, \sqrt{\mathsf{\widehat M}_{55}})$.
Eq.~\eqref{eq:IRRT} allows us to identify, in between the parenthesis, a complex $5 \times 3$ matrix, $\mathsf{R}$, satisfying that $\mathsf{R}^T \mathsf{R} = \mathbb{I}_{3\times 3}$.
Hence, 
\begin{equation}
    \mathsf{M}_D =e^{-i \pi/2} \mathsf{U}_L^*\sqrt{\mathsf{\widehat M}_L} \mathsf{R}^T \sqrt{\mathsf{\widehat M}_H} \mathsf{U}_H^\dagger = e^{i \pi/2}\mathsf{U}_L \sqrt{\mathsf{\widehat M}_L} \mathsf{R^\dagger} \sqrt{\mathsf{\widehat M}_H} \mathsf{U}_H^T ,
\end{equation}
where in the last equality we have made use of the reality of the Dirac mass matrix $\mathsf{M}_D$.
This allows us to write the right-hand-side of Eq.~\eqref{eq:equation} as a function of $\mathsf{R}$, 
\begin{equation}
 -(2/v_H^2)^2  \text{ Im} \left \{ \left [\sqrt{\mathsf{\widehat M}_H} \mathsf{R}^*  \mathsf{\widehat M}_L^2  \mathsf{R}^\dagger \sqrt{\mathsf{\bar M}_H} \right ]_{11} \right \} =  (2/v_H^2)^2 M_{N_1} \text{Im}\{ (\mathsf{R} \mathsf{\widehat M}_L^2  \mathsf{R}^T)_{11} \}.
\end{equation}
In the last step, we have used again that $\text{Im}\{ z \} = - \text{Im} \{ z^* \}$. 

On the other hand, the denominator in Eq.~\eqref{eq:B1} can also be written in terms of $\mathsf{R}$, 
\begin{equation}\label{eq:B7}
[\mathsf{h}^\dagger \mathsf{h}]_{11} =   \frac{2}{v_H^2}  \left [\sqrt{\mathsf{\widehat M}_H} \mathsf{R} \mathsf{\widehat M}_L \mathsf{R}^\dagger \sqrt{\mathsf{\widehat M}_H} \right ]_{11} = \frac{2}{v_H^2} M_{N_1} [\mathsf{R} \mathsf{\widehat M}_L \mathsf{R}^\dagger]_{11} .
\end{equation}
Altogether, combining the numerator and denominator, we get an expression of $\epsilon_{\rm CP}$ as a function of $\mathsf{R}$, the light neutrino masses and the lightest heavy neutrino mass:
\begin{equation}\label{eq:DIapp}
\epsilon_{\rm CP} = -\frac{3 M_{N_1}}{8\pi v_H^2} \frac{ \text{Im} \left \{ \sum_{i=1}^3 \mathsf{R}_{1i}^2 m_{\nu_i}^2 \right \}}{\sum_{i=1}^3 |\mathsf{R}_{1i}|^2 m_{\nu_i}}.
\end{equation}
Until this step, the result maps to the well-known DI bound~\cite{Davidson:2002qv}. However, since $\mathsf{R}$ is not a square matrix, in our case $\mathsf{R} \mathsf{R}^T$ is an idempotent matrix rather than the identity. Let us take a different path to show that $\epsilon_{\rm CP}$ is indeed bounded in this case as well.

Using that $\text{Im}\{z\} < |z|$, and the triangle inequality $|\sum_{i=1}^3 R_{1i}^2 m_{\nu_i}^2| \leq \sum_{i=1}^3 |R_{1i}^2|m_{\nu_i}^2$, we can bound Eq.~\eqref{eq:DIapp} as follows
\begin{equation}
    |\epsilon_{\rm CP}| \leq \frac{3 M_{N_1}}{8\pi v_H^2} \frac{m_{\nu_3}^2(|\mathsf{R}_{11}^2|+|\mathsf{R}_{12}^2|+|\mathsf{R}_{13}^2|) - (m_{\nu_3}^2 - m_{\nu_1}^2)|\mathsf{R}_{11}^2| - (m_{\nu_3}^2-m_{\nu_2}^2)|\mathsf{R}_{12}^2|}{m_{\nu_3}(|\mathsf{R}_{13}^2|+|\mathsf{R}_{12}^2|+|\mathsf{R}_{11}^2|) - (m_{\nu_3} - m_{\nu_1})|\mathsf{R}_{11}^2| - (m_{\nu_3} - m_{\nu_2})|\mathsf{R}_{12}^2|},
\end{equation}
Using the following relations $m_{\nu_3}^2 - m_{\nu_i}^2 = (m_{\nu_3}-m_{\nu_i})(m_{\nu_3}+m_{\nu_i})\geq(m_{\nu_3}-m_{\nu_i})m_{\nu_3}$ for $i=1,2$,
the previous expression simplifies to
\begin{equation}\label{eq:DIappendix}
    |\epsilon_{\rm CP}| \leq \frac{3 M_{N_1}}{8\pi v_H^2} m_{\nu_3}.
\end{equation}
We note that this bound applies generically, regardless of the number of neutral singlets in the theory~\cite{Eisele:2007ws}. 
As the plots in Sec.~\ref{sec:results} show, Eq.~\eqref{eq:DIappendix} is satisfied by our numerical solutions.

Although having a rectangular matrix, we could still bound $|\epsilon_{\rm CP}|$ thanks to the ratio between weighted sums of $\mathsf{R}$ entries. This is not the case, however, for the washout factor $K$, defined in Eq.~\eqref{eq:Kwash}. Writing $K$ {\it à la} Casas-Ibarra, i.e. in terms of $\mathsf{R}$, according to Eq.~\eqref{eq:B7} we have
\begin{equation}\label{eq:B11}
  m_\star K = \frac{[\mathsf{h}^\dagger \mathsf{h}]_{11} v_H^2}{2M_{N_1}} =  [\mathsf{R} \mathsf{\widehat M}_L \mathsf{R}^\dagger]_{11} = \sum_{i=1}^3 m_{\nu_i} |\mathsf{R}_{1i}^2| . 
\end{equation}
Ref.~\cite{Buchmuller:2003gz} derives an upper bound on the washout parameter, $K \lesssim m_{\nu_3} / m_\star$, by assuming that there are no strong phase cancellations so that $\sum_i |\mathsf{R}_{1i}^2| \simeq |\sum_i \mathsf{R}_{1i}^2|$. In our case, because $\mathsf{R}\mathsf{R}^T \neq \mathbb{I}$, we cannot claim the same. We note that, if $\mathsf{R}$ were a real matrix, $\mathsf{R}\mathsf{R}^T$ would be a projection matrix, and hence $(\mathsf{R}\mathsf{R}^T)_{11} = \sum_i \mathsf{R}_{1i}^2 \leq 1$, recovering the bound $K \lesssim m_3 / m_\star \sim 10^2$. However, due to the phases of $\mathsf{R}$, we do not expect this bound to be strictly respected. 

\bibliographystyle{JHEP}
\bibliography{BSCPV.bib}

\providecommand{\href}[2]{#2}\begingroup\raggedright\begin{thebibliography}{100}

\bibitem{Planck:2018vyg}
{\scshape Planck} collaboration, \emph{{Planck 2018 results. VI. Cosmological parameters}}, \href{https://doi.org/10.1051/0004-6361/201833910}{\emph{Astron. Astrophys.} {\bfseries 641} (2020) A6} [\href{https://arxiv.org/abs/1807.06209}{{\ttfamily 1807.06209}}].

\bibitem{Cyburt:2015mya}
R.H.~Cyburt, B.D.~Fields, K.A.~Olive and T.-H.~Yeh, \emph{{Big Bang Nucleosynthesis: 2015}}, \href{https://doi.org/10.1103/RevModPhys.88.015004}{\emph{Rev. Mod. Phys.} {\bfseries 88} (2016) 015004} [\href{https://arxiv.org/abs/1505.01076}{{\ttfamily 1505.01076}}].

\bibitem{Workman:2022ynf}
{\scshape Particle Data Group} collaboration, \emph{{Review of Particle Physics}}, \href{https://doi.org/10.1093/ptep/ptac097}{\emph{PTEP} {\bfseries 2022} (2022) 083C01}.

\bibitem{Sakharov:1967dj}
A.D.~Sakharov, \emph{{Violation of CP Invariance, C asymmetry, and baryon asymmetry of the universe}}, \href{https://doi.org/10.1070/PU1991v034n05ABEH002497}{\emph{Pisma Zh. Eksp. Teor. Fiz.} {\bfseries 5} (1967) 32}.

\bibitem{Gonzalez-Garcia:2021dve}
M.C.~Gonzalez-Garcia, M.~Maltoni and T.~Schwetz, \emph{{NuFIT: Three-Flavour Global Analyses of Neutrino Oscillation Experiments}}, \href{https://doi.org/10.3390/universe7120459}{\emph{Universe} {\bfseries 7} (2021) 459} [\href{https://arxiv.org/abs/2111.03086}{{\ttfamily 2111.03086}}].

\bibitem{SNO:2002tuh}
{\scshape SNO} collaboration, \emph{{Direct evidence for neutrino flavor transformation from neutral current interactions in the Sudbury Neutrino Observatory}}, \href{https://doi.org/10.1103/PhysRevLett.89.011301}{\emph{Phys. Rev. Lett.} {\bfseries 89} (2002) 011301} [\href{https://arxiv.org/abs/nucl-ex/0204008}{{\ttfamily nucl-ex/0204008}}].

\bibitem{Super-Kamiokande:1998kpq}
{\scshape Super-Kamiokande} collaboration, \emph{{Evidence for oscillation of atmospheric neutrinos}}, \href{https://doi.org/10.1103/PhysRevLett.81.1562}{\emph{Phys. Rev. Lett.} {\bfseries 81} (1998) 1562} [\href{https://arxiv.org/abs/hep-ex/9807003}{{\ttfamily hep-ex/9807003}}].

\bibitem{KamLAND:2002uet}
{\scshape KamLAND} collaboration, \emph{{First results from KamLAND: Evidence for reactor anti-neutrino disappearance}}, \href{https://doi.org/10.1103/PhysRevLett.90.021802}{\emph{Phys. Rev. Lett.} {\bfseries 90} (2003) 021802} [\href{https://arxiv.org/abs/hep-ex/0212021}{{\ttfamily hep-ex/0212021}}].

\bibitem{Abel:2020pzs}
C.~Abel et~al., \emph{{Measurement of the Permanent Electric Dipole Moment of the Neutron}}, \href{https://doi.org/10.1103/PhysRevLett.124.081803}{\emph{Phys. Rev. Lett.} {\bfseries 124} (2020) 081803} [\href{https://arxiv.org/abs/2001.11966}{{\ttfamily 2001.11966}}].

\bibitem{Pospelov:1999mv}
M.~Pospelov and A.~Ritz, \emph{{Theta vacua, QCD sum rules, and the neutron electric dipole moment}}, \href{https://doi.org/10.1016/S0550-3213(99)00817-2}{\emph{Nucl. Phys. B} {\bfseries 573} (2000) 177} [\href{https://arxiv.org/abs/hep-ph/9908508}{{\ttfamily hep-ph/9908508}}].

\bibitem{Liang:2023jfj}
{\scshape \ensuremath{\chi}QCD} collaboration, \emph{{Nucleon electric dipole moment from the \ensuremath{\theta} term with lattice chiral fermions}}, \href{https://doi.org/10.1103/PhysRevD.108.094512}{\emph{Phys. Rev. D} {\bfseries 108} (2023) 094512} [\href{https://arxiv.org/abs/2301.04331}{{\ttfamily 2301.04331}}].

\bibitem{Manohar:2018aog}
A.V.~Manohar, \emph{{Introduction to Effective Field Theories}},  \href{https://arxiv.org/abs/1804.05863}{{\ttfamily 1804.05863}}.

\bibitem{Peccei:1977hh}
R.D.~Peccei and H.R.~Quinn, \emph{{CP Conservation in the Presence of Instantons}}, \href{https://doi.org/10.1103/PhysRevLett.38.1440}{\emph{Phys. Rev. Lett.} {\bfseries 38} (1977) 1440}.

\bibitem{Peccei:1977ur}
R.D.~Peccei and H.R.~Quinn, \emph{{Constraints Imposed by CP Conservation in the Presence of Instantons}}, \href{https://doi.org/10.1103/PhysRevD.16.1791}{\emph{Phys. Rev. D} {\bfseries 16} (1977) 1791}.

\bibitem{Nelson:1983zb}
A.E.~Nelson, \emph{{Naturally Weak CP Violation}}, \href{https://doi.org/10.1016/0370-2693(84)92025-2}{\emph{Phys. Lett. B} {\bfseries 136} (1984) 387}.

\bibitem{Barr:1984qx}
S.M.~Barr, \emph{{Solving the Strong CP Problem Without the Peccei-Quinn Symmetry}}, \href{https://doi.org/10.1103/PhysRevLett.53.329}{\emph{Phys. Rev. Lett.} {\bfseries 53} (1984) 329}.

\bibitem{Nelson:1984hg}
A.E.~Nelson, \emph{{Calculation of $\theta$ Barr}}, \href{https://doi.org/10.1016/0370-2693(84)90827-X}{\emph{Phys. Lett. B} {\bfseries 143} (1984) 165}.

\bibitem{Wilczek:1977pj}
F.~Wilczek, \emph{{Problem of Strong $P$ and $T$ Invariance in the Presence of Instantons}}, \href{https://doi.org/10.1103/PhysRevLett.40.279}{\emph{Phys. Rev. Lett.} {\bfseries 40} (1978) 279}.

\bibitem{Weinberg:1977ma}
S.~Weinberg, \emph{{A New Light Boson?}}, \href{https://doi.org/10.1103/PhysRevLett.40.223}{\emph{Phys. Rev. Lett.} {\bfseries 40} (1978) 223}.

\bibitem{Kim:1979if}
J.E.~Kim, \emph{{Weak Interaction Singlet and Strong CP Invariance}}, \href{https://doi.org/10.1103/PhysRevLett.43.103}{\emph{Phys. Rev. Lett.} {\bfseries 43} (1979) 103}.

\bibitem{Shifman:1979if}
M.A.~Shifman, A.I.~Vainshtein and V.I.~Zakharov, \emph{{Can Confinement Ensure Natural CP Invariance of Strong Interactions?}}, \href{https://doi.org/10.1016/0550-3213(80)90209-6}{\emph{Nucl. Phys. B} {\bfseries 166} (1980) 493}.

\bibitem{Dine:1981rt}
M.~Dine, W.~Fischler and M.~Srednicki, \emph{{A Simple Solution to the Strong CP Problem with a Harmless Axion}}, \href{https://doi.org/10.1016/0370-2693(81)90590-6}{\emph{Phys. Lett. B} {\bfseries 104} (1981) 199}.

\bibitem{Zhitnitsky:1980tq}
A.R.~Zhitnitsky, \emph{{On Possible Suppression of the Axion Hadron Interactions. (In Russian)}}, {\emph{Sov. J. Nucl. Phys.} {\bfseries 31} (1980) 260}.

\bibitem{Georgi:1981pu}
H.M.~Georgi, L.J.~Hall and M.B.~Wise, \emph{{Grand Unified Models With an Automatic {Peccei-Quinn} Symmetry}}, \href{https://doi.org/10.1016/0550-3213(81)90433-8}{\emph{Nucl. Phys. B} {\bfseries 192} (1981) 409}.

\bibitem{Kamionkowski:1992mf}
M.~Kamionkowski and J.~March-Russell, \emph{{Planck scale physics and the Peccei-Quinn mechanism}}, \href{https://doi.org/10.1016/0370-2693(92)90492-M}{\emph{Phys. Lett. B} {\bfseries 282} (1992) 137} [\href{https://arxiv.org/abs/hep-th/9202003}{{\ttfamily hep-th/9202003}}].

\bibitem{Holman:1992us}
R.~Holman, S.D.H.~Hsu, T.W.~Kephart, E.W.~Kolb, R.~Watkins and L.M.~Widrow, \emph{{Solutions to the strong CP problem in a world with gravity}}, \href{https://doi.org/10.1016/0370-2693(92)90491-L}{\emph{Phys. Lett. B} {\bfseries 282} (1992) 132} [\href{https://arxiv.org/abs/hep-ph/9203206}{{\ttfamily hep-ph/9203206}}].

\bibitem{Bonnefoy:2022vop}
Q.~Bonnefoy, \emph{{Heavy fields and the axion quality problem}}, \href{https://doi.org/10.1103/PhysRevD.108.035023}{\emph{Phys. Rev. D} {\bfseries 108} (2023) 035023} [\href{https://arxiv.org/abs/2212.00102}{{\ttfamily 2212.00102}}].

\bibitem{Witten:1984dg}
E.~Witten, \emph{{Some Properties of O(32) Superstrings}}, \href{https://doi.org/10.1016/0370-2693(84)90422-2}{\emph{Phys. Lett. B} {\bfseries 149} (1984) 351}.

\bibitem{Wen:1985qj}
X.-G.~Wen and E.~Witten, \emph{{Electric and Magnetic Charges in Superstring Models}}, \href{https://doi.org/10.1016/0550-3213(85)90592-9}{\emph{Nucl. Phys. B} {\bfseries 261} (1985) 651}.

\bibitem{Svrcek:2006yi}
P.~Svrcek and E.~Witten, \emph{{Axions In String Theory}}, \href{https://doi.org/10.1088/1126-6708/2006/06/051}{\emph{JHEP} {\bfseries 06} (2006) 051} [\href{https://arxiv.org/abs/hep-th/0605206}{{\ttfamily hep-th/0605206}}].

\bibitem{Reece:2024wrn}
M.~Reece, \emph{{Extra-Dimensional Axion Expectations}},  \href{https://arxiv.org/abs/2406.08543}{{\ttfamily 2406.08543}}.

\bibitem{Craig:2024dnl}
N.~Craig and M.~Kongsore, \emph{{High-quality axions from higher-form symmetries in extra dimensions}}, \href{https://doi.org/10.1103/PhysRevD.111.015047}{\emph{Phys. Rev. D} {\bfseries 111} (2025) 015047} [\href{https://arxiv.org/abs/2408.10295}{{\ttfamily 2408.10295}}].

\bibitem{Agrawal:2024ejr}
P.~Agrawal, M.~Nee and M.~Reig, \emph{{Axion couplings in heterotic string theory}}, \href{https://doi.org/10.1007/JHEP02(2025)188}{\emph{JHEP} {\bfseries 02} (2025) 188} [\href{https://arxiv.org/abs/2410.03820}{{\ttfamily 2410.03820}}].

\bibitem{Asadi:2022vys}
P.~Asadi, S.~Homiller, Q.~Lu and M.~Reece, \emph{{Chiral Nelson-Barr models: Quality and cosmology}}, \href{https://doi.org/10.1103/PhysRevD.107.115012}{\emph{Phys. Rev. D} {\bfseries 107} (2023) 115012} [\href{https://arxiv.org/abs/2212.03882}{{\ttfamily 2212.03882}}].

\bibitem{Perez:2023zin}
P.F.~Perez, C.~Murgui and M.B.~Wise, \emph{{Automatic Nelson-Barr solutions to the strong CP puzzle}}, \href{https://doi.org/10.1103/PhysRevD.108.015010}{\emph{Phys. Rev. D} {\bfseries 108} (2023) 015010} [\href{https://arxiv.org/abs/2302.06620}{{\ttfamily 2302.06620}}].

\bibitem{Preskill:1982cy}
J.~Preskill, M.B.~Wise and F.~Wilczek, \emph{{Cosmology of the Invisible Axion}}, \href{https://doi.org/10.1016/0370-2693(83)90637-8}{\emph{Phys. Lett. B} {\bfseries 120} (1983) 127}.

\bibitem{Abbott:1982af}
L.F.~Abbott and P.~Sikivie, \emph{{A Cosmological Bound on the Invisible Axion}}, \href{https://doi.org/10.1016/0370-2693(83)90638-X}{\emph{Phys. Lett. B} {\bfseries 120} (1983) 133}.

\bibitem{Dine:1982ah}
M.~Dine and W.~Fischler, \emph{{The Not So Harmless Axion}}, \href{https://doi.org/10.1016/0370-2693(83)90639-1}{\emph{Phys. Lett. B} {\bfseries 120} (1983) 137}.

\bibitem{Dine:2024bxv}
M.~Dine, G.~Perez, W.~Ratzinger and I.~Savoray, \emph{{Nelson-Barr ultralight dark matter}},  \href{https://arxiv.org/abs/2405.06744}{{\ttfamily 2405.06744}}.

\bibitem{Kuzmin:1992up}
V.A.~Kuzmin, M.E.~Shaposhnikov and I.I.~Tkachev, \emph{{Strong CP violation, electroweak baryogenesis, and axionic dark matter}}, \href{https://doi.org/10.1103/PhysRevD.45.466}{\emph{Phys. Rev. D} {\bfseries 45} (1992) 466}.

\bibitem{Servant:2014bla}
G.~Servant, \emph{{Baryogenesis from Strong $CP$ Violation and the QCD Axion}}, \href{https://doi.org/10.1103/PhysRevLett.113.171803}{\emph{Phys. Rev. Lett.} {\bfseries 113} (2014) 171803} [\href{https://arxiv.org/abs/1407.0030}{{\ttfamily 1407.0030}}].

\bibitem{Co:2019wyp}
R.T.~Co and K.~Harigaya, \emph{{Axiogenesis}}, \href{https://doi.org/10.1103/PhysRevLett.124.111602}{\emph{Phys. Rev. Lett.} {\bfseries 124} (2020) 111602} [\href{https://arxiv.org/abs/1910.02080}{{\ttfamily 1910.02080}}].

\bibitem{Lee:1973iz}
T.D.~Lee, \emph{{A Theory of Spontaneous T Violation}}, \href{https://doi.org/10.1103/PhysRevD.8.1226}{\emph{Phys. Rev. D} {\bfseries 8} (1973) 1226}.

\bibitem{McNamara:2022lrw}
J.~McNamara and M.~Reece, \emph{{Reflections on Parity Breaking}},  \href{https://arxiv.org/abs/2212.00039}{{\ttfamily 2212.00039}}.

\bibitem{ParticleDataGroup:2020ssz}
{\scshape Particle Data Group} collaboration, \emph{{Review of Particle Physics}}, \href{https://doi.org/10.1093/ptep/ptaa104}{\emph{PTEP} {\bfseries 2020} (2020) 083C01}.

\bibitem{T2K:2023smv}
{\scshape T2K} collaboration, \emph{{Measurements of neutrino oscillation parameters from the T2K experiment using $3.6\times10^{21}$ protons on target}},  \href{https://arxiv.org/abs/2303.03222}{{\ttfamily 2303.03222}}.

\bibitem{NOvA:2021nfi}
{\scshape NOvA} collaboration, \emph{{Improved measurement of neutrino oscillation parameters by the NOvA experiment}}, \href{https://doi.org/10.1103/PhysRevD.106.032004}{\emph{Phys. Rev. D} {\bfseries 106} (2022) 032004} [\href{https://arxiv.org/abs/2108.08219}{{\ttfamily 2108.08219}}].

\bibitem{NOvA:2023iam}
{\scshape NOvA} collaboration, \emph{{Expanding neutrino oscillation parameter measurements in NOvA using a Bayesian approach}}, \href{https://doi.org/10.1103/PhysRevD.110.012005}{\emph{Phys. Rev. D} {\bfseries 110} (2024) 012005} [\href{https://arxiv.org/abs/2311.07835}{{\ttfamily 2311.07835}}].

\bibitem{Mikola:2024jnj}
{\scshape T2K, NOvA} collaboration, \emph{{Results from the T2K+NOvA Joint Analysis}}, \href{https://doi.org/10.1051/epjconf/202431202002}{\emph{EPJ Web Conf.} {\bfseries 312} (2024) 02002}.

\bibitem{Bento:1991ez}
L.~Bento, G.C.~Branco and P.A.~Parada, \emph{{A Minimal model with natural suppression of strong CP violation}}, \href{https://doi.org/10.1016/0370-2693(91)90530-4}{\emph{Phys. Lett. B} {\bfseries 267} (1991) 95}.

\bibitem{Branco:2003rt}
G.C.~Branco, P.A.~Parada and M.N.~Rebelo, \emph{{A Common origin for all CP violations}},  \href{https://arxiv.org/abs/hep-ph/0307119}{{\ttfamily hep-ph/0307119}}.

\bibitem{Murai:2024alz}
K.~Murai and K.~Nakayama, \emph{{Revisiting the minimal Nelson-Barr model}}, \href{https://doi.org/10.1007/JHEP11(2024)098}{\emph{JHEP} {\bfseries 11} (2024) 098} [\href{https://arxiv.org/abs/2407.16202}{{\ttfamily 2407.16202}}].

\bibitem{Suematsu:2023jqa}
D.~Suematsu, \emph{{CP issues in the SM from a viewpoint of spontaneous CP violation}}, \href{https://doi.org/10.1103/PhysRevD.108.095046}{\emph{Phys. Rev. D} {\bfseries 108} (2023) 095046} [\href{https://arxiv.org/abs/2309.04783}{{\ttfamily 2309.04783}}].

\bibitem{Alves:2023ufm}
J.a.M.~Alves, G.C.~Branco, A.L.~Cherchiglia, C.C.~Nishi, J.T.~Penedo, P.M.F.~Pereira et~al., \emph{{Vector-like singlet quarks: A roadmap}}, \href{https://doi.org/10.1016/j.physrep.2023.12.004}{\emph{Phys. Rept.} {\bfseries 1057} (2024) 1} [\href{https://arxiv.org/abs/2304.10561}{{\ttfamily 2304.10561}}].

\bibitem{Valenti:2021rdu}
A.~Valenti and L.~Vecchi, \emph{{The CKM phase and $ \overline{\theta} $ in Nelson-Barr models}}, \href{https://doi.org/10.1007/JHEP07(2021)203}{\emph{JHEP} {\bfseries 07} (2021) 203} [\href{https://arxiv.org/abs/2105.09122}{{\ttfamily 2105.09122}}].

\bibitem{Dine:2015jga}
M.~Dine and P.~Draper, \emph{{Challenges for the Nelson-Barr Mechanism}}, \href{https://doi.org/10.1007/JHEP08(2015)132}{\emph{JHEP} {\bfseries 08} (2015) 132} [\href{https://arxiv.org/abs/1506.05433}{{\ttfamily 1506.05433}}].

\bibitem{FileviezPerez:2023rxn}
P.~Fileviez~P\'erez, C.~Murgui, S.~Patrone, A.~Testa and M.B.~Wise, \emph{{Finite naturalness and quark-lepton unification}}, \href{https://doi.org/10.1103/PhysRevD.109.015011}{\emph{Phys. Rev. D} {\bfseries 109} (2024) 015011} [\href{https://arxiv.org/abs/2308.07367}{{\ttfamily 2308.07367}}].

\bibitem{Chivukula:1987py}
R.S.~Chivukula and H.~Georgi, \emph{{Composite Technicolor Standard Model}}, \href{https://doi.org/10.1016/0370-2693(87)90713-1}{\emph{Phys. Lett. B} {\bfseries 188} (1987) 99}.

\bibitem{DAmbrosio:2002vsn}
G.~D'Ambrosio, G.F.~Giudice, G.~Isidori and A.~Strumia, \emph{{Minimal flavor violation: An Effective field theory approach}}, \href{https://doi.org/10.1016/S0550-3213(02)00836-2}{\emph{Nucl. Phys. B} {\bfseries 645} (2002) 155} [\href{https://arxiv.org/abs/hep-ph/0207036}{{\ttfamily hep-ph/0207036}}].

\bibitem{Dvali:1995cc}
G.R.~Dvali and G.~Senjanovic, \emph{{Is there a domain wall problem?}}, \href{https://doi.org/10.1103/PhysRevLett.74.5178}{\emph{Phys. Rev. Lett.} {\bfseries 74} (1995) 5178} [\href{https://arxiv.org/abs/hep-ph/9501387}{{\ttfamily hep-ph/9501387}}].

\bibitem{Dvali:1996zr}
G.R.~Dvali, A.~Melfo and G.~Senjanovic, \emph{{Nonrestoration of spontaneously broken P and CP at high temperature}}, \href{https://doi.org/10.1103/PhysRevD.54.7857}{\emph{Phys. Rev. D} {\bfseries 54} (1996) 7857} [\href{https://arxiv.org/abs/hep-ph/9601376}{{\ttfamily hep-ph/9601376}}].

\bibitem{Haber:2012np}
H.E.~Haber and Z.~Surujon, \emph{{A Group-theoretic Condition for Spontaneous CP Violation}}, \href{https://doi.org/10.1103/PhysRevD.86.075007}{\emph{Phys. Rev. D} {\bfseries 86} (2012) 075007} [\href{https://arxiv.org/abs/1201.1730}{{\ttfamily 1201.1730}}].

\bibitem{Farina:2013mla}
M.~Farina, D.~Pappadopulo and A.~Strumia, \emph{{A modified naturalness principle and its experimental tests}}, \href{https://doi.org/10.1007/JHEP08(2013)022}{\emph{JHEP} {\bfseries 08} (2013) 022} [\href{https://arxiv.org/abs/1303.7244}{{\ttfamily 1303.7244}}].

\bibitem{Minkowski:1977sc}
P.~Minkowski, \emph{{$\mu \to e\gamma$ at a Rate of One Out of $10^{9}$ Muon Decays?}}, \href{https://doi.org/10.1016/0370-2693(77)90435-X}{\emph{Phys. Lett. B} {\bfseries 67} (1977) 421}.

\bibitem{Yanagida:1979as}
T.~Yanagida, \emph{{Horizontal gauge symmetry and masses of neutrinos}}, {\emph{Conf. Proc. C} {\bfseries 7902131} (1979) 95}.

\bibitem{Gell-Mann:1979vob}
M.~Gell-Mann, P.~Ramond and R.~Slansky, \emph{{Complex Spinors and Unified Theories}}, {\emph{Conf. Proc. C} {\bfseries 790927} (1979) 315} [\href{https://arxiv.org/abs/1306.4669}{{\ttfamily 1306.4669}}].

\bibitem{Mohapatra:1979ia}
R.N.~Mohapatra and G.~Senjanovic, \emph{{Neutrino Mass and Spontaneous Parity Nonconservation}}, \href{https://doi.org/10.1103/PhysRevLett.44.912}{\emph{Phys. Rev. Lett.} {\bfseries 44} (1980) 912}.

\bibitem{Davidson:2008bu}
S.~Davidson, E.~Nardi and Y.~Nir, \emph{{Leptogenesis}}, \href{https://doi.org/10.1016/j.physrep.2008.06.002}{\emph{Phys. Rept.} {\bfseries 466} (2008) 105} [\href{https://arxiv.org/abs/0802.2962}{{\ttfamily 0802.2962}}].

\bibitem{Buchmuller:2004nz}
W.~Buchmuller, P.~Di~Bari and M.~Plumacher, \emph{{Leptogenesis for pedestrians}}, \href{https://doi.org/10.1016/j.aop.2004.02.003}{\emph{Annals Phys.} {\bfseries 315} (2005) 305} [\href{https://arxiv.org/abs/hep-ph/0401240}{{\ttfamily hep-ph/0401240}}].

\bibitem{Fukugita:1986hr}
M.~Fukugita and T.~Yanagida, \emph{{Baryogenesis Without Grand Unification}}, \href{https://doi.org/10.1016/0370-2693(86)91126-3}{\emph{Phys. Lett. B} {\bfseries 174} (1986) 45}.

\bibitem{Kuzmin:1985mm}
V.A.~Kuzmin, V.A.~Rubakov and M.E.~Shaposhnikov, \emph{{On the Anomalous Electroweak Baryon Number Nonconservation in the Early Universe}}, \href{https://doi.org/10.1016/0370-2693(85)91028-7}{\emph{Phys. Lett. B} {\bfseries 155} (1985) 36}.

\bibitem{Davidson:2002qv}
S.~Davidson and A.~Ibarra, \emph{{A Lower bound on the right-handed neutrino mass from leptogenesis}}, \href{https://doi.org/10.1016/S0370-2693(02)01735-5}{\emph{Phys. Lett. B} {\bfseries 535} (2002) 25} [\href{https://arxiv.org/abs/hep-ph/0202239}{{\ttfamily hep-ph/0202239}}].

\bibitem{Grimus:2000vj}
W.~Grimus and L.~Lavoura, \emph{{The Seesaw mechanism at arbitrary order: Disentangling the small scale from the large scale}}, \href{https://doi.org/10.1088/1126-6708/2000/11/042}{\emph{JHEP} {\bfseries 11} (2000) 042} [\href{https://arxiv.org/abs/hep-ph/0008179}{{\ttfamily hep-ph/0008179}}].

\bibitem{Autonne1915}
L.~Autonne, \emph{Sur les matrices hypohermitiennes et sur les matrices unitaires}, Annales de l'Université de Lyon, Série I: Sciences, Médecine, Fasc. 38, A. Rey and Librairie Gauthier-Villars, Lyon and Paris (1915).

\bibitem{TAKAGI}
T.~TAKAGI, \emph{On an algebraic problem related to an analytic theorem of carathéodory and fejér and on an allied theorem of landau}, \href{https://doi.org/10.4099/jjm1924.1.0_83}{\emph{Japanese journal of mathematics :transactions and abstracts} {\bfseries 1} (1924) 83}.

\bibitem{KATRIN:2021uub}
{\scshape KATRIN} collaboration, \emph{{Direct neutrino-mass measurement with sub-electronvolt sensitivity}}, \href{https://doi.org/10.1038/s41567-021-01463-1}{\emph{Nature Phys.} {\bfseries 18} (2022) 160} [\href{https://arxiv.org/abs/2105.08533}{{\ttfamily 2105.08533}}].

\bibitem{DESI:2024mwx}
{\scshape DESI} collaboration, \emph{{DESI 2024 VI: Cosmological Constraints from the Measurements of Baryon Acoustic Oscillations}},  \href{https://arxiv.org/abs/2404.03002}{{\ttfamily 2404.03002}}.

\bibitem{Pilaftsis:2003gt}
A.~Pilaftsis and T.E.J.~Underwood, \emph{{Resonant leptogenesis}}, \href{https://doi.org/10.1016/j.nuclphysb.2004.05.029}{\emph{Nucl. Phys. B} {\bfseries 692} (2004) 303} [\href{https://arxiv.org/abs/hep-ph/0309342}{{\ttfamily hep-ph/0309342}}].

\bibitem{Blanchet:2006dq}
S.~Blanchet and P.~Di~Bari, \emph{{Leptogenesis beyond the limit of hierarchical heavy neutrino masses}}, \href{https://doi.org/10.1088/1475-7516/2006/06/023}{\emph{JCAP} {\bfseries 06} (2006) 023} [\href{https://arxiv.org/abs/hep-ph/0603107}{{\ttfamily hep-ph/0603107}}].

\bibitem{Blanchet:2012bk}
S.~Blanchet and P.~Di~Bari, \emph{{The minimal scenario of leptogenesis}}, \href{https://doi.org/10.1088/1367-2630/14/12/125012}{\emph{New J. Phys.} {\bfseries 14} (2012) 125012} [\href{https://arxiv.org/abs/1211.0512}{{\ttfamily 1211.0512}}].

\bibitem{Borsanyi:2016ksw}
S.~Borsanyi et~al., \emph{{Calculation of the axion mass based on high-temperature lattice quantum chromodynamics}}, \href{https://doi.org/10.1038/nature20115}{\emph{Nature} {\bfseries 539} (2016) 69} [\href{https://arxiv.org/abs/1606.07494}{{\ttfamily 1606.07494}}].

\bibitem{Giudice:2003jh}
G.F.~Giudice, A.~Notari, M.~Raidal, A.~Riotto and A.~Strumia, \emph{{Towards a complete theory of thermal leptogenesis in the SM and MSSM}}, \href{https://doi.org/10.1016/j.nuclphysb.2004.02.019}{\emph{Nucl. Phys. B} {\bfseries 685} (2004) 89} [\href{https://arxiv.org/abs/hep-ph/0310123}{{\ttfamily hep-ph/0310123}}].

\bibitem{Plumacher:1996kc}
M.~Plumacher, \emph{{Baryogenesis and lepton number violation}}, \href{https://doi.org/10.1007/s002880050418}{\emph{Z. Phys. C} {\bfseries 74} (1997) 549} [\href{https://arxiv.org/abs/hep-ph/9604229}{{\ttfamily hep-ph/9604229}}].

\bibitem{FileviezPerez:2021hbc}
P.~Fileviez~Perez, C.~Murgui and A.D.~Plascencia, \emph{{Baryogenesis via leptogenesis: Spontaneous B and L violation}}, \href{https://doi.org/10.1103/PhysRevD.104.055007}{\emph{Phys. Rev. D} {\bfseries 104} (2021) 055007} [\href{https://arxiv.org/abs/2103.13397}{{\ttfamily 2103.13397}}].

\bibitem{Barbieri:1999ma}
R.~Barbieri, P.~Creminelli, A.~Strumia and N.~Tetradis, \emph{{Baryogenesis through leptogenesis}}, \href{https://doi.org/10.1016/S0550-3213(00)00011-0}{\emph{Nucl. Phys. B} {\bfseries 575} (2000) 61} [\href{https://arxiv.org/abs/hep-ph/9911315}{{\ttfamily hep-ph/9911315}}].

\bibitem{Abada:2006ea}
A.~Abada, S.~Davidson, A.~Ibarra, F.X.~Josse-Michaux, M.~Losada and A.~Riotto, \emph{{Flavour Matters in Leptogenesis}}, \href{https://doi.org/10.1088/1126-6708/2006/09/010}{\emph{JHEP} {\bfseries 09} (2006) 010} [\href{https://arxiv.org/abs/hep-ph/0605281}{{\ttfamily hep-ph/0605281}}].

\bibitem{Nardi:2006fx}
E.~Nardi, Y.~Nir, E.~Roulet and J.~Racker, \emph{{The Importance of flavor in leptogenesis}}, \href{https://doi.org/10.1088/1126-6708/2006/01/164}{\emph{JHEP} {\bfseries 01} (2006) 164} [\href{https://arxiv.org/abs/hep-ph/0601084}{{\ttfamily hep-ph/0601084}}].

\bibitem{Davidson:2007xu}
S.~Davidson, \emph{{Flavoured Leptogenesis}},  in \emph{{12th International Workshop on Neutrinos Telescopes: Twenty Years after the Supernova 1987A Neutrino Bursts Discovery}}, pp.~531--545, 5, 2007 [\href{https://arxiv.org/abs/0705.1590}{{\ttfamily 0705.1590}}].

\bibitem{Blanchet:2006be}
S.~Blanchet and P.~Di~Bari, \emph{{Flavor effects on leptogenesis predictions}}, \href{https://doi.org/10.1088/1475-7516/2007/03/018}{\emph{JCAP} {\bfseries 03} (2007) 018} [\href{https://arxiv.org/abs/hep-ph/0607330}{{\ttfamily hep-ph/0607330}}].

\bibitem{Josse-Michaux:2007alz}
F.X.~Josse-Michaux and A.~Abada, \emph{{Study of flavour dependencies in leptogenesis}}, \href{https://doi.org/10.1088/1475-7516/2007/10/009}{\emph{JCAP} {\bfseries 10} (2007) 009} [\href{https://arxiv.org/abs/hep-ph/0703084}{{\ttfamily hep-ph/0703084}}].

\bibitem{Buchmuller:2001sr}
W.~Buchmuller and M.~Plumacher, \emph{{Spectator processes and baryogenesis}}, \href{https://doi.org/10.1016/S0370-2693(01)00614-1}{\emph{Phys. Lett. B} {\bfseries 511} (2001) 74} [\href{https://arxiv.org/abs/hep-ph/0104189}{{\ttfamily hep-ph/0104189}}].

\bibitem{Nardi:2005hs}
E.~Nardi, Y.~Nir, J.~Racker and E.~Roulet, \emph{{On Higgs and sphaleron effects during the leptogenesis era}}, \href{https://doi.org/10.1088/1126-6708/2006/01/068}{\emph{JHEP} {\bfseries 01} (2006) 068} [\href{https://arxiv.org/abs/hep-ph/0512052}{{\ttfamily hep-ph/0512052}}].

\bibitem{Harvey:1990qw}
J.A.~Harvey and M.S.~Turner, \emph{{Cosmological baryon and lepton number in the presence of electroweak fermion number violation}}, \href{https://doi.org/10.1103/PhysRevD.42.3344}{\emph{Phys. Rev. D} {\bfseries 42} (1990) 3344}.

\bibitem{Buchmuller:2003gz}
W.~Buchmuller, P.~Di~Bari and M.~Plumacher, \emph{{The Neutrino mass window for baryogenesis}}, \href{https://doi.org/10.1016/S0550-3213(03)00449-8}{\emph{Nucl. Phys. B} {\bfseries 665} (2003) 445} [\href{https://arxiv.org/abs/hep-ph/0302092}{{\ttfamily hep-ph/0302092}}].

\bibitem{n2EDM:2021yah}
{\scshape n2EDM} collaboration, \emph{{The design of the n2EDM experiment: nEDM Collaboration}}, \href{https://doi.org/10.1140/epjc/s10052-021-09298-z}{\emph{Eur. Phys. J. C} {\bfseries 81} (2021) 512} [\href{https://arxiv.org/abs/2101.08730}{{\ttfamily 2101.08730}}].

\bibitem{EuropwanEDMprojects:2025okn}
{\scshape Europwan EDM projects} collaboration, \emph{{Community input to the European Strategy on particle physics: Searches for Permanent Electric Dipole Moments}},  \href{https://arxiv.org/abs/2505.22281}{{\ttfamily 2505.22281}}.

\bibitem{nEDM:2019qgk}
{\scshape nEDM} collaboration, \emph{{A New Cryogenic Apparatus to Search for the Neutron Electric Dipole Moment}}, \href{https://doi.org/10.1088/1748-0221/14/11/P11017}{\emph{JINST} {\bfseries 14} (2019) P11017} [\href{https://arxiv.org/abs/1908.09937}{{\ttfamily 1908.09937}}].

\bibitem{TUCAN:2022koi}
{\scshape TUCAN} collaboration, \emph{{The Precision nEDM Measurement with UltraCold Neutrons at TRIUMF}}, \href{https://doi.org/10.7566/JPSCP.37.020701}{\emph{JPS Conf. Proc.} {\bfseries 37} (2022) 020701} [\href{https://arxiv.org/abs/2207.09880}{{\ttfamily 2207.09880}}].

\bibitem{Alexander:2022rmq}
J.~Alexander et~al., \emph{{The storage ring proton EDM experiment}},  \href{https://arxiv.org/abs/2205.00830}{{\ttfamily 2205.00830}}.

\bibitem{PhysRevD.19.2227}
V.~Baluni, \emph{$\mathrm{CP}$-nonconserving effects in quantum chromodynamics}, \href{https://doi.org/10.1103/PhysRevD.19.2227}{\emph{Phys. Rev. D} {\bfseries 19} (1979) 2227}.

\bibitem{Crewther:1979pi}
R.J.~Crewther, P.~Di~Vecchia, G.~Veneziano and E.~Witten, \emph{{Chiral Estimate of the Electric Dipole Moment of the Neutron in Quantum Chromodynamics}}, \href{https://doi.org/10.1016/0370-2693(79)90128-X}{\emph{Phys. Lett. B} {\bfseries 88} (1979) 123}.

\bibitem{pEDM:2022ytu}
{\scshape pEDM} collaboration, \emph{{The storage ring proton EDM experiment}},  \href{https://arxiv.org/abs/2205.00830}{{\ttfamily 2205.00830}}.

\bibitem{CPEDM:2019nwp}
{\scshape CPEDM} collaboration, \emph{{Storage ring to search for electricdipole moments of charged particles: Feasibility study}}, CERN, Geneva (6, 2021), \href{https://doi.org/10.23731/CYRM-2021-003}{10.23731/CYRM-2021-003}, [\href{https://arxiv.org/abs/1912.07881}{{\ttfamily 1912.07881}}].

\bibitem{FileviezPerez:2011pt}
P.~Fileviez~Perez and M.B.~Wise, \emph{{Breaking Local Baryon and Lepton Number at the TeV Scale}}, \href{https://doi.org/10.1007/JHEP08(2011)068}{\emph{JHEP} {\bfseries 08} (2011) 068} [\href{https://arxiv.org/abs/1106.0343}{{\ttfamily 1106.0343}}].

\bibitem{Duerr:2013dza}
M.~Duerr, P.~Fileviez~Perez and M.B.~Wise, \emph{{Gauge Theory for Baryon and Lepton Numbers with Leptoquarks}}, \href{https://doi.org/10.1103/PhysRevLett.110.231801}{\emph{Phys. Rev. Lett.} {\bfseries 110} (2013) 231801} [\href{https://arxiv.org/abs/1304.0576}{{\ttfamily 1304.0576}}].

\bibitem{Super-Kamiokande:2020wjk}
{\scshape Super-Kamiokande} collaboration, \emph{{Search for proton decay via $p\to e^+\pi^0$ and $p\to \mu^+\pi^0$ with an enlarged fiducial volume in Super-Kamiokande I-IV}}, \href{https://doi.org/10.1103/PhysRevD.102.112011}{\emph{Phys. Rev. D} {\bfseries 102} (2020) 112011} [\href{https://arxiv.org/abs/2010.16098}{{\ttfamily 2010.16098}}].

\bibitem{Casas:2001sr}
J.A.~Casas and A.~Ibarra, \emph{{Oscillating neutrinos and $\mu \to e, \gamma$}}, \href{https://doi.org/10.1016/S0550-3213(01)00475-8}{\emph{Nucl. Phys. B} {\bfseries 618} (2001) 171} [\href{https://arxiv.org/abs/hep-ph/0103065}{{\ttfamily hep-ph/0103065}}].

\bibitem{Eisele:2007ws}
M.-T.~Eisele, \emph{{Leptogenesis with many neutrinos}}, \href{https://doi.org/10.1103/PhysRevD.77.043510}{\emph{Phys. Rev. D} {\bfseries 77} (2008) 043510} [\href{https://arxiv.org/abs/0706.0200}{{\ttfamily 0706.0200}}].

\end{thebibliography}\endgroup

\end{document}